\documentclass[manuscript]{aastex62}
\usepackage{epsfig}
\usepackage{alltt,color}
\usepackage{graphicx}
\usepackage{epstopdf}
\usepackage{float}
\usepackage{palatino}
\usepackage{amsmath}
\usepackage{cases}
\usepackage{pifont}
\usepackage{txfonts}
\usepackage{multirow}

\makeatletter

\newcommand{\Rmnum}[1]{\expandafter\@slowromancap\romannumeral #1@}
\makeatother

\DeclareMathOperator\erfc{erfc}
\DeclareMathOperator\IFFT{IFFT}
\DeclareMathOperator\FFT{FFT}

\received{***}
\revised{***}
\accepted{***}
\submitjournal{ApJ}

\shorttitle{Wave excitation by energetic ring-beam electrons}
\shortauthors{Zhou et al.}

\begin{document}

\title{Wave excitation by energetic ring-distributed electron beams in the solar corona}

\correspondingauthor{X. Zhou}
\email{zhouxw@pmo.ac.cn}

\author[0000-0002-0786-7307]{X. Zhou}
\affil{Key Laboratory of Dark Matter and Space Astronomy, Purple Mountain Observatory, Chinese Academy of Sciences, Nanjing, 210034, China}
\affiliation{Max Planck Institute for Solar System Research, G\"{o}ttingen, Germany}

\author{P. A. Mu\~{n}oz}
\affiliation{Center for Astronomy and Astrophysics, Technical University Berlin, Berlin, Germany}

\author{J. B\"{u}chner}
\affiliation{Center for Astronomy and Astrophysics, Technical University Berlin, Berlin, Germany}
\affiliation{Max Planck Institute for Solar System Research, G\"{o}ttingen, Germany}

\author{S. Liu}
\affiliation{Key Laboratory of Dark Matter and Space Astronomy, Purple Mountain Observatory, Chinese Academy of Sciences, Nanjing, 210034, China}



\begin{abstract}
We analyzed properties of waves excited by mildly relativistic electron beams propagating
along magnetic field with a ring-shape perpendicular momentum distribution in neutral and
current-free solar coronal plasmas.
These plasmas are subject to both the beam and the electron cyclotron maser (ECM) instabilities
driven by the positive momentum gradient of the ring-beam electron distribution in the
directions parallel and perpendicular to the ambient magnetic field, respectively.
To explore the related kinetic processes self-consistently, 2.5-dimensional fully kinetic particle-in-cell (PIC) simulations
were carried out.
To quantify excited wave properties in different coronal conditions, we investigated the dependence
of their energy and polarization on the ring-beam electron density and magnetic field.
In general, electrostatic waves dominate the energetics of waves and nonlinear waves are ubiquitous.
In weakly magnetized plasmas, where the electron cyclotron frequency
$\omega_{ce}$ is lower than the electron plasma frequency $\omega_{pe}$,
it is difficult to produce escaping electromagnetic waves with frequency
$\omega > \omega_{pe}$ and small refractive index $|c k / \omega| < 1$ ($k$ and $c$ are
the wavenumber and the light speed, respectively).
Highly polarized and anisotropic escaping electromagnetic waves can, however, be effectively excited
in strongly magnetized plasmas with $\omega_{ce}/\omega_{pe} \geq 1$.
The anisotropy of the energy, circular polarization degree (CPD), and spectrogram of these escaping electromagnetic waves
strongly depend on the number density ratio of the ring-beam electrons to the background electrons.
In particular, their CPDs can vary from left-handed to right-handed with the decrease of the
ring-beam density, which may explain some observed properties of solar radio bursts (e.g., radio spikes)
from the solar corona.
\end{abstract}


\keywords{Beam instabilities ---  Electron cyclotron maser instability ---
Magnetic reconnection --- Particle in cell simulations --- Solar corona ---
Electromagnetic waves --- Solar radio bursts}


\section{Introduction}
\label{Introduction}

The solar corona consists of a very dynamic, hot and dilute magnetized plasma in which eruptive energy
and mass releases take place, such as solar flares and coronal mass ejections (CMEs).
In the course of these solar activity, energetic particles can be accelerated by
magnetic reconnection~\citep{zhou_etal_2015ApJ...815....6Z, zhou_etal_2016ApJ...827...94Z, Munoz&Buechner_2016PhPl...23j2103M, MunozBuechner2018},
shocks~\citep{Aschwanden_2002SSRv..101....1A, Benz_2008LRSP....5....1B, Chen_etal_2015Sci...350.1238C},
as well as turbulence~\citep{Petrosian&Liu_2004ApJ...610..550P, Fletcher&Hudson_2008ApJ...675.1645F, Vlahos&Cargill_2009tsp..book.....V}.
These energetic particles can produce electromagnetic emissions from radio to $\gamma$-ray wavelengths.
Of particular interests are solar radio bursts (SRBs) characterized by
(a) high brightness temperatures,
(b) short, eruptive time scales,
(c) narrow frequency bands and
(d) strong polarization.
These characteristics indicate that the SRBs are likely due to coherent emissions
of plasma waves caused by plasma instabilities in the solar corona plasma~\citep{Melrose_2017RvMPP...1....5M}.
Therefore, SRBs carry rich information of plasma dynamics in the solar
corona and they may be used to remotely study the related plasma processes.

SRBs can be classified into many types by their distinctive
structures in the dynamical spectrum (or spectrogram, \citealp{Wild_etal_1963ARA&A...1..291W}).
Among all types of the SRBs, spikes immediately attracted the attention of researchers since their first
detections~\citep{Elgaroy_1961ApNr....7..123E, droege1961, degroot1962} due to their particular observed properties:
extremely short duration (down to and probably even less than a few milliseconds, limited by the time resolution of
radio telescopes), narrow bandwidth ($< 1\%$), and
mostly X-mode dominated high degree of polarization (can be  $\approx 100\%$, see, e.g.,
\citealp{Fleishman&Melniko_1998PhyU...41.1157F} for a review and references therein).
Solar radios spikes are closely related to particle acceleration and
primary energy release processes in solar flares. They might provide direct information
on the finest structure of these energy release processes~\citep{Benz_1985SoPh...96..357B, Benz_1994SSRv...68..135B}.
To deduce energy release information of solar flares from observations
of solar radio spikes, one first needs a reliable generation mechanism for these solar radio spikes.
Since right-handed polarized X-mode waves are predominant in most cases of solar radio spikes,
electron cyclotron maser (ECM) emission~\citep{Twiss_1958AuJPh..11..564T, Schneider_1959PhRvL...2..504S, Gaponov59}
has been widely accepted as the most likely coherent emission mechanism for their
generation~\citep{Dulk_1985ARA&A..23..169D, Vlahos_1987SoPh..111..155V, Vlahos&Sprangle_1987ApJ...322..463V, Melrose_1994SSRv...68..159M, Fleishman&Melniko_1998PhyU...41.1157F}.

The so called ECM mechanism, proposed by~\citealp{Twiss_1958AuJPh..11..564T},
is a consequence of a linear ECM instability, where electromagnetic waves absorb the energy of
energetic electrons, i.e., negative absorption of waves by energetic electrons, via wave-particle
interactions~\citep[see][Chap. 3.2]{Melrose_2017RvMPP...1....5M}.
Besides solar radio spikes, the ECM mechanism was also applied to the generation of the
Earth's auroral kilometric radiation (AKR, \citealp{Wu&Lee_1979ApJ...230..621W, Lee&Wu_1980PhFl...23.1348L, Lee_etal_1980P&SS...28..703L, Strangeway_etal_2001PCEC...26..145S})
and Jupiter's decametric emission (DAM, \citealp{Goldreich&Julian_1969ApJ...157..869G}).
For the ECM mechanism to operate on remote radio emissions, first the electron cyclotron frequency $\omega_{ce}$
needs to be greater than the plasma frequency $\omega_{pe}$ in the generation sites of radio emissions, since strong
wave excitations by the ECM instability are mainly located around $\omega_{ce}$ and waves with frequencies below
local $\omega_{pe}$ cannot escape from a plasma directly, i.e., the escape condition~\citep{Melrose_2017RvMPP...1....5M}.
The condition $\omega_{ce} > \omega_{pe}$, however, implies high local Alfv\'{e}n velocities
$\sim 0.02 c$~\citep{Wu_etal_2014A&A...566A.138W}, which
cannot be easily satisfied within the standard model of the solar
atmosphere~\citep{Wild_1985srph.book....3W, WuChingSheng2012, Wu_2014PhPl...21f4506W, Wu_etal_2014A&A...566A.138W}.
\citealp{Wu_etal_2014A&A...566A.138W, Chen_etal_2017JGRA..122...35C} suggested
that the condition $\omega_{ce} > \omega_{pe}$ can be fulfilled if local density cavities
form, e.g., due to fluctuations in the ubiquitous Alfv\'{e}nic turbulence.
Such density cavities are, indeed, found recently along the path of the electron
beam propagating parallel to the low-density separatrices of strong-guide-field magnetic
reconnection via 3D fully kinetic particle-in-cell (PIC) simulations~\citep{Drake_etal_2003Sci...299..873D, Pritchett&Coroniti_2004JGRA..109.1220P, Munoz2018PhRvE}.
Observation by~\citealp{Regnier_2015A&A...581A...9R, Morosan_etal_2016A&A...589L...8M} also
demonstrated that the condition of $\omega_{ce} > \omega_{pe}$ can be satisfied within some areas of the solar corona,
such as the core of a large active region.

Furthermore, to trigger the ECM emission (i.e., the ECM instability), a positive gradient
is required in the electron momentum distribution perpendicular to the ambient magnetic field,
i.e., $\partial f / \partial u_{\perp} > 0$, where $f$ is the electron momentum distribution
and $u_{\perp}$ is the perpendicular momentum of electrons.
This property, called population inversion, drives the maser instability.
Possible momentum distributions with $\partial f / \partial u_{\perp} > 0$ include
ring distributions~\citep{Pritchett1984JGR....89.8957P, Vandas&Hellinger_2015PhPl...22f2107V},
loss-cone distributions~\citep{Wu&Lee_1979ApJ...230..621W, Tsang_1984PhFl...27.1659T}
or horseshoe distributions~\citep{Melrose&Wheatland_2016SoPh..291.3637M, Pritchett_etal_1999JGR...10410317P}.

$\partial f / \partial u_{\perp} > 0$ was obtained by considering particle acceleration in the
outflow region of magnetic reconnection, where cup-like momentum distributions are found~\citep{BuchnerKuska:1996b}.
Energetic particles' magnetic gradient drifts can also cause a redistribution of
the energy of parallel flowing beam particles to the perpendicular direction~\citep{zhou_etal_2015ApJ...815....6Z},
forming ring and gyro-phase restricted as well as crescent-shaped
momentum distributions in the perpendicular direction~\citep{Voitcu&Echim_2012PhPl...19b2903V, Voitcu&Echim_angeo-2018-102}.
\citealp{Vlahos&Sprangle_1987ApJ...322..463V, Vlahos_1987SoPh..111..155V} mentioned that interaction of
quasi-perpendicular shocks with the ambient solar coronal plasma might lead to formation of a ring
momentum distribution in the direction perpendicular to the ambient magnetic field.
A quasi-perpendicular shock related SRB event during a solar flare
was reported by~\citealp{Chen_etal_2015Sci...350.1238C}.
Additionally, by means of fully kinetic PIC simulations, it has been proved that
ring momentum distribution in the direction perpendicular to the ambient magnetic field  can indeed be produced during
magnetic reconnection~\citep{Shuster_etal_2014GeoRL..41.5389S, Bessho_etal_2014GeoRL..41.8688B, Shuster_etal_2015GeoRL..42.2586S}.
Moreover electron holes in the electron exhaust regions at the X-points of magnetic reconnection could also
provide $\partial f / \partial u_{\perp} > 0$ for the ECM emissions
\citep{Treumann_etal_2011AnGeo..29.1885T, Treumann_etal_2012AnGeo..30..119T, Treumann&Baumjohann_2017AnGeo..35..999T}.
Note that gradients in the parallel direction $\partial f / \partial u_{\parallel}$
can also drive ECM emissions. This requires, however, extremely anisotropic
electron momentum distributions, e.g.,
$(\Delta u_{\perp} / c)^{2} \geq \Delta u_{\parallel} / c$ for a bi-Maxwellian electron momentum distribution, where
$\Delta u_{\perp}$, $\Delta u_{\parallel}$ and $c$ are the perpendicular, parallel thermal momenta of electrons
and the speed of light, respectively~\citep{Melrose_1973AuJPh..26..229M, Melrose_2017RvMPP...1....5M}.
There is, however, no observational evidence for the existence of such
strong anisotropy in the solar corona.

On the other hand, energetic electrons always follow a beam momentum distribution in the
direction along the coronal magnetic field based on some high-energy phenomena in the solar corona
\citep{Cairns_etal_2018NatSR...8.1676C, Chen_etal_2015Sci...350.1238C, Chen_etal_2018ApJ...866...62C}, e.g.,
Type \Rmnum{3} SRBs, hard X-ray bursts, solar energetic particle (SEP) events.
3D fully kinetic PIC and test particle simulations have also shown that strongly energized
electron beams can be generated by guide-field magnetic
reconnection~\citep{Buechner:2018-Mercury, MunozBuechner2018, zhou_etal_2016ApJ...827...94Z}.
The beam momentum distribution is unstable to the beam instability
driven by free energies from electrons with a momentum distribution
$f$ containing $u_{\parallel} \cdot \partial f / \partial u_{\parallel} > 0$
\citep[see][]{Melrose_1986islp.book.....M, Gary_1993tspm.book.....G}.
The classical theory of plasma emission, suggested by~\citealp{Ginzburg&Zhelezniakov_1958SvA.....2..653G},
is based on this beam instability.

The plasma emission mechanism contains nonlinear three-wave interaction processes.
The theory starts with the excitation of electrostatic Langmuir waves ($L$) via the beam instability.
Then backward-directed Langmuir ($L'$) waves can be generated
via the electrostatic decay or induced backscattering of forward-directed Langmuir waves
by fluctuations of ions ($L \rightarrow L' \pm S$, where $S$ represents ion-acoustic
wave, see~\citealp{Umeda_2010JGRA..115.1204U}).
Electromagnetic decay or coalescence of $L$ and $S$ waves will lead to the fundamental electromagnetic
emission ($T_{\omega_{pe}}$) at the electron plasma frequency $\omega_{pe}$ ($L  \rightarrow T_{\omega_{pe}} \pm S$).
While the second harmonic electromagnetic emission ($T_{2\omega_{pe}}$) at $2 \omega_{pe}$
can be produced by the coupling of $L$ and $L'$ waves
($L + L' \rightarrow T_{2\omega_{pe}}$)~\citep{Karlicky&Barta2011IAUS..274..252K, Melrose_2017RvMPP...1....5M, Henri_etal_2019_doi:10.1029/2018JA025707}.
Generally, the classical plasma emission processes will lead to excitations of the
$L$, $L'$, $T_{\omega_{pe}}$, $T_{2\omega_{pe}}$ as well as $S$ waves due to the beam instability.
Recently, \citealp{Umeda_2010JGRA..115.1204U} proposed an alternative
mechanism for the generation of the $L'$ waves with two symmetric counter-propagating electrons beams, where
the $L$ and $L'$ waves can be directly induced by the forward and backward-propagating electron
beams, respectively~\citep{Ganse_etal_2012ApJ...751..145G, Ganse_etal_2012SoPh..280..551G, Thurgood&Tsiklauri_2015A&A...584A..83T}.
The plasma emission theory has been widely used to explain the formations of
Type \Rmnum{1},  \Rmnum{2} and \Rmnum{3} SRBs~\citep{Aschwanden2005psci.book.....A, Melrose_2017RvMPP...1....5M}.
In situ spacecraft observations of the interplanetary Type \Rmnum{3} SRB have also confirmed the plasma emission theory
~\citep{Lin_etal_1981ApJ...251..364L, Ergun_etal_1998ApJ...503..435E}.
By considering many propagation
effects (e.g., wave scattering, decreasing magnetic field strength, interplanetary shocks) during the
transportation of energetic electrons from the solar corona to the interplanetary medium (IPM),
momentum distributions of the energetic electrons in the solar corona should be quite different from
those in the IPM.

Both the beam and ECM instabilities (driven by free energies in $u_{\parallel} \cdot \partial f / \partial u_{\parallel} > 0$
and $\partial f / \partial u_{\perp} > 0$ distributions, respectively) have been invoked separately to explain
the coherent emission mechanism of different types of SRBs~\citep[see, e.g.,][for reviews of SRBs]{Aschwanden2005psci.book.....A, Melrose_2017RvMPP...1....5M}.
For a more general application to the microscopic emission processes in plasmas, here we will generalize
these two distributions and characterize the properties of emission processes due to both instabilities.
On the other hand, based on the above mentioned theoretical studies and numerical simulations, both free
energies with population inversion $u_{\parallel} \cdot \partial f/\partial u_{\parallel} > 0$ and
$\partial f /\partial u_{\perp} > 0$ in the electron momentum distribution and density cavity with
$\omega_{ce} > \omega_{pe}$ can be realized simultaneously
in the dynamically evolving fast solar magnetic reconnection events in the solar corona.

In this paper, via 2.5D fully kinetic PIC simulations, we investigate
properties of waves excited by mildly relativistic ring-beam electrons
in neutral and current-free solar coronal plasmas. In this system, the ring-beam electrons together
with protons and background electrons support the global charge and current
neutralities, respectively.
Many theories and observations have proved that the majority of the current induced
by energetic beam electrons can be rapidly compensated by the return
current of drifting background electron, e.g.,
\citealp{Brown_1984A&A...131L..11B, van_den_Oord_1990A&A...234..496V, Melrose_1990SoPh..130....3M, Khodachenko_etal_2009SSRv..149...83K}.
Note that non-zero net current in plasmas can not only introduce a current instability
\citep[see][]{Melrose_1986islp.book.....M, Matsumoto1993, Wu_etal_2014A&A...566A.138W, Chen_etal_2017JGRA..122...35C}
but also generate strong magnetic field and oscillations leading to a very complex plasma system
\citep{Henri_etal_2019_doi:10.1029/2018JA025707}.

Some parametric dependence of the wave excitations resulting from the
electron ring-beam momentum distribution
have been investigated by~\citealp{Lee_etal_2011PhPl...18i2110L}
utilizing 2.5D fully kinetic PIC simulations.
In particular, these authors explored the influence of the
average kinetic energy and pitch angle of the ring-beam electrons
on the wave excitations, keeping the density ratio of ring-beam and
background electrons fixed ($n_{rb}/n_{bg} = 1:19$) as well
as the frequency ratio $\omega_{ce}/\omega_{pe} = 5$.
In order to derive properties of waves generated
by energetic ring-beam electrons at different locations along the beam trajectory in the solar corona,
we utilize a 2.5D version of the fully kinetic PIC
code ACRONYM to explore the dependence of the nonlinear wave generation and
saturation by energetic ring-beam electrons on the frequency ratio of $\omega_{ce}/\omega_{pe}$ and
the density ratio of the ring-beam $n_{rb}$ to background $n_{bg}$ electrons with the average kinetic energy and
pitch angle of the ring-beam electrons fixed.

Compared with previous studies (e.g.,~\citealp{Pritchett1984JGR....89.8957P, Lee_etal_2009PhRvL.103j5101L, Lee_etal_2011PhPl...18i2110L}),
we have developed precise diagnostics to investigate the nonlinear evolution, saturation, and
anisotropy of different electromagnetic wave modes guided by the dispersion relations of a magnetized cold
plasma (see Sect.\ref{mode_energy_method}).
Since only electromagnetic waves with frequency $\omega > \omega_{pe}$ and small refractive index $|c k / \omega| < 1$
can escape from their generation sites and might be detected by remote detectors,
polarization, spectrogram and anisotropies of these escaping electromagnetic waves
are explored to compare with the ground-based observations
of solar radio spikes (see Sect.\ref{polarization_method}).

This paper is organized as follows:
after the introduction, we present the numerical simulation model in Sect.\ref{Numerical_Simulation}.
Sect.\ref{Results} contains the key results of this study
and in Sect.\ref{Conclusions} we draw our conclusions and discuss the
application of our results.

\section{Numerical Simulation}
\label{Numerical_Simulation}

\subsection{Model and Setup}
\label{setup}
Since the excitation and growth of waves driven by plasma instabilities are, in general, kinetic and nonlinear
processes, self-consistent kinetic simulations are required to investigate it. In these simulations, particle
interactions through the electromagnetic fields and effects of particles' motions
on the electromagnetic fields as well as nonlinear wave-wave, wave-particle interactions
can be correctly incorporated. Our tool of choice is a fully kinetic PIC code, which can model all
those processes from a first-principle approach.
In a fully kinetic PIC code, generally, the electromagnetic fields are calculated from the Maxwell's equations
with the charge and current densities by knowing the positions and velocities of all particles. And then
the particles move due to these electromagnetic fields from the Newton-Lorentz equation of motion and the
new positions and velocities of particles lead to a new state of the electromagnetic fields.
This step is equivalent to solve the effective Vlasov equation for the (numerical) particles.
These steps are repeated until the end of a simulation.
Therefore, the fully kinetic PIC algorithm solves the full set of the
Vlasov-Maxwell equations,~\citep[see, e.g.,][for reviews of the basic theories and applications of fully kinetic PIC codes]{Birdsall&Langdon_1991ppcs.book.....B, Tskhakaya_etal_2007CoPP...47..563T, Lapenta_2012JCoPh.231..795L, Vay&Godfrey_2014CRMec.342..610V}.

We performed this study with the fully kinetic PIC code --- ACRONYM
(http://plasma.nerd2nerd.org/, \citealp[][]{Kilian_etal_2017_PoP}),
a fully relativistic electromagnetic code tuned for the study of kinetic-scale plasma wave phenomena and
interactions in collisionless plasmas in a wide variety of physical
environments~\citep[see, e.g.,][]{Ganse_etal_2012ApJ...751..145G, Kempf_etal_2016A&A...585A.132K, Munoz&Buechner_2016PhPl...23j2103M, Schreiner_etal_2017ApJ...834..161S, Munoz2018PhRvE, Buechner:2018-Mercury}.
We use its version in two spatial dimensions and three dimensions (i.e., 2.5D) in momentum and
components of the electromagnetic fields.

The 2D simulation box contains $1024 \times 1024$ grid points in the $x-y$
plane. Periodic boundaries are applied in both directions for both fields and
particles.
In our simulations, to keep the global charge neutrality, three species of particles are
employed, one for the mildly relativistic ring-beam electrons and other two
species for the protons and the background electrons. For physically realistic
results, the proton-to-electron mass ratio has been chosen as the physical
$m_{p}/m_{e} = 1836$.
Initially, 2000 particles per cell are implemented to reduce the numerical noise
\citep{Hockney_1971JCoPh...8...19H, Dawson_1983RvMP...55..403D, Birdsall&Langdon_1991ppcs.book.....B}.
These particles are homogeneously distributed in the whole
simulation domain with a constant ambient magnetic field
$\vec{B_{0}} = B_{0} \vec{x}$ along the x-axis, since typical domain sizes modelled by
fully kinetic PIC simulations are much smaller than the typical length scale
of the density gradient in the solar corona.

The initial momentum distributions of all particles (both electrons and protons) are characterized in terms of momentum
per unit mass, $\vec{u} = \gamma \vec{v}$, where $\gamma = 1/\sqrt{1-v^2/c^2} = \sqrt{1+u^2/c^2}$.
Note that, hereafter, we will simply call "momentum per unit mass" as "momentum".
Correspondingly, the momentum distribution for the mildly relativistic ring-beam electrons
is \citep{Umeda_etal_2007JGRA..112.4212U, Lee_etal_2011PhPl...18i2110L, Kainer_etal_1996JGR...101..495K}:
\begin{eqnarray}
F_{rb}(u_{\parallel}, u_{\perp}) = F_{rb\parallel}(u_{\parallel}) F_{rb\perp}(u_{\perp})
\nonumber \\
F_{rb\parallel}(u_{\parallel}) = \displaystyle\frac{1}{\sqrt{2\pi} u_{th\parallel}} \exp\left[-\displaystyle\frac{(u_{\parallel}-u_{rb\parallel})^2}{2 u_{th\parallel}^2}\right]
\nonumber \\
F_{rb\perp}(u_{\perp}) = \displaystyle\frac{1}{2\pi u_{th\perp}^2 Q_{\perp}} \exp\left[-\displaystyle\frac{(u_{\perp}-u_{rb\perp})^2}{2 u_{th\perp}^2}\right]
\label{rb_dis}
\end{eqnarray}
where $u_{\parallel}, u_{\perp}$ are the particle momenta along and
perpendicular to the ambient magnetic field $\vec{B_{0}}$, respectively.
($u_{rb\parallel}$, $u_{rb\perp}$) and ($u_{th\parallel}$, $u_{th\perp}$) are their corresponding
bulk drift and thermal momenta ($v_{th\parallel} = u_{th\parallel} / \gamma_{th} =\sqrt{k_{B}T_{e,\parallel}/m_e}$
and $v_{th\perp}=u_{th\perp}/ \gamma_{th}=\sqrt{k_{B}T_{e,\perp}/m_e}$,
where $T_{e,\parallel}$ ($T_{e,\perp}$) is the parallel (perpendicular) electron temperature,
$k_{B}$ is the Boltzmann's constant and $\gamma_{th} = \sqrt{1+(u_{th\parallel}^2+u_{th\perp}^2)/c^2}$).
According to our simulations, the parallel and
perpendicular directions are along the x-axis and y-axis, respectively.
Considering the typical velocity of energetic beam electrons
\citep{Wild_etal_1959AuJPh..12..369W, Alvarez&Haddock_1973SoPh...29..197A, Suzuki&Dulk_1985srph.book..289S, Reid&Ratcliffe_2014RAA....14..773R}
and typical temperature in the solar corona, initially we take
$\gamma = \sqrt{1+(u_{rb\parallel}^2 + u_{rb\perp}^2)/c^2} = 1.2$
($\sim 100~keV$ and $\sqrt{u_{rb\parallel}^2 + u_{rb\perp}^2} = 0.67~c$) as the average
initial kinetic energy of the ring-beam electrons. Their averaged pitch
angle is $\phi_{0} =\tan^{-1}(u_{rb\perp} / u_{rb\parallel}) = 30^{\circ}$,
which indicates that the ring-beam electrons have more energies in the $\vec{B_{0}}$ parallel direction.
And $u_{th\parallel} = u_{th\perp} = u_{rbth} = u_{th} = 0.025 c$.
$Q_{\perp}$ in Eq.\eqref{rb_dis} is the normalization constant
\begin{equation}
Q_{\perp} = \exp\left[-\displaystyle\frac{u_{rb\perp}^2}{2 u_{th\perp}^2}\right] +
            \sqrt{\displaystyle\frac{\pi}{2}}\displaystyle\frac{u_{rb\perp}}{u_{th\perp}} \erfc\left[-\displaystyle\frac{u_{rb\perp}}{\sqrt{2}u_{th\perp}} \right]
\end{equation}

For a current-free system, all the background electrons drift in the opposite direction to that of the
ring-beam electrons with a momentum $u_{bg\parallel} = - u_{rb\parallel} n_{rb} /n_{bg}$, here $n_{rb}$
and $n_{bg}$ are the number density of the ring-beam and background electrons,
respectively~\citep{Karlicky&Barta_2009NPGeo..16..525K, Ganse_etal_2012SoPh..280..551G}.

The background electrons, hence, follow a drifting Maxwellian momentum
distribution with a thermal spread $u_{bgth} = 0.05 c$ along each dimension. The backward-drifting
background electrons, hence, also contain free energies $u_{\parallel} \cdot \partial f / \partial u_{\parallel} > 0$
for the beam instability.
Protons are used for the global charge neutrality. They are assumed to follow
an isotropic Maxwellian momentum distribution with the same temperature as the background electrons.
Note that protons are quite important for the generation of the ion-acoustic waves in the plasma emission
mechanism, so that we let them move freely, i.e., our simulations also solve the equations of motion for the
protons, even though they respond to electromagnetic forces at much larger timescales than electrons.

Due to the free energies provided by the ring-beam and backward-drifting
background electrons, both the beam and ECM instabilities can be driven due to the
positive gradients $u_{\parallel} \cdot \partial f / \partial u_{\parallel} > 0$ and $\partial f /\partial u_{\perp} > 0$
in the electron momentum distribution, respectively.
To distinguish contributions from the beam and ECM instabilities,
we also carry out simulations with only either a pure-beam or a pure-ring
momentum distribution for the energetic electrons
($u_{rb\perp} = 0$ or $u_{rb\parallel} = 0$, respectively), while other parameters are the
same as these ring-beam simulations.

In this study, all quantities are solved in real (spatial-temporal) space and
all simulations have the same spatial and time resolution.
In particular, the grid cell size is $\Delta x = \Delta y \simeq \lambda_{De}$, where $\lambda_{De} = u_{th}/ \omega_{pe}$
is the electron Debye length and $\omega_{pe}$ is the total electron plasma frequency,
i.e., $ \omega_{pe} = \sqrt{\omega_{prb}^2+\omega_{pbg}^2}  = \sqrt{4 \pi n_{t} e^2/m_{e}}$,
where $n_{t} = n_{rb} + n_{bg}$ and $e$ are the total electron number density and
charge of electrons, respectively. And $\omega_{prb}$ ($\omega_{pbg}$) is the
plasma frequency of the ring-beam (background) electrons.
The timestep in our simulations is determined by the inherent length and timescale requirements in
a fully kinetic PIC code, i.e., the Courant-Friedrichs-Lewy (CFL) condition for the speed of light $c$.
Correspondingly, our simulations can cover $|k_{x,y}/(\omega_{pe}/c)| < 92.2$ and $|\omega/ \omega_{pe}| < 12.3$
with resolutions $\Delta k_{x, y} = 0.18 \omega_{pe}/c$ and $\Delta \omega =  0.015 \omega_{pe}$, respectively,
in the wavevector-frequency ($\vec{k} - \omega$) space.

Variable parameters are $n_{rb}/n_{t}$ (being
equal to $5\%$, $10\%$ , $20\%$, $30\%$, $40\%$, $50\%$ with fixed
$\omega_{ce} / \omega_{pe} = 5$, see Sect.\ref{density_dependence}) and
$\omega_{ce}/\omega_{pe}$ (being equal to $0.2$, $0.3$, $0.5$, $1$, $2$, $3$
with fixed $n_{rb}/n_{t} = 5\%$, see Sect.\ref{frequency_dependence}).
Note that simulation with $n_{rb}/n_{t} = 5\%$ and $\omega_{ce} / \omega_{pe} = 5$
can be compared with the Case B in the study of~\citealp{Lee_etal_2011PhPl...18i2110L}.
The ambient magnetic field $\vec{B_{0}}$ is initialized based on the frequency ratio
$\omega_{ce} / \omega_{pe}$.
Note that the values of the beam to total density ratio are probably much higher than
those thought to exist in the solar corona, but they could be considered
appropriate for density cavities, where the background density drops considerably.
In addition, fully kinetic PIC simulations of magnetic reconnection tend to generate
electron beams, propagating through the low density separatrices, with similar density ratios
\citep{Munoz&Buechner_2016PhPl...23j2103M}.

Normalizations used throughout this paper are as follows:
$\omega_{norm} = 5.0 \omega_{pe}$ is the normalization of frequency.
Time, momentum and distance are normalized by $1/\omega_{norm}$, $c$ and $c/\omega_{norm}$, respectively.
$B_{norm}$ is the normalization of the electric and magnetic field strengths and corresponds
to the ambient magnetic field $\vec{B_{0}}$ for $\omega_{ce}/\omega_{pe} = 5.0$.
Energy is normalized by the magnetic field energy $\varepsilon_{norm}$ corresponding to a
homogeneous and uniform $B_{norm}$ in the whole simulation domain.

\subsection{Diagnostic Method}
\label{method}

\subsubsection{Energy in wave modes}
\label{mode_energy_method}

Since all quantities in our simulations are given in real (spatial-temporal) space, to
characterize the plasma waves, one applies fast Fourier transforms (FFTs) on the
electromagnetic fields over the space-time domain of the simulations.
Meanwhile, different wave modes are distinguished by their own dispersion
relation. To estimate the energy contained by different wave modes,
we should consider their dispersion relation.
As a simplification, we take the wave dispersion relations in the magnetized cold
plasma as an approximation~\citep[see, e.g.,][]{Andre1985JPlPh..33....1A, Melrose_1986islp.book.....M, Stix_1992wapl.book.....S},
despite the criteria for the validity of the cold plasma approximation~\citep[see][]{Melrose_1986islp.book.....M, Stupp2000MNRAS.311..251S} could not be always satisfied in our simulations.
For a numerical simulation, different from theoretical studies, these criteria are, however, difficult to
adopt since the effective electron temperature in simulated plasmas are quite inhomogeneous and dynamic.
Generally, the cold plasma dispersion relation constitutes a
good approximation to the full hot plasma dispersion relation
in many conditions~\citep{Chen_etal_2013JGRA..118.2185C}.

We also assume that energy spectral density of a wave mode $M$, i.e., $I_{M}(\vec{k}, \omega)$
follows a Gaussian frequency distribution around its dispersion
surface in the $\vec{k} - \omega$ space \citep{Comisel_etal_2013PhPl...20i0701C}:
\begin{equation}
I_{M}(\vec{k}, \omega) = \sum\limits_{m} \left|A_{m, M}(\vec{k}, \omega)\right|^2
\label{GF_1}
\end{equation}
\begin{equation}
A_{m, M}(\vec{k}, \omega) = A_{m}(\vec{k}, \omega) \cdot \left\{
                                          \displaystyle\frac{1}{\sqrt{2 \pi} \sigma}
                                          \exp\left[-\displaystyle\frac{\left(\omega - \omega_{Mcold}(\vec{k})\right)^2}{2 \sigma^2}\right]
                              \right\}^{1/2}
\label{GF_2}
\end{equation}

where $\omega_{Mcold}(\vec{k})$ denote the frequency of the wave mode $M$ at the given wavevector
$\vec{k}$ (dispersion relation) in the magnetized cold plasma approximation. $\sigma$ characterizes the frequency
broadening for the wave mode $M$ around its corresponding cold-plasma dispersion relation surface. Here we use
 $\sigma = 0.05 \omega_{norm}$%
for each wave mode as a simplification. And $m$ indicates different components of electromagnetic field.
We apply FFTs on the electromagnetic field components $a_{m}(x, y, t)$ over the entire
space and time domain of our simulations to get their fluctuations $A_{m}(\vec{k}, \omega) = \FFT_{x, y, t}[a_{m}(x, y, t)]$
in $\vec{k} - \omega$ space. And then Gaussian filter is applied on $A_{m}(\vec{k}, \omega)$ to get
the electromagnetic field component fluctuations of the wave mode $M$, i.e., $A_{m, M}(\vec{k}, \omega)$.
This method is hereinafter called Gaussian filter method.

For the (temporal) evolution of energy of the wave mode $M$, an inverse fast Fourier transform (IFFT) is implemented
on $A_{m, M}(\vec{k}, \omega)$ (Eq.\ref{GF_2}) in the frequency $\omega$ space.
An integration over the wavevector $\vec{k}$ space is applied on the IFFT results to get the energy evolution of the
the electromagnetic field component $m$, i.e.,
\begin{equation}
\varepsilon_{M}(t) = \sum\limits_{m} \sum\limits_{\vec{k}} \left|\IFFT_{\omega}\left[A_{m, M}(\vec{k}, \omega)\right]\right|^{2} \Delta\vec{k}
\end{equation}
For the study of the wave energy along different wave propagation directions, 
we integrate $I_{M}(\vec{k}, \omega)$ in the $\vec{k} - \omega$ space
only if $\cos \theta = k_{\parallel}/k$ is satisfied, where $\theta$ is the
pitch angle between the $\vec{k}$ and $\vec{B_{0}}$ and $k_{\parallel}$ is
the $\vec{B_{0}}$ parallel component of the wave vector $\vec{k}$, then
the energy of the wave mode M along the direction $\theta$:
\begin{equation}
\varepsilon_{M}(\theta) = \sum\limits_{\omega} \sum\limits_{\vec{k}} I_{M}(\vec{k}, \omega) \delta(k_{\parallel}/k - \cos \theta) \Delta \vec{k} \Delta \omega
\label{anisotropic_emission_Eq}
\end{equation}
where $\delta(\ast)$ is the Dirac delta function.

\subsubsection{Polarization}
\label{polarization_method}

To get the polarization of waves propagating along each direction, the polarization vector $\vec{e_{p}}$
is defined with respect to the wave propagation vector $\vec{k}$ in the $x-y$ plane
\citep{Melrose_1986islp.book.....M, Bittencourt_2004fopp.book.....B, Willes&Cairns_2000PhPl....7.3167W}:
\begin{equation}
\vec{e_{p}} = \left(
                  \begin{array}{lr}
                        \vec{e_{1}} = \vec{k}/ |k|
                        \\
                        \vec{e_{2}} = \vec{e_{3}} \times \vec{e_{1}}
                        \\
                        \vec{e_{3}} = \vec{e_{z}}
                  \end{array}
\right)
\label{polarization_vector}
\end{equation}
where $\vec{e_{z}} = \vec{e_{x}} \times \vec{e_{y}}$ is the unit vector in
the direction perpendicular to the $x-y$ plane.
To separate the left and right-handed polarized components
($E_{l}(\vec{k}, \omega), E_{r}(\vec{k}, \omega)$) of the transverse
electric fields, i.e., perpendicular components of $\vec{E}(\vec{k}, \omega)$ respecting to $\vec{k}$, a circular basis
($\vec{e_{l}}, \vec{e_{r}}$) is defined based on the
polarization vector $\vec{e_{p}}$ in Eq.\eqref{polarization_vector}:
\begin{equation}
\begin{array}{c}
      \vec{e_{l}} =
                  \left\{
                        \begin{array}{lr}
                              (\vec{e_{2}} + i \vec{e_{3}})/\sqrt{2}  \qquad  [\omega \cdot k_{\parallel} > 0 \quad or \quad (k_{\parallel} = 0 \quad and \quad \omega \cdot k_{\perp} > 0)]  \\
                              (\vec{e_{2}} - i \vec{e_{3}})/\sqrt{2}  \qquad  [\omega \cdot k_{\parallel} < 0 \quad or \quad  (k_{\parallel} = 0 \quad and \quad \omega \cdot k_{\perp} < 0)]
                        \end{array}
                  \right.
      \\
      \vec{e_{r}} =
                  \left\{
                        \begin{array}{lr}
                              (\vec{e_{2}} - i \vec{e_{3}})/\sqrt{2}  \qquad  [\omega \cdot k_{\parallel} > 0 \quad or \quad (k_{\parallel} = 0 \quad and \quad \omega \cdot k_{\perp} > 0)]       \\
                              (\vec{e_{2}} + i \vec{e_{3}})/\sqrt{2}  \qquad  [\omega\cdot k_{\parallel} < 0 \quad or \quad (k_{\parallel} = 0 \quad and \quad \omega \cdot k_{\perp} < 0)]
                        \end{array}
                  \right.
      \\
      E_{l} =  \vec{E}(\vec{k}, \omega) \cdot \vec{e_{l}}
      \qquad \qquad
      E_{r} = \vec{E}(\vec{k}, \omega) \cdot \vec{e_{r}}
\end{array}
\label{polarization_vector_2}
\end{equation}
where $\vec{E}(\vec{k}, \omega)$ is the electric field in the $\vec{k} - \omega$ space and it is obtained via the FFT.
With the definition of the Eq.\eqref{polarization_vector_2},
the polarization state of a wave also refers to the ambient magnetic
field~\citep{Stix_1962tpw..book.....S, Gary_1993tspm.book.....G}.
Hence, the right and left-handed polarized waves rotate in the same sense as an electron and a proton, respectively,
as far as they propagate along (either parallel or antiparallel to) the ambient magnetic field.
Also note that wave and its
polarization make no sense when $\omega = 0$ and/or $|k| = 0$. Hence the contribution of $\omega = 0$
and/or $|k| = 0$ to the polarization is not considered in our calculations.

Following the definition of the Stokes parameters
\citep{McMaster1954AmJPh..22..351M, Carozzi_etal_2001JGR...10621395C}, for a wave at a given time,
its circular polarization degree (CPD, $P$) can be calculated as:
\begin{equation}
      P = \displaystyle\frac{|E_{r}|^2 - |E_{l}|^2}{|E_{r}|^2 + |E_{l}|^2}
\label{circle_polarization_0}
\end{equation}
where the vertical bars $|*|$ indicate the amplitude of the respective quantity. 
In order to determine the CPD, $P$ in a plasma (with many waves) at a given time or a wave over a period,
instead of taking the average value of CPDs from different waves, we use:
\begin{equation}
      P = \displaystyle\frac{\langle|E_{r}|^2\rangle - \langle|E_{l}|^2\rangle}{\langle|E_{r}|^2\rangle + \langle|E_{l}|^2\rangle}
\label{circle_polarization}
\end{equation}
where the angle brackets $\langle*\rangle$ indicate the average value of each corresponding quantity.
Note that for different studies, averages are calculated in different spaces, i.e.,
(1) averages over the $\vec{k}$ space are considered for the evolution of the CPD,
(2) for CPD along a wave propagation direction $\theta$ respect to $\vec{B_{0}}$,
we take averages in both $\omega$ and $k_{\parallel}/k_{\perp} = \cos \theta$ spaces.
Definition in Eq.\eqref{circle_polarization}, hence, can give us a direct idea that
which polarization is energetically dominant.
We thus can verify that the polarization is circular with a right- or
left-hand sense according to $P > 0$ or $P < 0$, respectively.
A value of $P = 1$ $(-1)$ corresponds to fully right- (left-)hand circular
polarization and $P = 0$ indicates a linear polarization.

In our diagnostics, magnetic fields are used to determine
the energies of electromagnetic wave modes. Adopting magnetic fields
can automatically filter out electrostatic waves since an electrostatic wave does not contain
magnetic fluctuations.
For calculations related to polarization, however, electric fields are used.
Note that we will not investigate evolutions of the anisotropy
and CPD, since the whole time duration of our simulations
(dozens of microseconds) are much shorter
than the time resolutions of the remote detectors (more than milliseconds).

\section{Simulation Results}
\label{Results}

As mentioned in Sect.\ref{mode_energy_method}, the dispersion relation of magnetized cold plasmas
will be applied to identify which wave has been excited.
There are 5 different wave branches in the cold plasma limit.
While each mode branch (or surface) can be differently named for different frequencies and/or
wave propagation directions \citep{Andre1985JPlPh..33....1A}.
For an example, when waves propagate along the ambient magnetic field,
the X mode is usually called (right-handed polarized) R-mode , while
the O mode is associated with the (left-handed polarized) L-mode.
In this paper, however, we will simply call them as ion-cyclotron, whistler, slow extraordinary (Z), ordinary (O) as well as
fast extraordinary (X) modes from the low to high frequencies, respectively.
For the applied physical proton-to-electron mass ratio $m_{p}/m_{e} =
1836$, the frequencies in the ion-cyclotron branch are marginally resolved
in our calculations.
In the following, we will ignore the ion-cyclotron branch.

\subsection{$n_{rb}/n_{t}$ Dependence}
\label{density_dependence}

In this section,  dependence of excited wave properties on the number density ratio
between the ring-beam and total electrons $n_{rb}/n_{t}$ are discussed.
While the ratio between the electron cyclotron frequency $\omega_{ce}$
and the electron plasma frequency $\omega_{pe}$ is fixed $\omega_{ce} / \omega_{pe}= 5$.

Both beam and ECM instabilities can occur with the ring-beam momentum distribution.
Note that many instabilities can fit with the description of the beam instability, since
their free energy sources come from the drifting beam population, e.g.,
the reactive beam instability, kinetic Langmuir beam (or bump-in-tail) instability, whistler heat flux
instability, firehose instability, etc~\citep{Melrose_1986islp.book.....M, Gary_1993tspm.book.....G}.
With the initial setup in this study (see Sect.\ref{setup}), these instabilities may occur
at the different stages of the free energy release. We will not distinguish these instabilities and
call them simply as the beam instability in this study.

\subsubsection{Statistics of particles}
\label{particles}

\begin{figure*}[htbp]
      \begin{center}
            \includegraphics[width=1.0\textwidth]{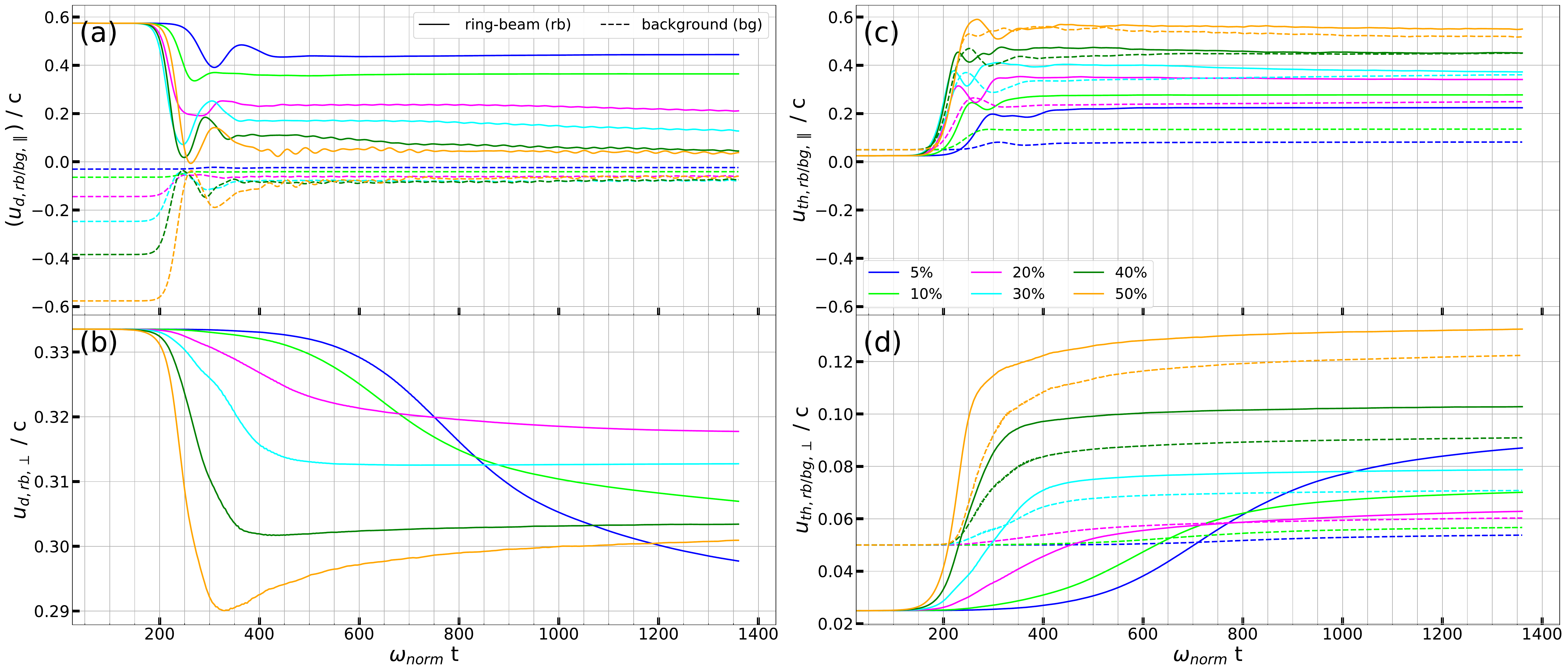}
            \caption{Evolutions of the bulk (or average, $u_{d}$) drift momenta and thermal spreads ($u_{th}$) in the
                            directions along ($//$, panel \textbf{a} and \textbf{c})
                            and perpendicular ($\perp$, panel \textbf{b} and \textbf{d}) to the
                            ambient magnetic field $\vec{B_{0}}$ for
                            both the ring-beam ($rb$, solid lines) and the background ($bg$, dashed lines) electrons, except for the
                            perpendicular bulk drift momenta of the background electrons ($u_{d, bg, \perp}$), which is close to 0.
                            In each panel, different colors are used to distinguish the different number density ratio
                            between the ring-beam and total electrons $n_{rb}/n_{t}$. Here $\omega_{ce}/\omega_{pe} = 5.0$ and all
                            momenta are normalized by the speed of light $c$.
                   }
          \label{average_velocity}
      \end{center}
      \end{figure*}

The beam and ECM instabilities, in general, are triggered by
the electron free energy in the directions along and perpendicular to the ambient
magnetic field $\vec{B_{0}}$, respectively.
Evolution of the electron momentum along each direction can, hence, give us insights on
the growth and saturation of these instabilities.
Panels (\textbf{a}) to (\textbf{d}) of Fig.\ref{average_velocity} show the evolution of the bulk
(or average) drift momenta and thermal spreads in the
directions along and perpendicular to $\vec{B_{0}}$ for
both the ring-beam and background electrons, respectively.
The bulk drift momentum and thermal spread of different electron species ($s$) along
different directions ($t$) are defined as $u_{d,s,t} = \left(\sum\limits_{i} u_{s,t,i}\right) / N$ and
$u_{th,s,t} =\sqrt{\left[\sum\limits_{i} (u_{s,t,i}^{2} - u_{d,s,t}^{2})\right]/N}$, where
$s = rb$ or $bg$ for the ring-beam or background electrons
and $t = \parallel$ or $\perp$ for the direction along or perpendicular to $\vec{B_{0}}$,  respectively.
While $u_{s,t,i}$ is the parallel or perpendicular momentum of a single electron $i$ and $N$ is the total electron
number in species $s$.
The evolution of the perpendicular bulk drift momentum of the background electrons is not shown in
panel (\textbf{b}) of Fig.\ref{average_velocity}, since it is negligible compared to that of the ring-beam population.
Note that we stopped our simulations when these quantities reach quasi-steady values, i.e.,
there is no obvious energy exchange between electromagnetic fields and particles.

Panels (\textbf{a}) and (\textbf{b}) of Fig.\ref{average_velocity} show the evolution of the bulk drift momenta
in the parallel and perpendicular directions, respectively. The first minima of these curves
indicate the saturation of their corresponding instabilities.
One can see that the reduction of the free energy, generally, is faster and larger in the parallel
than in the perpendicular direction.
The faster free energy reduction in the parallel direction implies that
waves excited by the beam instability will saturate earlier than those by the ECM instability.
(Note that, in this paper, the saturation of a wave corresponds to the end of the growth phase in its
energy evolution profile.)

While the free energy release rates of the beam instability do not vary significantly among
cases with dense ring-beam electrons ($n_{rb}/n_{t} \geq 20\%$, panel \textbf{a} of Fig.\ref{average_velocity}),
the free energy for the ECM instability decrease faster with the increase of the ring-beam
electron population (panel \textbf{b} of Fig.\ref{average_velocity}).
Moreover, the greater free energy reduction in the parallel direction implies that
waves induced by the beam instability should contain more energy
than those due to the ECM instability. This difference
becomes larger with the increase of $n_{rb}/n_{t}$.

Panel (\textbf{a}) of Fig.\ref{average_velocity} shows that both the ring-beam and background electrons
simultaneously lose their bulk drift energies along $\vec{B_{0}}$, making
contributions to the wave excitation driven by the beam instability.
Generally, in both the parallel and antiparallel directions,
the release of the electron drift energy increases monotonically
with the increase of  $n_{rb}/n_{t}$.
Evolution of the bulk drift momentum of the ring-beam electrons in the
direction perpendicular to $\vec{B_{0}}$  (panel \textbf{b} of Fig.\ref{average_velocity})
is, however, more complicated than the
parallel direction, i.e.,  the decrease of the perpendicular bulk drift momentum in the cases with
$n_{rb}/n_{t} = 5\%$ and $10\%$ are slower but even more significant
than the cases with $n_{rb}/n_{t} \geq 20\%$.
This indicates that different dynamic processes are underway between cases
with $n_{rb}/n_{t} \leq 10\%$ and $n_{rb}/n_{t} \geq 20\%$, which we will clarify later.

Part of the released energies from the bulk drift motion are, however, absorbed again by electrons
themselves via wave-particle interactions, leading to electron heating and acceleration.
Electron thermal spread is, hence,
strongly enhanced and have opposite behavior to their corresponding bulk drift motion
in both parallel and perpendicular directions, see panels (\textbf{c}) and (\textbf{d}) of Fig.\ref{average_velocity}.
Especially in the direction along $\vec{B_{0}}$ (panel \textbf{c}), the final thermal
spread of the ring-beam electrons already reach relativistic regime ($> 0.4 c$)
in the cases with $n_{rb}/n_{t} > 30\%$. For the case with $n_{rb}/n_{t} = 50\%$,
this is almost equal to its initial parallel drift momentum.
In the final quasi-steady state, the thermal spread of the background electrons is, in general, smaller than
that of the ring-beam electrons, and the thermal spread of all electrons is much wider in the parallel
direction than in the perpendicular direction, which agrees with the
distributions of the parallel and perpendicular momenta shown in Fig.\ref{velocity_energy_distribution}.

Fig.\ref{velocity_energy_distribution} shows the evolution of the parallel, perpendicular momentum
and energy distributions of all electrons. One can see that when the plasma system is close to
its quasi-steady state ($t = 1275\omega_{norm}^{-1}$, column \textbf{f}), the initial free energies
for the beam ($u_{\parallel} \cdot \partial f/\partial u_{\parallel} > 0$, row \textbf{a}) and
ECM ($\partial f / \partial u_{\perp} > 0$, row \textbf{b}) instabilities are almost totally dissipated and plateau momentum
distribution forms in all directions.
Meanwhile, with panels in rows (\textbf{a}) and (\textbf{b}), strong electron acceleration
can also be seen along each direction, particularly, in the cases with larger $n_{rb}/n_{t}$. Note that
the high momentum tail in the antiparallel direction contains reflected ring-beam electrons and
reflection of the ring-beam electrons is suppressed in cases with $n_{rb}/n_{t} \leq 10\%$.
The reflection of the ring-beam electrons makes the wave generation more
symmetric with respect to the plane perpendicular to $\vec{B_{0}}$.
\citealp{Petrosian&Liu_2004ApJ...610..550P} found that acceleration of particles
via resonant wave-particle interactions can be enhanced significantly if particle can
resonate with multi-waves simultaneously. Hence reflection of the
ring-beam electrons will increase their acceleration efficiency.
This may explain the correlation between
the ring-beam electron reflection and their acceleration
in the perpendicular direction.
Strong perpendicular acceleration also leads to the late increase of perpendicular drift momentum in
the case with the maximum ring-beam electron density ($n_{rb}/n_{t} = 50\%$ in panel \textbf{b}
of Fig.\ref{average_velocity}).
Interestingly, in each $n_{rb}/n_{t}$ case, a double power-law distribution forms
in the high energy tail with $\gamma-1 > 0.1 \sim 50~keV$ when the plasma system is close to
its quasi-steady state, i.e., after the release of the free energy for both the beam and ECM
instabilities (see the right-bottom panel \textbf{c-f}). These break energies are located around
the initial energy ($\gamma \sim 1.2$) of the ring-beam electrons.

\subsubsection{Excited electrostatic waves}
\label{Excited_electrostatic_waves}

Based on the coordinates of our simulations, electric component $E_{z}$ is purely
transverse, while the character of the $E_{x}$ and $E_{y}$ components change with wave
propagation direction, i.e., $E_{x}$ is a purely longitudinal
(transverse) component when waves propagate along (perpendicular to)
the ambient magnetic field $\vec{B_{0}}$, i.e., $\vec{k} \parallel \vec{B_{0}}$
and $\theta = 0^{\circ}$ or $180^{\circ}$ ($\vec{k} \perp \vec{B_{0}}$ and $\theta = 90^{\circ}$ or $270^{\circ}$).
But, in general, $E_{x}$ and $E_{y}$ represent a mixture of both longitudinal and transverse
electric field components.
Note that $\theta \leq 90^{\circ}$ together with the sign of $\pm k$
gives two supplementary wave propagation directions
in the wave  $\vec{k} - \omega$ (or dispersion) spectra of Figs.\ref{Electrostatic_Wave}, \ref{dispersion},
\ref{Polarization} and \ref{Dispersion_Relation_2}.

Hence, in row (\textbf{a}) of Fig.\ref{Electrostatic_Wave},
one can mainly find excited electrostatic modes, i.e., Langmuir and
(electron) beam modes. Hereinafter, we define a wave mode being excited
if its spectral intensity is significantly higher than that of an isotropic equilibrium Maxwellian
plasma, which has the same thermal spread and $\omega_{pe}$ as the background
and total electrons in the ring-beam simulations (see Sect.\ref{setup}), respectively, corresponding to
$n_{rb}/n_{t} = 0\%$ in Figs.\ref{energy_profile} and \ref{Emission_Symmetry}.
Since the ECM instability, in general, mainly excites electromagnetic
modes, the excitation of the electrostatic Langmuir and beam modes should be mostly due to
the beam instability.

Similar to~\citealp{Karlicky&Barta_2009NPGeo..16..525K, Ganse_etal_2012SoPh..280..551G},
antiparallel-propagating Langmuir waves ($L'$) are also excited in all ring-beam plasmas
(see left-half panels in row \textbf{a} of Fig.\ref{Electrostatic_Wave}, where $\theta = 0^{\circ}$ and $k < 0$).
As mentioned in Sect.\ref{Introduction}, generation of the $L'$ waves could be attributed to
electrostatic decay of the parallel-propagating Langmuir waves ($L \rightarrow L' + S$) and/or
free energies for the beam instability in the antiparallel-drifting background electrons.
In our simulations, we indeed found both intensity enhancement in the ion density fluctuation spectra
for plasmas with dense ring-beam electrons $n_{rb}/n_{t} \geq 20\%$
(similar to Fig.4 of~\citealp{Thurgood&Tsiklauri_2015A&A...584A..83T}) as well as reduction of the free energy
in the antiparallel-drifting background electrons (see row \textbf{a} of Fig.\ref{velocity_energy_distribution}).

Furthermore, one can see the rise of these excited $L'$ branch toward larger
$\omega$ at a given wavenumber $k<0$ with the increase of  $n_{rb}/n_{t}$, which
agrees with a higher effective electron temperature of the antiparallel-moving electrons
in plasmas with a larger $n_{rb}/n_{t}$ (panel \textbf{c} of Fig.\ref{average_velocity}),
since higher electron temperature will lead to a larger slope ($d\omega/dk$) in the dispersion
relation of the Langmuir wave, i.e.,
$\omega^2 = \omega_{pe}^2 + 3 k^2 v_{the}^{2}$, where $v_{the} \propto \sqrt{T_{e}}$ is the
effective thermal velocity of electrons.
Correspondingly, excitation of the $L'$ waves appears at increasingly smaller
wavenumbers $|k|$. That could be due to electron Landau
damping~\citep{Landau1946On, Tsurutani&Lakhina1997RvGeo..35..491T}
of longitudinal electric fluctuations with large $|k|$ in hot plasmas, where
Langmuir waves will get damped when their $|k|$ become larger than
$1/\lambda_{De}$ ($\lambda_{De} \propto \sqrt{T_{e}}$ and $T_{e}$ is the effective electron temperature).

In the direction antiparallel to $\vec{B_{0}}$,
except for the $L'$ wave, intensity of the beam mode $\omega = k v_{b} < \omega_{pe}$ is also enhanced.
The typical drift velocity $v_{b}$ of these excited antiparallel-propagating beam modes also increases with
the increase of $n_{rb}/n_{t}$, since initially we have $u_{bg\parallel} = - u_{rb\parallel} n_{rb} / n_{bg}$, i.e.,
the initial bulk drift momentum of the background electrons $u_{bg\parallel}$ increases
with the increase of $n_{rb}/n_{t}$ (see dashed lines at $t=0$ in panel
\textbf{a} of Fig.\ref{average_velocity}).

In the direction parallel to $\vec{B_{0}}$ ($\theta = 0^{\circ}$ with $k>0$, right-half panels in row
\textbf{a} of Fig.\ref{Electrostatic_Wave}), similar to the conditions in the antiparallel direction,
enhanced intensity of both the parallel-propagating Langmuir ($L$) and beam modes can be found
in each $n_{rb}/n_{t}$ case.
And also, due to the Landau damping in hot plasmas, these excited $L$ waves are confined
to smaller $|k|$ with the increase of $n_{rb}/n_{t}$.
Landau damping of small scale longitudinal electric fluctuations will lead to electron heating 
discussed in Sect.\ref{particles}.
While different from those antiparallel-propagating beam modes, the typical drift velocity $v_{b}$ of the excited
parallel-propagating beam modes, however,
decrease with the increase of $n_{rb}/n_{t}$ due to the stronger
reduction of the parallel bulk drift energy of the ring-beam electrons in
cases with larger $n_{rb}/n_{t}$ (see solid lines in panel
\textbf{a} of Fig.\ref{average_velocity}).

By comparing these electrostatic waves in the $\theta = 0^{\circ}$ and $180^{\circ}$ directions
(row \textbf{a} of Fig.\ref{Electrostatic_Wave}),
one can see that intensity of the parallel-propagating Langmuir and beam modes are, generally, stronger than
the antiparallel-propagating ones for each $n_{rb}/n_{t}$ case.
This difference is, however, reduced with the increase of $n_{rb}/n_{t}$,
since the free energy for the beam instability from the antiparallel-moving
background electrons becomes comparable to that from the
parallel-moving ring-beam electrons with the increase of $n_{rb}/n_{t}$.
Moreover, for high ring-beam density cases with $n_{rb}/n_{t} \geq 20\%$, a significant
fraction of the ring-beam electrons can be reflected, which also makes the wave excitation
more symmetric with respect to the perpendicular plane.
Additionally, besides linear waves as indicated by the cold plasma dispersion relations, diffusive nonlinear
electrostatic waves are also excited.

Panel (\textbf{b}) of Fig.\ref{Electrostatic_Wave} shows energy evolutions of the total electric fields
($\sum\limits_{x, y}\left[\vec{E}(x, y)\right]^{2} \Delta x \Delta y$)
and total longitudinal electric fields
($\sum\limits_{\vec{k}}\left[(\vec{E}(\vec{k}) \cdot \vec{k})/|k|\right]^{2} \Delta\vec{k}$) of all waves in the simulation domain,
where $\vec{E}(\vec{k})$ is the electric field vector of waves with wavevector $\vec{k}$.
Note that the total longitudinal electric fields contain not only electric fields of electrostatic waves but also the longitudinal
electric component of the electromagnetic waves.
One can see that, for each $n_{rb}/n_{t}$ case, the total longitudinal electric component
occupies most of the total electric field energy (their energy evolution profiles
are almost overlapped with each other): it is over one order of magnitude larger than
the transverse electric field energy, which is shown in panel (\textbf{e})
of Fig.\ref{Polarization}. Similar result had also been found by~\citealp{Lee_etal_2009PhRvL.103j5101L}.

Energy evolutions for the electric component $E_{x}(x, y)$
($\sum\limits_{x, y}\left[E_{x}(x, y)\right]^{2} \Delta x \Delta y$) and the
longitudinal electric fields of waves propagating along $\vec{B_{0}}$
($\sum\limits_{k_{\perp} = 0}\left[\left(\vec{E}(\vec{k}) \cdot \vec{k}\right)/|k|\right]^2$, i.e., waves
shown in row \textbf{a} of Fig.\ref{Electrostatic_Wave}) are presented in panel (\textbf{c}) of Fig.\ref{Electrostatic_Wave}.
One can see that the energy of $E_{x}(x, y)$ is comparable to the total longitudinal electric fields. And the energy of
the $\vec{B_{0}}$ aligned longitudinal electric fields is a factor of a few lower except in the early phase of these
simulations, when the energy of the total longitudinal electric fields appears to be isotropic (especially at $t = 0$)
and dominated by nonparallel-propagating waves.
To validate our simulations, insert of panel (\textbf{c}), we compare
the growth rate of these $\vec{B_{0}}$ aligned longitudinal electric fields
to the maximum growth rate of electrostatic waves by the beam instability in the nonresonant fluid or
reactive regime of unmagnetized plasmas (i.e., in cold unmagnetized plasmas).
These two growth rates should agree with each other,
since the $\vec{B_{0}}$ aligned longitudinal electric fields are
dominated by electrostatic waves (see row \textbf{a} of Fig.\ref{Electrostatic_Wave}) and
these electrostatic waves are mainly excited by the beam instability.
Moreover in magnetized plasmas,
excitation of the electrostatic waves propagating along the ambient magnetic field is exactly the same as that in
unmagnetized plasmas~\citep[see][Chap. 3.3]{Gary_1993tspm.book.....G}.
Additionally, the setup of our simulations are also located in the reactive regime
with $(n_{rb}/n_{t})^{1/3} (u_{rb\parallel}/u_{th\parallel}) \gg 1$, where
$(n_{rb}/n_{t})^{1/3} (u_{rb\parallel}/u_{th\parallel})$ is a measure of the reactive ($\geq 1$) and kinetic ($< 1$)
nature of the beam instability~\citep{Melrose_1986islp.book.....M, Gary_1993tspm.book.....G, Melrose_2017RvMPP...1....5M}.

The maximum growth rate of the electrostatic waves due to the reactive beam instability was
obtained from the dispersion equation for unmagnetized plasmas by setting the longitudinal dielectric
element $K^{L}(\omega, \vec{k})$ to be zero. In the unmagnetized cold plasma limit ($u_{rb\parallel} \gg u_{th\parallel}$,
Eq.2.16 in \citealp{Melrose_1986islp.book.....M}), that is :
\begin{equation}
      K^{L}(\omega, \vec{k}) = 1 - \Sigma_{s} \displaystyle\frac{q^2 n_{s}}{\varepsilon_{0} \gamma_{s} m_{s} \omega^{2}}\left[
                                                      1 + \displaystyle\frac{2\vec{k} \cdot \overrightarrow{v_{ds}}}{\omega - \vec{k} \cdot \overrightarrow{v_{ds}}}
                                                      + \displaystyle\frac{1 - \omega^{2}/(k^{2} c^2)}{(\omega - \vec{k} \cdot \overrightarrow{v_{ds}})^{2}}
                                                         (\vec{k} \cdot \overrightarrow{v_{ds}})^2
                                                      \right]
                                                  = 0
\label{growth_rate_two-stream_instability}
\end{equation}
where $\Sigma_{s}$ is for a summing over all particle species ($s$) in plasma and $v_{ds}$,
$\gamma_{s} = (1 - v_{ds}^{2}/c^{2})^{-1/2}$ are the bulk drift velocity and its
corresponding gamma factor of particle species $s$, respectively. When $n_{rb} \ll n_{t}$, one can get the growth rate of the classical
weak-beam instability from Eq.\eqref{growth_rate_two-stream_instability}:
$\Gamma/\omega_{pe} = \sqrt{3} (n_{rb}/n_{bg})^{1/3}/2^{4/3}$ (see Eq.3.2.9 in~\citealp{Gary_1993tspm.book.....G}).
Growth rate of the electrostatic waves propagating along $\vec{B_{0}}$ in our simulations
is evaluated via a linear fit in the range indicated by "o" and "x" points in panel (\textbf{e}).
One can see that the growth rate of the electrostatic waves propagating along $\vec{B_{0}}$ is
generally slightly smaller than the theoretical maximum ones by the reactive beam instability.
Similar results were also found in the study of
\citealp{Karlicky&Barta_2009NPGeo..16..525K}.
On the one hand, this could be due to the free energy reduction of the energetic ring-beam electrons, i.e.,
effective $n_{rb}$ and $u_{rb\parallel}$ for the reactive beam instability will
decrease with the wave excitation. Small effective $n_{rb}$ and $u_{rb\parallel}$ leads to
smaller growth rate for the reactive beam instability~\citep{Gary_1993tspm.book.....G}.
On the other hand, particles can simultaneously absorb some waves during the wave excitations
(i.e., plasma heating by Landau damping, see panel \textbf{c} of Fig.\ref{average_velocity}),
and increased electron momentum spread (or temperature) can also reduce the growth
rate of the reactive beam instability~\citep[see Sect.3.4 in][]{Melrose_1986islp.book.....M}.
Additionally, not all electrostatic waves propagating along $\vec{B_{0}}$ grow with the theoretical
maximum rate of the reactive beam instability.
Generally, values of the growth rate of the electrostatic waves propagating along $\vec{B_{0}}$
are quite similar to those of the whistler mode (panel \textbf{e} of Fig.\ref{energy_profile}),
which is also consistent with the study of~\citealp{Lee_etal_2011PhPl...18i2110L}.

\subsubsection{Excited electromagnetic waves}
\label{Excited_electomagnetic_waves}

Fig.\ref{dispersion} shows the electric field dispersion spectra of electromagnetic waves
along different wave propagation directions $\theta$ ($= 0^{\circ}$ or  $20^{\circ}$ or $90^{\circ}$).
Due to the rotation symmetry in the direction perpendicular to the ambient magnetic field $\vec{B_{0}}$,
dispersion spectra of $E_{y}$ and $E_{z}$ are very similar for waves with $k_{\perp} = 0$ (row \textbf{a})
and the dispersion spectra of waves with $k_{\parallel} = 0$ are symmetric with respect to $k_{\perp} = 0$
(rows \textbf{c} to \textbf{e}).

Excitation of all electromagnetic whistler, Z, O and X modes can be found
in the purely transverse electric component $E_{z}$ spectra
along $\vec{B_{0}}$ (row \textbf{a} of Fig.\ref{dispersion}).
Similar to the electrostatic component $E_{x}$ along $\vec{B_{0}}$ (row \textbf{a} of Fig.\ref{Electrostatic_Wave}),
there is also an asymmetry on the transverse electric intensity of waves
oppositely propagating along $\vec{B_{0}}$, especially for plasmas with tenuous ring-beam electrons.
It is interesting to note that the X (Z) mode dominates the transverse electric field spectra in the
direction parallel (antiparallel) to $\vec{B_{0}}$ with tenuous ring-beam electrons. Between the excited X and Z modes,
there are also diffusive nonlinear waves that don't follow the dispersion relations of
the linear waves in the cold plasma limit.
With the increase of $n_{rb}/n_{t}$, excitations of the X and Z modes as well as the whistler and O modes
become more and more symmetric with respect to the $\vec{B_{0}}$ perpendicular plane.
The enhanced excitation of the antiparallel-propagating X mode in plasmas with dense ring-beam electrons
may be caused by reflected ring-beam electrons (see row \textbf{a} of Fig.\ref{velocity_energy_distribution}),
while the intensity of the parallel-propagating X-mode waves appears to be saturated.
Along $\vec{B_{0}}$, the excitation of the whistler, Z and O modes are inefficient
for plasmas with tenuous ring-beam electrons in contrast to the dense ring-beam cases.

For obliquely propagating electromagnetic waves with $\theta = 20^{\circ}$ and $160^{\circ}$
(rows \textbf{b} of Fig.\ref{dispersion}), their transverse electric component $E_{z}$ spectra
have similar properties to those of the parallel and antiparallel-propagating electromagnetic waves (in row \textbf{a}).
$E_{z}$ intensity of these excited obliquely propagating electromagnetic waves are, however, enhanced comparing to
those of the $\vec{B_{0}}$ aligned electromagnetic waves, especially in the whistler and Z modes.

Rows (\textbf{c}) to (\textbf{e}) of Fig.\ref{dispersion} show the dispersion spectra of
the electric components $E_{x}$, $E_{y}$ and $E_{z}$, respectively, for perpendicular
propagating electromagnetic waves.
The whistler branch is absent in these panels, since its resonance or maximum frequency
$\omega_{W}^{res} \rightarrow 0$ at $\theta = 90^{\circ}$ and $270^{\circ}$ with the
physical proton-to-electron mass ratio $m_{p}/m_{e} = 1836$ in the magnetized cold plasma limit
\citep{Melrose_1986islp.book.....M, Stix_1992wapl.book.....S}. %
It is well known that the electric field of the O (Z and X) mode is parallel (perpendicular) to $\vec{B_{0}}$,
when they propagate in the direction perpendicular to $\vec{B_{0}}$, i.e., $\vec{k} \perp \vec{B_{0}}$.
In row (\textbf{c}), hence, one can
find a strong O-mode excitation. The strong Z and X-mode excitations, on the other hand, appear
in the $E_{y}$ and $E_{z}$ spectra (rows \textbf{d} and \textbf{e}).
The transverse electric component $E_{z}$ of the Z and X modes contains more energies than their
longitudinal electric component $E_{y}$.
Generally, similar to the electrostatic modes (row \textbf{a} of Fig.\ref{Electrostatic_Wave}),
intensity of the O, Z and X modes also increase with
the increase of  $n_{rb}/n_{t}$ in the plane perpendicular to $\vec{B_{0}}$.

In rows (\textbf{d}) and (\textbf{e}) of Fig.\ref{dispersion}, the X mode appears to be enhanced just below the
second harmonic of $\omega_{ce}$ and there is, additionally,
an excited horizontal band located around $\omega_{ce}$ and below the cutoff frequency of the X mode
$\omega_{X}^{cut} = (\omega_{ce}+\sqrt{\omega_{ce}^{2} + 4 \omega_{pe}^{2}})/2 \approx 1.04 \omega_{norm}$.
Following \citealp{Pritchett1984JGR....89.8957P}, we call this horizontal band as (electromagnetic)
relativistic Bernstein mode. This mode result from the relativistic corrections to the
classical dispersion of the magnetized cold plasma approximation, see \citealp{Pritchett1984JGR....89.8957P} for more details.
This relativistic Bernstein mode is, however, evident only in the cases with $n_{rb}/n_{t} = 5\% - 30\%$ and
the central frequency of this excited relativistic Bernstein mode increases with the increase of
$n_{rb}/n_{t}$.
The absence of this horizontal mode in the cases with
$n_{rb}/n_{t} > 30\%$ could be due to merging of the X and Bernstein modes.
Based on Fig.1 in \citealp{Pritchett1984JGR....89.8957P}, one can see that
the cutoff frequencies of the X and Bernstein modes can be the same and above $\omega_{ce}$ in
plasmas with electron temperature above $0.1~c$. In other words,
the merging of the X and Bernstein modes indicates that the effective electron temperature
could be higher than $0.1~c$ (i.e., efficient heating occurs) in the cases with $n_{rb}/n_{t} > 30\%$.

\subsubsection{Electromagnetic wave energy}
\label{energy_result}

Since solar radio emissions are electromagnetic waves,
we will mainly concentrate on properties of the electromagnetic whistler,
Z, O and X modes in the following.
Sect.\ref{Excited_electomagnetic_waves} shows that these four mode branches dominate the excited
electromagnetic waves and they roughly follow the dispersion relations of a magnetized cold plasma.
Following the Gaussian filter method described in Sect.\ref{mode_energy_method}, we
extract the magnetic energy carried by each branch of these four electromagnetic modes.

Panels (\textbf{a}) to (\textbf{d}) of Fig.\ref{energy_profile}
show the evolution of the magnetic energy of the whistler, X, Z and O modes, respectively,
where the case with $n_{rb}/n_{t} = 0\%$,  an equivalent
isotropic thermal plasma, shows how much these whistler, Z, O and X-mode waves are
enhanced with respect to their corresponding thermal levels.
Note that, in numerical simulations, the whistler, Z, O and X-mode waves can also be seen
in isotropic thermal plasmas without source of free energy,
which is due to the thermal noise numerically enhanced by the finite number of
macroparticles \citep{Kilian_etal_2017_PoP}.
Moreover, in Fig.\ref{energy_profile}, the solid (dashed) lines represent the plasmas with a ring-beam (pure-beam)
momentum distribution of energetic electrons.
Differences in the magnetic energy evolution between the ring-beam and the associated pure-beam
momentum distributions are used to assess the effects of the ring feature in the ring-beam
momentum distribution.

By comparing the magnetic energy evolution between the ring-beam and the equivalent isotropic thermal plasmas,
one can see that the saturation of each mode is over three orders of magnitude larger than
their corresponding thermal levels.
However, these magnetic energies are about one order of magnitude lower than
the electric energy of the electrostatic waves propagating along $\vec{B_{0}}$ shown
in panel (\textbf{c}) of Fig.\ref{Electrostatic_Wave}.
All excited waves in the simulation domain are therefore dominated by electrostatic waves.
Comparing the results of the ring-beam and pure-beam simulations, one can see that
while the whistler mode appears to be mostly driven by the beam instability,
the growth of the Z, O and X modes have two components, especially in plasmas with
tenuous ring-beam electrons.
Moreover, in ring-beam plasmas, the onset of the Z, O, X-mode growth
appear to be dominated by the ECM instability. The beam instability has delayed
contributions to the excitations of the Z, O, X modes. For each wave mode,
this delay decreases with the increase of the ring-beam electron density.
In a plasma, this delay appears to increase with the increase of wave frequency.
The beam instability, hence, tends to excite low frequency waves first.

The saturation of the Z mode is, however, dominated by the beam instability in ring-beam plasmas.
The same is true for O modes with dense ring-beam electrons ($n_{rb}/n_{t} \geq 20\%$).
For $n_{rb}/n_{t} \leq 10\%$, the O-mode saturation is governed by the ECM instability and similar to
the saturation time of the X mode, which corresponds well to the slow dissipation rate
of the free energies in the direction perpendicular to the ambient magnetic field $\vec{B_{0}}$ (see panel \textbf{b}
of Fig.\ref{average_velocity}) in the cases with $n_{rb}/n_{t} \leq 10\%$.
For the X mode, as predicted by the classical plasma emission
theory, pure-beam distribution alone cannot lead to an efficient excitation of the X mode in
plasmas with tenuous pure-beam electrons.  On the other hand, with dense pure-beam electrons (e.g., $n_{pb}/n_{t} = 50\%$,
here $n_{pb}$ is the number density of the pure-beam electrons),
generation of the X mode can also saturate at a quite high energy.

Note that the magnetic energy envelope of the X mode in the case
with ring-beam electrons $n_{rb}/n_{t} = 5\%$ and $\omega_{ce} / \omega_{pe} = 5$ (solid blue line in panel
\textbf{b}) is almost the same as those (panel \textbf{i} of Fig.4) in the study
of~\citealp{Lee_etal_2011PhPl...18i2110L}.

Generally, the magnetic energy saturation of the whistler, Z and O modes are enhanced with increasing
ring-beam electron population (i.e., larger $n_{rb}/n_{t}$).
But for the X mode with the ring-beam momentum distribution,
its magnetic energy saturation in the cases with $n_{rb}/n_{t} = 5\%$ and $10\%$ are not the smallest ones,
since the free energies released from the perpendicular bulk drift momenta are, correspondingly, not the
least in these two cases (see panel \textbf{b} of Fig.\ref{average_velocity}).
Furthermore, in the ring-beam plasmas with the same $n_{rb}/n_{t}$, the whistler mode has
larger magnetic energy saturation than the other three (Z, O and X) modes.
Saturation of the Z, O and X-mode waves
decrease in that order when $n_{rb}/n_{t} \geq 20\%$, which implies that wave excitation
is more efficient at lower frequencies in plasmas with dense ring-beam electrons.
For $n_{rb}/n_{t}  \leq 10\%$, the X mode can, however, has a larger saturation than the Z and
O modes and the saturation of the Z mode becomes the least among the Z, O and X modes
due to contributions from the ECM instability.

\begin{figure*}[htbp]
      \begin{center}
         \includegraphics[width=1.0\textwidth]{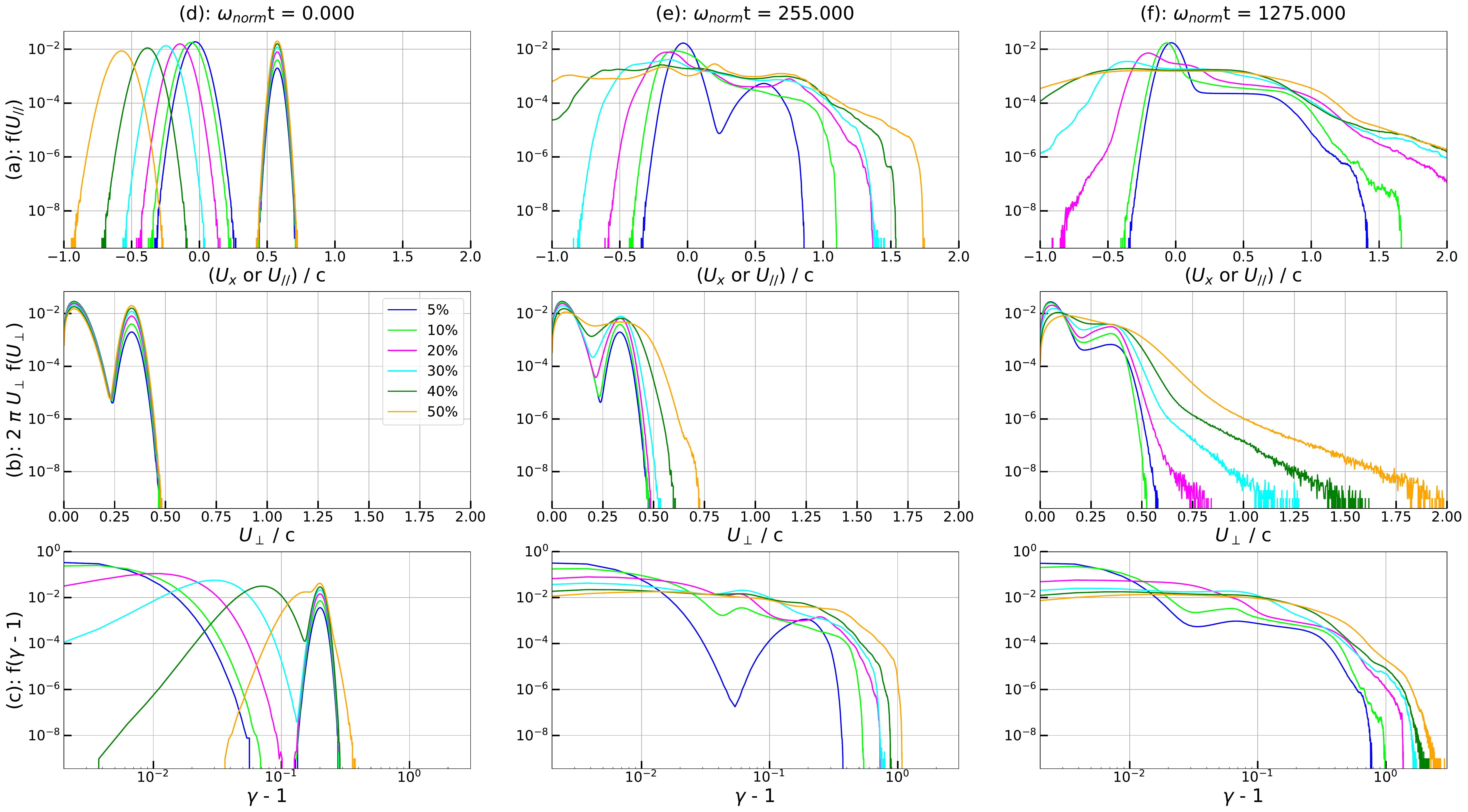}
          \caption{Distributions of the parallel momentum ($f(u_{//})$, row \textbf{a}), perpendicular momentum
                            ($2 \pi U_{\perp} f(u_{\perp})$, row \textbf{b})
                            and total kinetic energy ($f(\gamma-1)$, row \textbf{c}) of all electrons at $t = 0.0$  (column \textbf{d}),
                            $255\omega_{norm}^{-1}$  (column \textbf{e}), and $1275\omega_{norm}^{-1}$  (column \textbf{f}),
                            corresponding to the initial condition,
                            the time around when the parallel bulk drift momenta of the ring-beam electrons
                            reach their minima (see panel \textbf{a} in Fig.\ref{average_velocity}),
                            and the time close to the end of simulations.
                            In each panel, different colors are used to distinguish the different number density ratio
                            between the ring-beam and total electrons $n_{rb}/n_{t}$. Here $\omega_{ce}/\omega_{pe} = 5.0$.
                            All distributions are normalized by the number of all electrons.
                   }
          \label{velocity_energy_distribution}
      \end{center}
      \end{figure*}

\begin{figure*}[htbp]
      \begin{center}
            \includegraphics[width=1.0\textwidth]{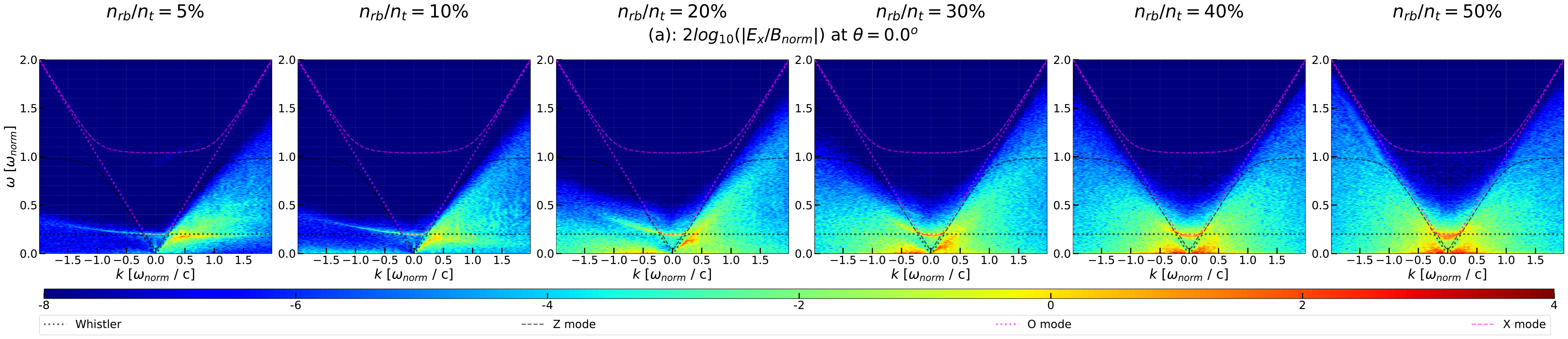}\\
            \includegraphics[width=1.0\textwidth]{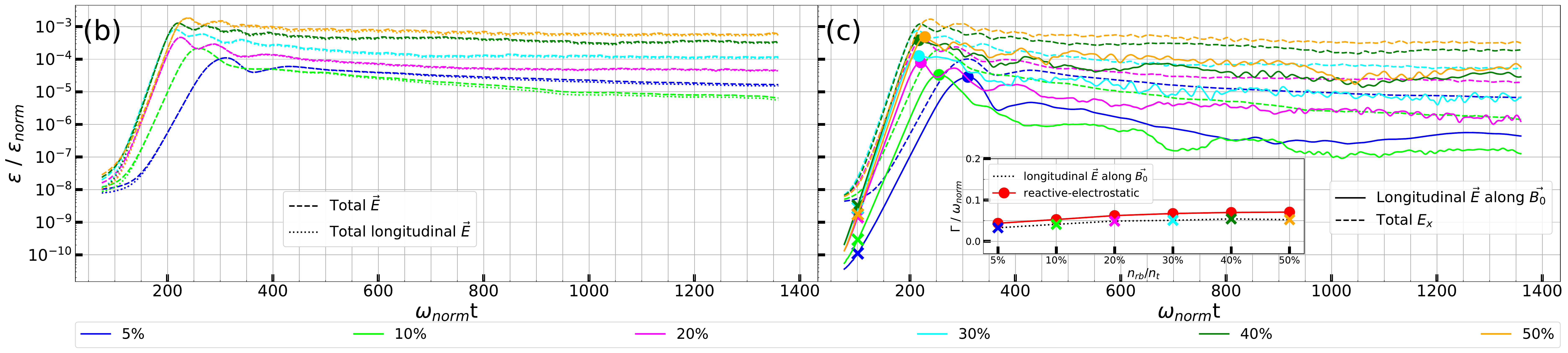}
            \caption{Row (\textbf{a}): wavevector-frequency ($\vec{k} - \omega$ or dispersion) spectra of
                              electric component $E_{x}$ of waves propagating along (either parallel $k > 0$ or antiparallel $k < 0$ to)
                              the ambient magnetic field $\vec{B_{0}}$ ($\theta = 0^{\circ}$).
                              Different panels in row (\textbf{a}) are for different $n_{rb}/n_{t}$
                              (from left to right column $n_{rb}/n_{t} = 5\%, 10\%, 20\%, 30\%, 40\%, 50\%$, respectively)
                              and share the same contour scale, normalization $B_{norm}$ (Sect.\ref{setup}).
                              In each panel of row (\textbf{a}) from the bottom to top, the overplotted lines are
                              whistler (black dotted lines), Z (black dashed lines), O (magenta dotted lines)
                              and X (magenta lines) modes in magnetized cold plasmas, respectively.
                              These $\vec{k} - \omega$ spectra are obtained via the fast
                              Fourier transform (FFT) over the entire space-time domain of our simulations.
                              Panel (\textbf{b}) presents energy evolutions of the total electric fields
                              (dashed lines) and total longitudinal electric fields (dotted lines)
                              of all waves in the simulation domain.
                              Panel (\textbf{c}) shows energy evolution of the longitudinal electric fields of waves
                              propagating along $\vec{B_{0}}$ (solid lines) as well as
                              that of the electric component $E_{x}$ (dashed lines) in the whole simulation domain.
                              Insert of panel (\textbf{c}) shows the fitted exponential growth rate of these $\vec{B_{0}}$ aligned longitudinal electric fields
                              (black dotted line) as well as the theoretical maximum growth rate of the electrostatic waves
                              by the reactive beam instability in unmagnetized cold plasmas (red-dot solid line,
                              see Eq.\ref{growth_rate_two-stream_instability}).
                              Fitted ranges for these growth rates are indicated by "o" and "x" points in the main part of panel (\textbf{c}) .
                              Different colors in panels (\textbf{b}) and (\textbf{c}) are used to distinguish cases with
                              different $n_{rb}/n_{t}$ but $\omega_{ce}/\omega_{pe} = 5.0$.
                  }
          \label{Electrostatic_Wave}
      \end{center}
      \end{figure*}

In general, Fig.\ref{average_velocity} and Fig.\ref{energy_profile} are well corrected and can be used to
study energy exchange between waves and electrons. Due to the presence of waves
and dynamic energy exchanges between particles and waves,
the magnetic energy saturation of these electromagnetic wave modes are not exactly
the same as the saturation time of their dominating instabilities (indicated by the formation of
a plateau in their corresponding momentum distribution functions).
In particular, a small positive gradient still remains in the perpendicular momentum distribution
close to the end of simulations for plasmas with tenuous ring-beam
electrons (panel \textbf{b-f} of Fig.\ref{velocity_energy_distribution}).
Since the growth rates of the beam and ECM instabilities are proportional to
the positive gradients of their corresponding distributions, their growth rates will become smaller
when those gradients (free energy sources) are reduced (dissipated).
When the gain of wave energy (due to instabilities) is equal to its losses
(to heat plasma or accelerate particles), the wave energy will stop increasing
and its energy saturation will be reached, no matter whether free energies for the instabilities remain
or not.
Moreover, the energy exchange between particles and waves can also lead to electron acceleration and heating:
increase of the electron perpendicular momentum in plasmas with dense ring-beam electrons
(row \textbf{b} of Fig.\ref{velocity_energy_distribution}) is
likely caused by cyclotron resonances, while the spread of the electron distribution in the parallel direction
is dominated by Landau damping (row \textbf{a} of Fig.\ref{velocity_energy_distribution}).
The nonlinear dynamic coupling between waves and particles are simplified or ignored in most of theoretical models.
But they can be self-consistently recovered in fully kinetic PIC simulations.

Panel (\textbf{e}) of Fig.\ref{energy_profile} shows the fitted growth rates of all four electromagnetic
(whistler, X, Z and O) modes in the ring-beam plasmas with different $n_{rb}/n_{t}$, although these magnetic energy does not
increase exactly exponentially with the time.
The fitted ranges for these growth rates are shown in their corresponding panels  of Fig.\ref{energy_profile}.
As one can see, the growth rate of each wave mode
monotonously increases with the increase of $n_{rb}/n_{t}$
in agreement with the theoretical predictions for the growth rate of the
O and X modes, e.g., \citealp{Freund_etal_1983PhFl...26.2263F, Wu&Freund_1984RaSc...19..519W}.
In addition, with the fitted ranges we used, the growth rates between the whistler and Z (as well as O and X) modes
are quite similar probably due to the same dominant beam (ECM) instability during their growth phases.
And the whistler mode always has a larger growth rate than the X mode, which is consistent with
the study of~\citealp{Lee_etal_2011PhPl...18i2110L}.

Additionally, we also study anisotropy of the whistler, Z, O and X-mode magnetic energies,
as shown in Fig.\ref{Emission_Symmetry}.
In each panel, the total magnetic energy of an electromagnetic mode, covering the whole simulation domain and time series,
is divided among the different wave propagation directions $\theta$ (Eq.\ref{anisotropic_emission_Eq}).
Considering the axis-symmetry of the system, we only need to investigate dependence of the energy
on $\theta$ from $0^{\circ}$ to $180^{\circ}$.
Note that the magnetic energy of the whistler mode at $\theta = 90^{\circ}$ is not included in panel (\textbf{a})
due to its resonance frequency $\omega_{W}^{res} \rightarrow 0$ at $\theta = 90^{\circ}$ with the
physical proton-to-electron mass ratio $m_{p}/m_{e} = 1836$ in the magnetized cold plasma limit.

Contrary to the electrostatic waves (Fig.\ref{Electrostatic_Wave}), magnetic energies are
dominated by non-parallel electromagnetic waves.
In plasmas with tenuous ring-beam electrons, the energy dominated waves of all mode branches propagate in the
same side as the ring-beam electron propagating (i.e., $\theta < 90^{\circ}$).
The anisotropy of the whistler, Z and O modes decrease with the increase of the ring-beam electron density,
while the anisotropy of the X mode is always high.
In plasmas with dense ring-beam electrons, the X-mode magnetic energy has the strongest anisotropy
than the other three (whistler, Z and O) modes.
The X mode is the strongest around $\theta = 60^{\circ}$, indicating influences from the beam instability on
the excitation of the X mode, since the strongest energy of the X mode is exactly located at
$\theta = 90^{\circ}$ when the energetic electrons initially follow a pure-ring momentum
distribution~\citep{Pritchett1984JGR....89.8957P}.
In isotropic thermal plasmas (i.e., with $n_{rb}/n_{t} = 0\%$),
the magnetic energy of the whistler, Z, O, and X-mode waves are much smaller than those excited
ones and more or less isotropic except of the whistler mode, which has (about one order of magnitude) less magnetic energy
than the other three wave modes but its anisotropy is the strongest.

\begin{figure*}[htbp]
      \begin{center}
            \includegraphics[width=1.0\textwidth]{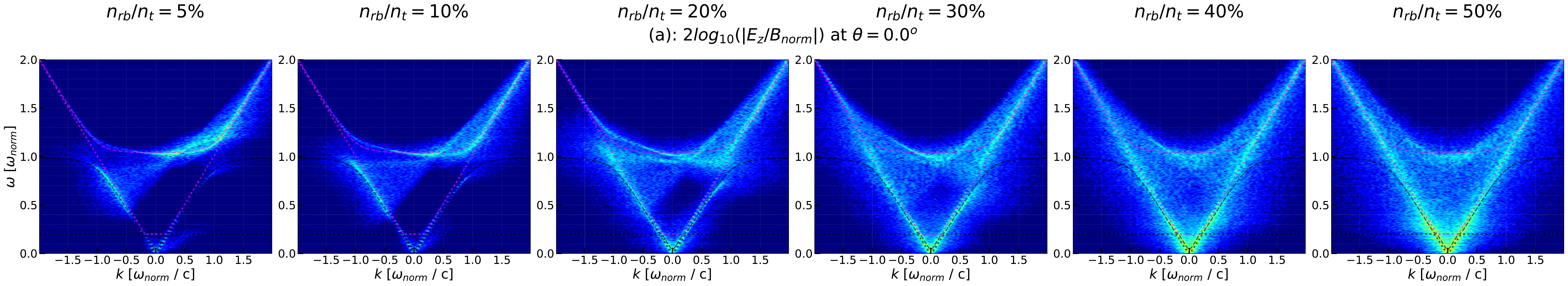}\\
            \includegraphics[width=1.0\textwidth]{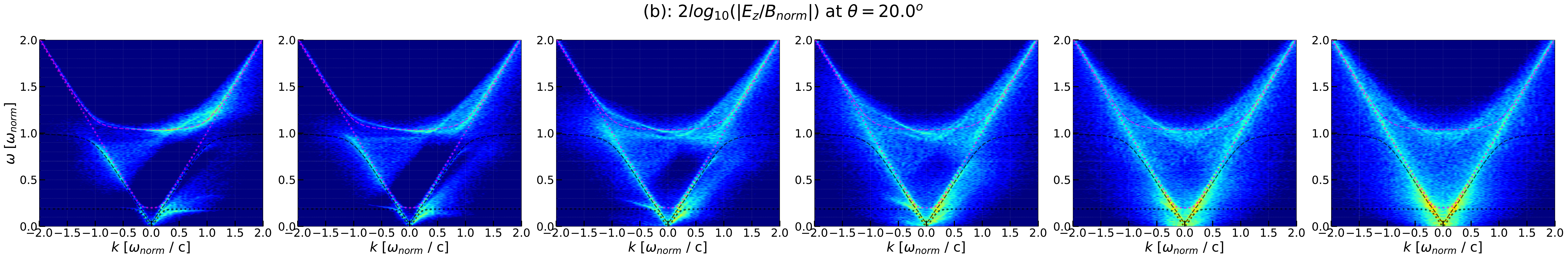}\\
            \includegraphics[width=1.0\textwidth]{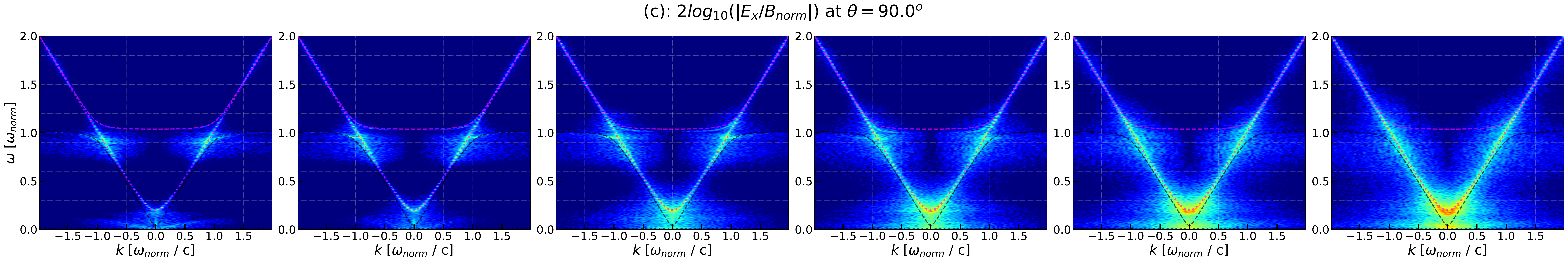}\\
            \includegraphics[width=1.0\textwidth]{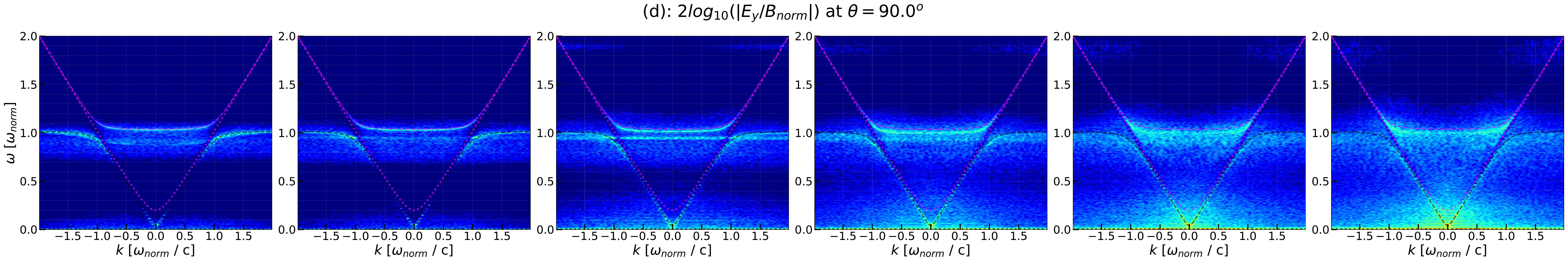}\\
            \includegraphics[width=1.0\textwidth]{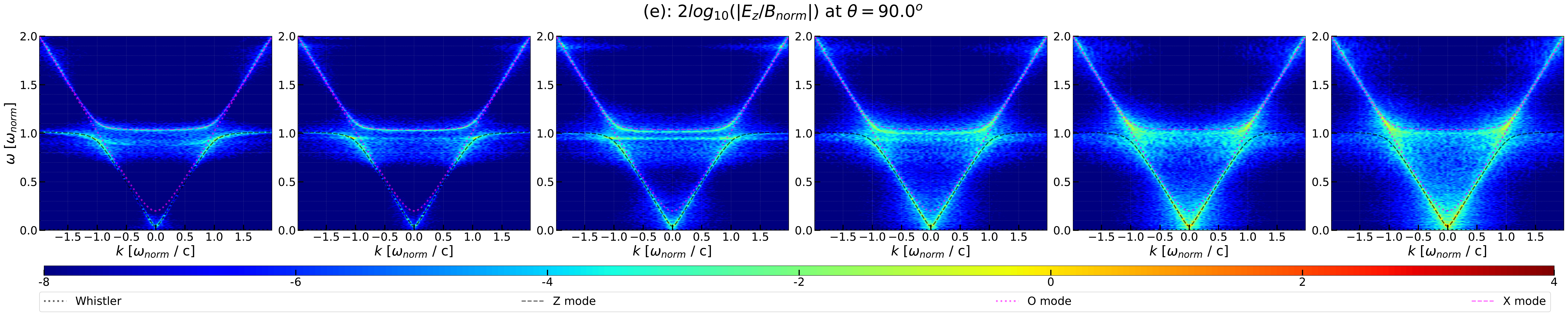}
            \caption{$\vec{k} - \omega$ spectra of different electric field components for different $n_{rb}/n_{t}$
                            (from left to right column $n_{rb}/n_{t} = 5\%, 10\%, 20\%, 30\%, 40\%, 50\%$, respectively) and
                            waves propagation directions $\theta$
                            (row \textbf{a}: $E_{z}$ with $\theta = 0^{\circ}$,
                             row \textbf{b}: $E_{z}$ with $\theta = 20^{\circ}$,
                             row \textbf{c}: $E_{x}$  with $\theta = 90^{\circ}$,
                             row \textbf{d}: $E_{y}$ with $\theta = 90^{\circ}$,
                             row \textbf{e}: $E_{z}$ with $\theta = 90^{\circ}$) with $\omega_{ce}/\omega_{pe} = 5.0$.
                           In each panel, overplotted lines, contour scale, and normalization are the same as those in raw
                           (\textbf{a}) of Fig.\ref{Electrostatic_Wave}.
                  }
          \label{dispersion}
      \end{center}
      \end{figure*}

\begin{figure*}[htbp]
      \begin{center}
          \includegraphics[width=1.0\textwidth]{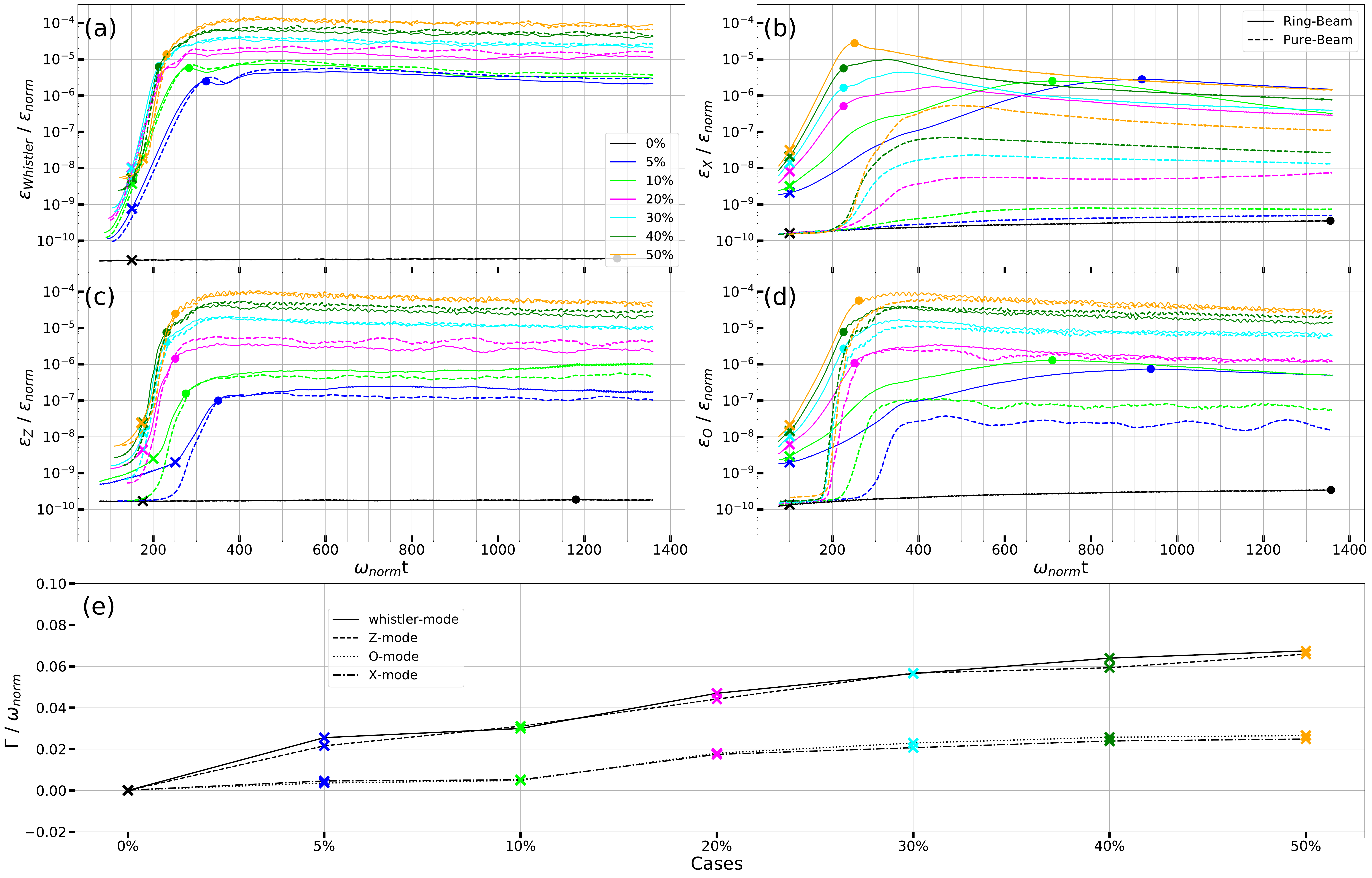}
          \caption{Magnetic energy evolution of the electromagnetic whistler ($\varepsilon_{Whistler}$, panel \textbf{a}),
                            X ($\varepsilon_{X}$, panel \textbf{b}),
                            Z ($\varepsilon_{Z}$, panel \textbf{c}) and O ($\varepsilon_{O}$, panel \textbf{d}) modes,
                            where solid (dashed) lines are for
                            plasmas with energetic ring-beam (pure-beam) electrons and $\omega_{ce}/\omega_{pe} = 5.0$.
                           Panel (\textbf{e}) shows the fitted exponential growth rates of these four electromagnetic wave modes for plasmas with
                           energetic ring-beam electrons but different $n_{rb}/n_{t}$ (distinguished with different colors).
                           The solid, dashed, dotted and dash-dot lines in panel (\textbf{e}) are
                           for the whistler, X , Z and O modes, respectively.
                           And fitted ranges for these growth rates are indicated by "o" and "x"
                           points in their corresponding panels (\textbf{a} to \textbf{d}).
                   }
          \label{energy_profile}
      \end{center}
      \end{figure*}

\begin{figure*}[htbp]
      \begin{center}
           \includegraphics[width=1.0\textwidth]{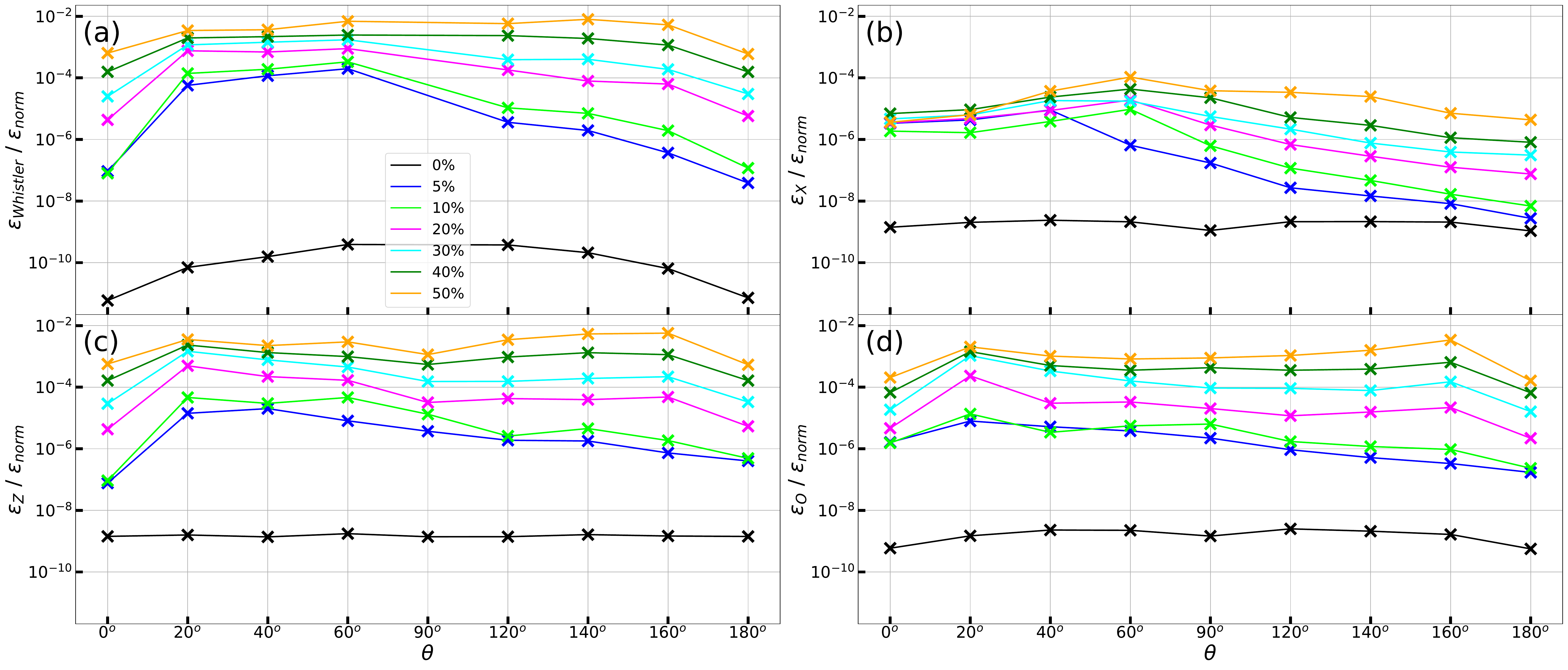}
           \caption{Anisotropic magnetic energy  (Eq.\ref{anisotropic_emission_Eq})
                             of the electromagnetic whistler ($\varepsilon_{Whistler}$, panel \textbf{a}), X ($\varepsilon_{X}$, panel \textbf{b}),
                             Z($\varepsilon_{Z}$, panel \textbf{c}) and O ($\varepsilon_{O}$, panel \textbf{d}) modes with
                             $\omega_{ce}/\omega_{pe} = 5.0$.
                             Different colors in each panel are for different $n_{rb}/n_{t}$.
                  }
          \label{Emission_Symmetry}
      \end{center}
      \end{figure*}

Consistent with row (\textbf{a}) of Fig.\ref{dispersion}, in the direction along $\theta = 0^{\circ}$ and $180^{\circ}$,
wave excitation is dominated by the X and Z modes, respectively, in plasmas with tenuous ring-beam electrons.
And energies of the obliquely propagating whistler and Z modes with $\theta = 20^{\circ}$ and $160^{\circ}$
are larger than those of the $\vec{B_{0}}$ aligned ones (rows \textbf{a} and \textbf{b} of Fig.\ref{dispersion}, respectively).
As well as with the increase of the ring-beam electron density, magnetic energy of each electromagnetic mode increases
in directions both along (row \textbf{a} of Fig.\ref{dispersion}) and perpendicular (rows \textbf{c} and \textbf{e} of
Fig.\ref{dispersion}) to the ambient magnetic field $\vec{B_{0}}$, except for the quasi-parallel-propagating X-mode
waves, which appears to be saturated as discussed for rows (\textbf{a}) and (\textbf{b}) of Fig.\ref{dispersion}
in Sect.\ref{Excited_electomagnetic_waves}.

We note that, with the Gaussian filter method, magnetic energy for each wave
mode in Fig.\ref{energy_profile} and \ref{Emission_Symmetry} might contaminate each other
when the dispersion relations of two wave modes are close to each other.
For example, row (\textbf{a}) of Fig.\ref{dispersion} shows that, in plasmas with tenuous
ring-beam electrons $n_{rb}/n_{t} \leq 10\%$,  the energy of the O mode propagating
in the parallel (antiparallel) direction can have contributions from the X (Z) mode.
The growth of the O mode in plasmas with tenuous
ring-beam electrons (panel \textbf{d} of Fig.\ref{energy_profile}),
therefore, may be influenced by these effects.
This can be clearly seen in the following section when we consider the
polarization of these waves.

\subsubsection{Polarization properties}
\label{polarization_result}

\begin{figure*}[htbp]
      \begin{center}
            \includegraphics[width=1.0\textwidth]{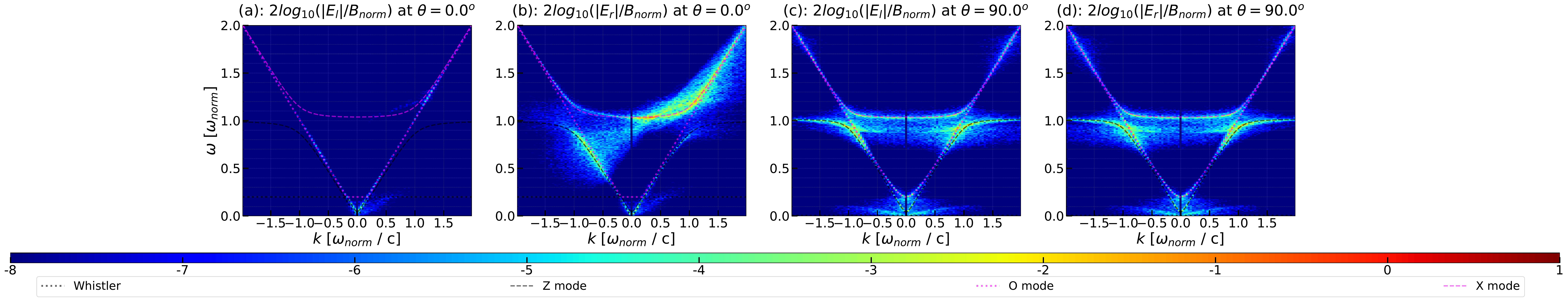}\\
            \includegraphics[width=1.0\textwidth]{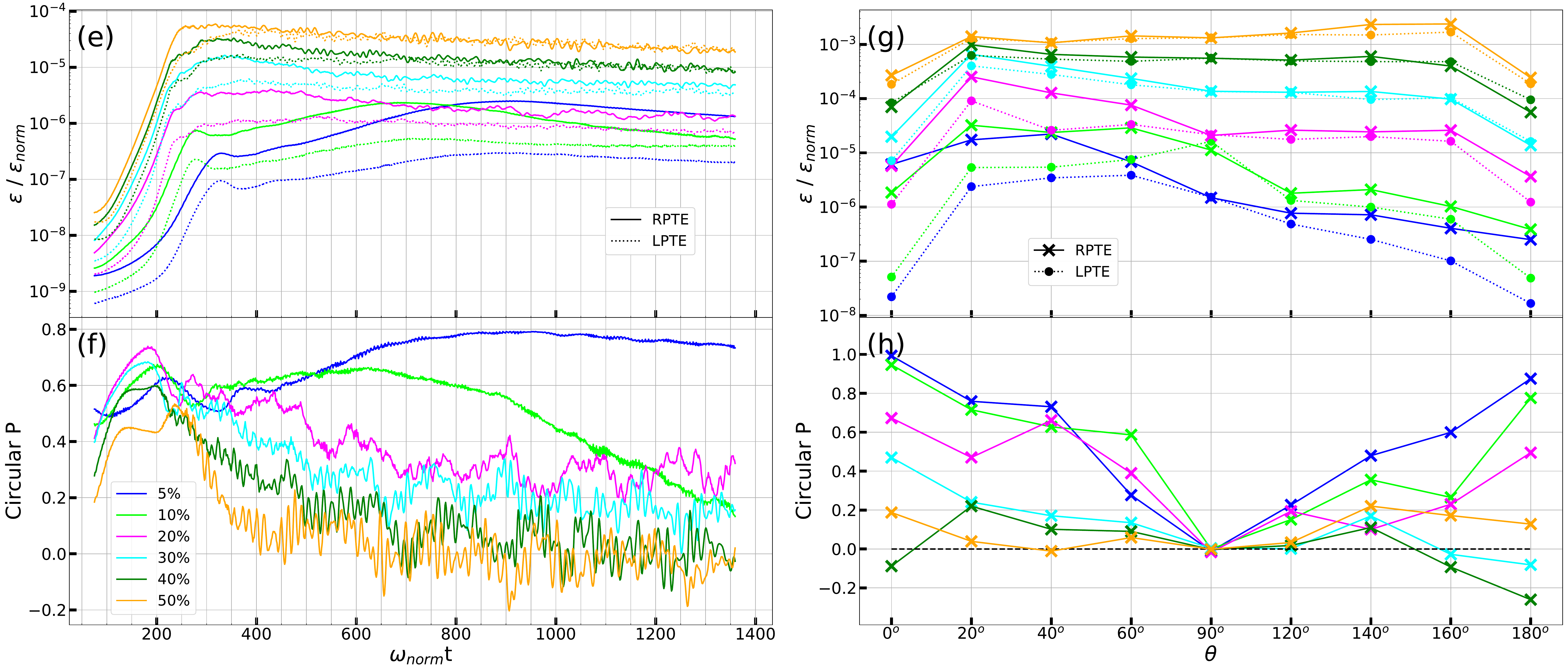}
            \caption{ Panels (\textbf{a}) - (\textbf{d}) show the $\vec{k} - \omega$
                               spectra of the LPTE ($E_{l}$, panels \textbf{a} and \textbf{c}) and
                               RPTE ($E_{r}$, panels \textbf{b} and \textbf{d}) of electromagnetic waves propagating along
                               $\theta = 0^{\circ}$ (panels \textbf{a} and \textbf{b}) and
                               $90^{\circ}$ (panels \textbf{c} and \textbf{d}), respectively, in plasmas
                               with $n_{rb}/n_{t}=5\%$. These four panels share the same color bar.
                               Overplotted lines and normalization in these four panels are the same as
                               those in Figs.\ref{Electrostatic_Wave} and \ref{dispersion}.
                               Energy evolutions of the RPTE (solid lines) and  LPTE (dotted lines) of all electromagnetic waves in
                               the simulation domain are presented in panel (\textbf{e}).
                               Panel (\textbf{f}) shows the circular polarization degree (CPD) evolution of these transverse electric feilds.
                               Dependence of the energy of the RPTE (cross-solid lines),  LPTE (dot-dotted lines) and the CPD on the wave propagation direction $\theta$ are shown in
                               panels (\textbf{\texttt{g}}) and (\textbf{h}) for transverse electric feilds of all electromagnetic waves
                               in the simulation domain.
                               Different colors in each panel of (\textbf{e} to \textbf{h}) are used to distinguish the different
                               $n_{rb}/n_{t}$ cases but all cases have $\omega_{ce}/\omega_{pe} = 5.0$.
                  }
          \label{Polarization}
      \end{center}
      \end{figure*}

The polarization of a wave depends on its propagation direction $\theta$~\citep{Melrose_1986islp.book.....M}.
When propagating parallel to $\vec{B_{0}}$
(i.e., $\theta = 0^{\circ}$), the O (X)-mode waves are fully left
(right)-handed circularly polarized and the Z-mode waves are fully left (right)-handed circularly  polarized when their
frequencies $\omega < (>)~\omega_{pe}$ (see panels \textbf{a} and \textbf{b} of Fig.\ref{Polarization}).
When $\theta = 90^{\circ}$, both O and X-mode waves are linearly polarized
(see panels \textbf{c} and \textbf{d} of Fig.\ref{Polarization}  as well as
rows \textbf{c} to \textbf{e} of Fig.\ref{dispersion}),
since the electric field of the O (X)-mode waves are parallel
(perpendicular) to $\vec{B_{0}}$.

Following the method described in Sect.\ref{polarization_method}, we separate the energy
contained by the left and right-handed polarized transverse electric fields (LPTE and RPTE) in electromagnetic waves.
Panel (\textbf{e}) of Fig.\ref{Polarization} shows the energy evolution of the LPTE and RPTE for all electromagnetic waves
in the simulation domain.
In the $n_{rb}/n_{t} \leq 10\%$ cases, energy evolution profiles of the LPTE and RPTE
contain two growth phases, indicating that both the beam and ECM instabilities play
roles on the excitation of the electromagnetic waves in the simulation domain. These two growth phases
correspond well to the obviously different dissipation rates of the free energies along the parallel and
perpendicular directions for the beam and ECM instabilities, respectively (see panels \textbf{a} and \textbf{b}
of Fig.\ref{average_velocity} and Fig.\ref{energy_profile}).
In general, in each $n_{rb}/n_{t}$ case, the RPTE dominates the transverse electric field energy
during most of the simulation time, due to the energy dominance of the
right-handed polarized whistler and X-mode waves, see Fig.\ref{energy_profile}.

Panel (\textbf{f}) of Fig.\ref{Polarization} shows
the evolutions of the CPD (Eq.\ref{circle_polarization}) for all electromagnetic waves in the simulation domain.
Due to the dominance of the RPTE in each $n_{rb}/n_{t}$ case,
all CPDs are positive at the beginning and increase during the wave growth phase.
They, however, start to decrease after the energy saturation of the transverse electric fields.
For larger $n_{rb}/n_{t}$  cases, the CPDs can be close to 0 and become negative at the end of the simulations.
The decreased CPD indicates more reduction of the right-handed polarized waves than
the left-handed polarized ones by electrons via wave-electron cyclotron resonance interactions.

Over the whole time-frequency domain and for all electromagnetic waves in the simulation domain,
energy anisotropy of their LPTE and RPTE (panel \textbf{\texttt{g}}) as well as the anisotropy of their CPDs
(panel \textbf{h}) are also presented in Fig.\ref{Polarization}.
Consistent with panels (\textbf{e}) and (\textbf{f}), RPTE predominates the total energy of
transverse electric fields along most of  the wave propagation directions in each $n_{rb}/n_{t}$ case.
Correspondingly, the CPDs at different wave propagation angles are, hence, mostly positive (right-handed polarized).
Small negative (or left-handed polarized) CPDs, however, also exist for the $n_{rb}/n_{t} = 30\%, 40\%$ cases,
e.g., along $\theta \geq 160^{\circ}$.
Furthermore, consistent with the classical definition of the perpendicular propagating electromagnetic waves,
their CPDs are always around 0 (linearly polarized) at $\theta = 90^{\circ}$ for all cases.
For parallel-propagating electromagnetic waves ($\theta = 0^{\circ}$) in $n_{rb}/n_{t} = 5\%, 10\%$ cases,
the resulting  CPDs  can reach 1.0, i.e, fully right-handed circularly polarized
(see panels \textbf{a} and \textbf{b} of Fig.\ref{Polarization} for $n_{rb}/n_{t} = 5\%$). There the left-handed polarized
O mode has negligible contributions to the total energies of the parallel-propagating electromagnetic waves.

\begin{figure*}[htbp]
      \begin{center}
            \includegraphics[width=1.0\textwidth]{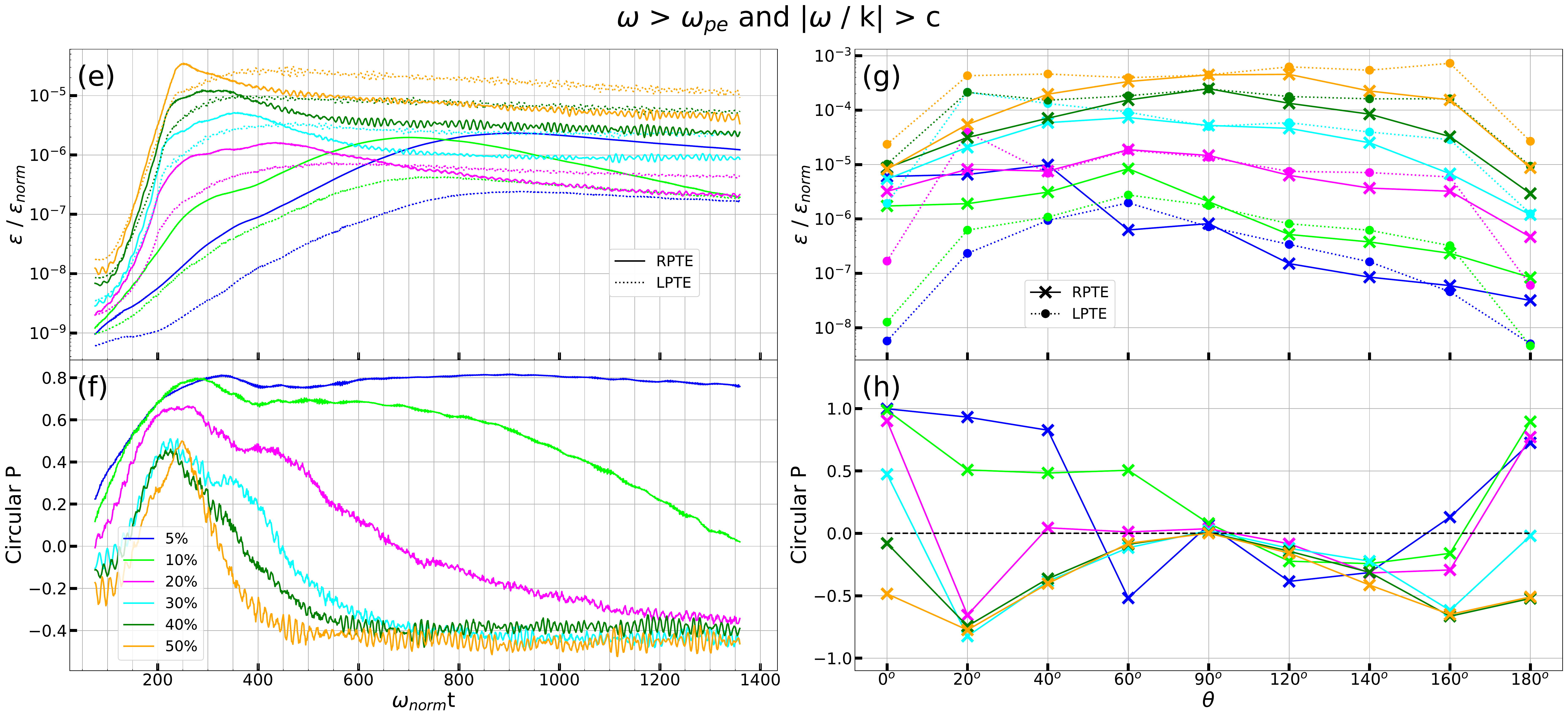}
            \caption{Same as panels (\textbf{e} to \textbf{h}) in Fig.\ref{Polarization}, but for transverse electric feilds of
                              the escaping electromagnetic waves with $\omega > \omega_{pe}$ and $|\omega / k| > c$.
                  }
          \label{Polarization_Escape}
      \end{center}
      \end{figure*}

All electromagnetic waves in the simulation domain are included in the above discussions. From observational
point of view, however, not all excited waves can be detected remotely.
It is known that an electromagnetic wave can escape from an astrophysical plasma only if its refractive index
is less than unity, i.e., $|c k / \omega| < 1$ and its frequency is larger than the local plasma frequency, i.e.,
$\omega > \omega_{pe}$~\citep{Melrose_1986islp.book.....M, Budden_1988prw..book.....B, Stix_1992wapl.book.....S, Benz_2002ASSL..279.....B, Bellan_2006fpp..book.....B}.
Escaping electromagnetic waves in plasmas, hence, are only the O and X modes.
Properties of the polarization and spectrogram are, hence, investigated for these escaping electromagnetic waves
with $\omega > \omega_{pe}$ and $|c k/\omega| < 1$ (the escape condition),
shown in Fig.\ref{Polarization_Escape} and Fig.\ref{spectrogram_Escape}, respectively.
Other waves, with larger refractive indices and low frequencies, are trapped and can be
absorbed or reflected (depending on their cutoff or/and resonance frequencies) during
wave propagations in the interplanetary medium (IPM) or interestellar plasmas.
The only way that those waves can be remotely detected is by means of conversion
to escaping electromagnetic waves via mechanisms such as wave-wave coupling,
coalescence or decay, antenna mechanisms or mode conversation in
inhomogeneous plasmas~\citep[see, e.g.,][and references therein]{Graham2017,Graham2018}, which
is, however, beyond the scope of this study.

Due to the removal of electromagnetic waves with $\omega \leq \omega_{pe}$ or $|c k/\omega| \geq 1$ (mostly the
whistler and Z modes), energy of both the escaping RPTE and LPTE are reduced
(panel \textbf{e}) in comparison with those of all electromagnetic waves in the simulation domain
(panel \textbf{e} of Fig.\ref{Polarization}).
Another obvious difference between the escaping RPTE, LPTE and the RPTE, LPTE of all electromagnetic waves
is that, when $n_{rb}/n_{t} \leq 10\%$, the first growth phase
(during $\omega_{norm} t < 325$, which associated with the beam instability) in the energy
evolution profile of the RPTE and LPTE of all electromagnetic waves
do not exist anymore in Fig.\ref{Polarization_Escape} for the escaping electromagnetic waves.
That indicates that the excitation of the high-frequency escaping electromagnetic waves
are mainly due to the the ECM instability in plasmas with tenuous ring-beam electrons
($n_{rb}/n_{t} \leq 10\%$) and the significant growth of the O mode associated with the beam instability
in Fig.\ref{energy_profile} is likely caused by contamination from the Z mode in these tenuous ring-beam cases.

Panel (\textbf{f}) of Fig.\ref{Polarization_Escape} shows the evolution of the
CPD for the escaping electromagnetic waves.
In the cases with larger $n_{rb}/n_{t} \geq 20\%$, one can see that
the CPDs of the escaping electromagnetic waves are always smaller than those of all electromagnetic waves and
can flip sign and be close to $-0.4$ at the end of simulations.
In contrast, the CPDs of the escaping electromagnetic waves in $n_{rb}/n_{t} \leq 10\%$ cases are larger
than those of all electromagnetic waves, especially around the saturation of the first growth phase
in the energy evolution profile for all waves ($\sim \omega_{norm} t < 325$ in
panel \textbf{e} of Fig.\ref{Polarization}).

Panel (\textbf{\texttt{g}}) of Fig.\ref{Polarization_Escape}
shows the energy anisotropy of the LPTE and RPTE for the escaping electromagnetic waves.
Different from those of all the electromagnetic waves in the simulation domain, dominance
of the RPTE or LPTE changes with the wave propagation direction $\theta$ in plasmas with tenuous
ring-beam electrons $n_{rb}/n_{t} \geq 10\%$.
For plasmas with dense ring-beam electrons (especially $n_{rb}/n_{t} \geq 30\%$),
LPTE dominates the escaping transverse electric field energy along most of the
wave propagation directions.
Correspondingly, the anisotropy of the CPD for the escaping electromagnetic waves
(panel \textbf{h} of Fig.\ref{Polarization_Escape}) is
also quite different from the one for all electromagnetic waves in the simulation domain
(panel \textbf{h} of Fig.\ref{Polarization}).
For the escaping electromagnetic waves, a left-handed CPD can be found at some directions for each $n_{rb}/n_{t} $ case.
Furthermore, the left-handed CPD dominates over all wave propagation
directions in plasmas with dense ring-beam electrons $n_{rb}/n_{t} \geq 40\%$.
At $\theta = 90^{\circ}$, escaping electromagnetic waves are still linearly polarized.
Additionally, with the increase of $n_{rb}/n_{t}$,
the CPDs of the escaping electromagnetic waves become increasingly
symmetric around $\theta = 90^{\circ}$, corresponding to their symmetric energies of the
RPTE and LPTE (panel \textbf{\texttt{g}} of Fig.\ref{Polarization_Escape}).

Fig.\ref{spectrogram_Escape} shows dependence of the spectrogram of the RPTE and LPTE in
escaping electromagnetic waves on the wave propagation direction.
These spectrograms have similar anisotropy and symmetry properties
(i.e., corresponds well) as their corresponding CPDs in panel \textbf{h} of Fig.\ref{Polarization_Escape}.
With the exception of waves propagating near $\theta = 90^{\circ}$,
the spectrogram of the RPTE and LPTE are quite different along
any other propagation directions. These differences increase with increasing value of  $|\theta - 90^{\circ}|$.
And with Fig.\ref{spectrogram_Escape},
one can also find that the CPD of these escaping electromagnetic waves depend on not only the wave propagation direction,
population of the ring-beam electrons but the wave frequency and time.
Furthermore, intense emission in these spectrograms are generally located
around the frequencies $\omega_{pe}$ and/or $\omega_{ce}$. Bandwidth and intensity
as well as pattern of these intense emissions vary a lot among different frequencies, wave
propagation directions and population of the ring-beam electrons.
Emission around $2\omega_{ce}$ can also be found in
these spectrograms, especially for the perpendicular propagation $\theta = 90^{\circ}$ and for
plasmas with a dense ring-beam electron population.


\subsection{$\omega_{ce}/\omega_{pe}$ dependence}
\label{frequency_dependence}

As mentioned in Sect.\ref{Introduction}, $\omega_{ce} > \omega_{pe}$ is required for an
efficient escaping ECM emission. Many previous numerical studies for the ECM emission
(e.g., \citealp{Pritchett1984JGR....89.8957P, Lee_etal_2009PhRvL.103j5101L, Lee_etal_2011PhPl...18i2110L}), hence,
considered situations with $\omega_{ce} > \omega_{pe}$.
Although $\omega_{ce} > \omega_{pe}$ can exist in some density cavities due to, e.g., turbulent magnetic field fluctuations
\citep{Wu_etal_2014A&A...566A.138W, Chen_etal_2017JGRA..122...35C, Melrose_2017RvMPP...1....5M},
based on the standard solar atmosphere model \citep{Wild_1985srph.book....3W}, however,
$\omega_{ce} < \omega_{pe}$ is typical for the solar coronal conditions.
In this section, we will focus on the wave excitation dependence on the
$\omega_{ce} / \omega_{pe}$ for ring-beam energetic electrons with fixed number density ratio
$n_{rb}/n_{t}$ to $5\%$ and fixed total electron plasma frequency $\omega_{pe}$.
This is justified considering that the typical gradient length of the particle number density is
usually larger than that of the magnetic field strength in the solar corona (see Eqs.1.6.1 and 1.4.2 in
\citealp{Aschwanden2005psci.book.....A}).
The ratio between the electron cyclotron frequency $\omega_{ce}$
and $\omega_{pe}$ takes $0.2$, $0.3$, $0.5$, $1$, $2$,
$3$, while the case $\omega_{ce}/\omega_{pe} = 5$ has been analyzed in Sect.\ref{density_dependence}.
Similar to Fig.\ref{dispersion}, the dispersion spectra dependence on the $\omega_{ce}/\omega_{pe}$
and wave propagation direction $\theta$ are presented in Fig.\ref{Dispersion_Relation_2}.

As the cases with $\omega_{ce}/\omega_{pe} = 5$, excitation of the beam, Langmuir (row \textbf{a}), whistler (row \textbf{b}),
O (row \textbf{c}), Z and X (row \textbf{d} and \textbf{e}) modes still exist in each $\omega_{ce}/\omega_{pe} < 5$ cases.
Intensity of the escaping electromagnetic waves from the weakly magnetized plasmas
$\omega_{ce}/\omega_{pe} < 1$ are, however, significantly suppressed and negligible
comparing with those from plasmas with $\omega_{ce}/\omega_{pe} > 1$,
which is consistent with the statement of~\citealp{Vlahos_1987SoPh..111..155V}.
With the decrease of the $\omega_{ce}/\omega_{pe}$
the dispersion relation surface of the O and X modes tend to overlap with each other in the cold plasma approximation
the intensity differences between the O and X modes can not be well resolved for small
$\omega_{ce}/\omega_{pe}$ cases with the limited resolution in the $\omega$ space in our simulations.
Quantitative investigation on the energy and polarization property dependence on the
$\omega_{ce}/\omega_{pe}$ ratio will be presented in following papers with a higher $\omega$ resolution.
Here, we will concentrate on discussing the harmonic excitation of the $\omega_{pe}$ and $\omega_{ce}$.

In Fig.\ref{Dispersion_Relation_2}, one can find that excitation at higher and higher harmonics $s_{h}$ of
both $\omega_{pe}$ (rows \textbf{a} and \textbf{b}) and $\omega_{ce}$ (rows \textbf{c} to \textbf{e})
appear with the decrease of the $\omega_{ce}/\omega_{pe}$.
However, excitation of harmonic $\omega_{pe}$ can only be found in cases with
$\omega_{ce}/\omega_{pe} <  1$, i.e., weakly magnetized plasmas.
In row (\textbf{b}), although frequencies of the excited harmonics of $\omega_{pe}$ is evidently
higher than $\omega_{pe}$, the refractive index in these harmonics of $\omega_{pe}$  are, however, much larger
than 1. These waves are likely reflected at the boundary of plasmas with distinct properties and they
therefore cannot be observed remotely.
Meanwhile, these non-escaping harmonics of $\omega_{pe}$ are mainly located
in the direction quasi-parallel to the ambient magnetic field $\vec{B_{0}}$, implying a beam instability
origin.
Note that these excited non-escaping harmonics of $\omega_{pe}$ are not centered exactly at the integer multiples of
$\omega_{pe}$. Instead, their frequencies increase with $k$, i.e., with a small
positive slope in each non-escaping harmonic $\omega_{pe}$ band. Similar results can also be found
in the study of~\citealp{Thurgood&Tsiklauri_2015A&A...584A..83T} for the plasma emission theory.
This frequency shifts in the fundamental $\omega_{pe}$ mode
have been attributed to deviations from the prediction of the cold plasma theory
in the case of dense beams~\citep{Fuselier1985, Cairns1989}, where
the beam-mode waves might affect the generation of the fundamental $\omega_{pe}$ mode.
And the frequency shifts at higher non-escaping harmonics of $\omega_{pe}$ are, perhaps,
due to the frequency shift of the fundamental $\omega_{pe}$ mode, since the
fundamental mode is responsible for the excitations of other higher
harmonics.
Enhanced harmonics of $\omega_{ce}$ can be found in each panel of rows (\textbf{c}) to (\textbf{e}) of Fig.\ref{Dispersion_Relation_2}.
In other words, excitation of harmonic $\omega_{ce}$ does not depend on the magnetized condition of plasma
 $\omega_{ce}/\omega_{pe}$. Additionally, these excited harmonics of $\omega_{ce}$ are likely excited by the ECM
instability, since excitation of them are mainly located in the direction perpendicular to the ambient magnetic field
$\vec{B_{0}}$.

Although the excitation mechanism for the harmonics of $\omega_{pe}$ and $\omega_{ce}$
are totally different, they still have some common characteristics.
For instance, all these harmonic waves contain
both longitudinal (row \textbf{a} for $s_{h} \omega_{pe}$,  row \textbf{d} for $s_{h} \omega_{ce}$) and
transverse (row \textbf{b} for $s_{h} \omega_{pe}$, rows \textbf{c} and \textbf{e} for $s_{h} \omega_{ce}$) components,
but the longitudinal component is stronger than the transverse one, which is opposite to that of the Z and X modes
in the perpendicular direction.
Harmonic excitation of $\omega_{pe}$ with a preferential longitudinal component has been found
previously by~\citealp{Klimas1983, Nishikawa1991, Yoon2003, Yi2007, Rhee_etal_2009ApJ...694..618R}.
Additionally, intensity in these excited harmonics of $\omega_{pe}$ and $\omega_{ce}$
decrease with the the increase of the harmonic number.
Moreover, all excited harmonics of $\omega_{pe}$ and $\omega_{ce}$ are non-escaping
modes in weakly magnetized plasmas with $\omega_{ce}/\omega_{pe} <  1$.

\section{Conclusions and discussion}
\label{Conclusions}
Using 2.5D fully kinetic PIC simulations, we investigated the energy and polarization
properties of electromagnetic waves excited by mildly relativistic ring-beam electrons
in neutral and current-free solar coronal plasmas.
These energetic ring-beam electrons could be produced by
magnetic reconnection, quasi-perpendicular shocks and/or electron beams propagating
in inhomogeneous magnetic fields in the solar corona.
These ring-beam electrons together with the background electrons and protons support the global current and
charge neutralities in these plasmas, where all background electrons drift
oppositely to the ring-beam electrons (i.e., return current) to fully compensate the current induced by
the ring-beam electrons, i.e., a neutral ring-beam-return current system.
To apply the simulation results to solar radio observations and
considering variations of the electron ring-beam density and
magnetic field strength along the path of electron propagation, we explore the
dependence of the electromagnetic wave excitations on
the number density ratio of  the ring-beam electrons over the total electrons
($n_{rb}/n_{t}$) and the ratio of the electron cyclotron frequency ($\omega_{ce}$) to the
electron plasma frequency ($\omega_{pe}$).

We found that the beam and electron cyclotron maser (ECM) instabilities
together can efficiently excite the whistler, Z, O, and X-mode electromagnetic waves as well as
harmonics of $\omega_{pe}$ (only when $\omega_{ce}/\omega_{pe} < 1$) and $\omega_{ce}$.
We also found the excitations of electrostatic waves, relativistic Bernstein waves and some diffusive nonlinear waves
that do not follow well defined dispersion relations.
Electrostatic waves always dominate the energetics of all excited waves.
These electrostatic waves can lead to a significant heating on the ring-beam and background electrons
due to Landau damping.

\begin{figure*}[htbp]
      \begin{center}
            \includegraphics[width=0.83\textwidth]{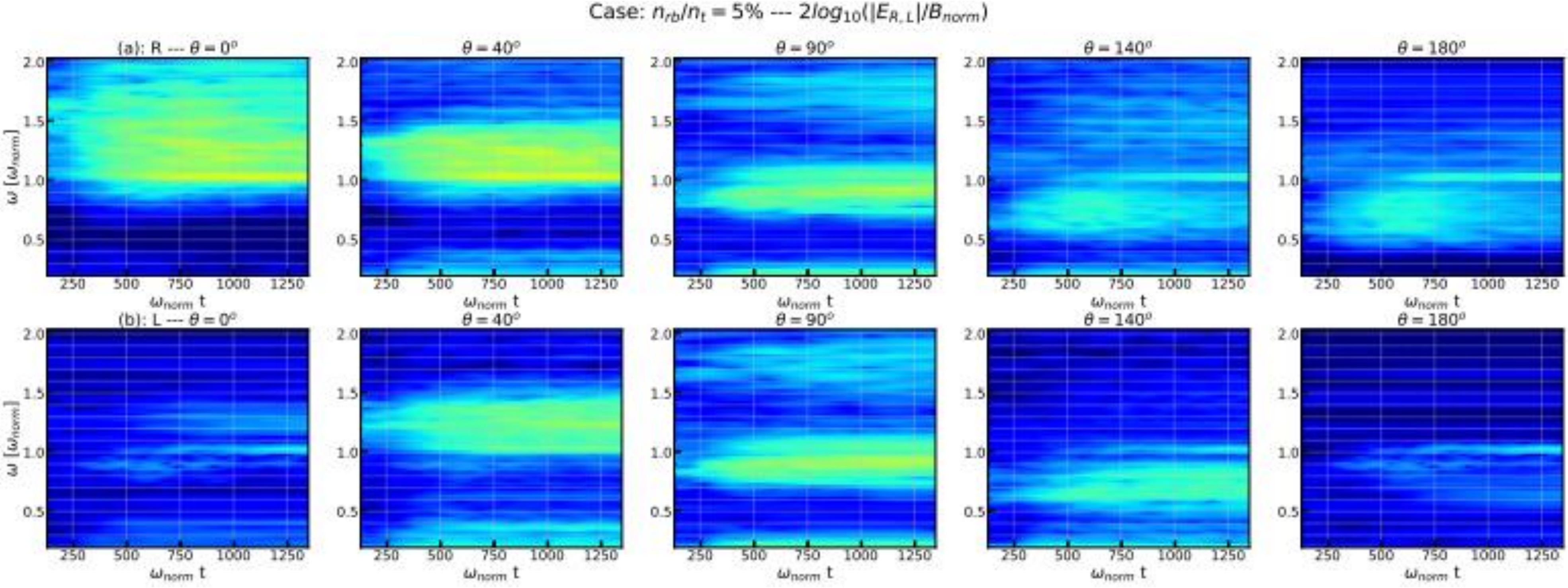}
            \includegraphics[width=0.83\textwidth]{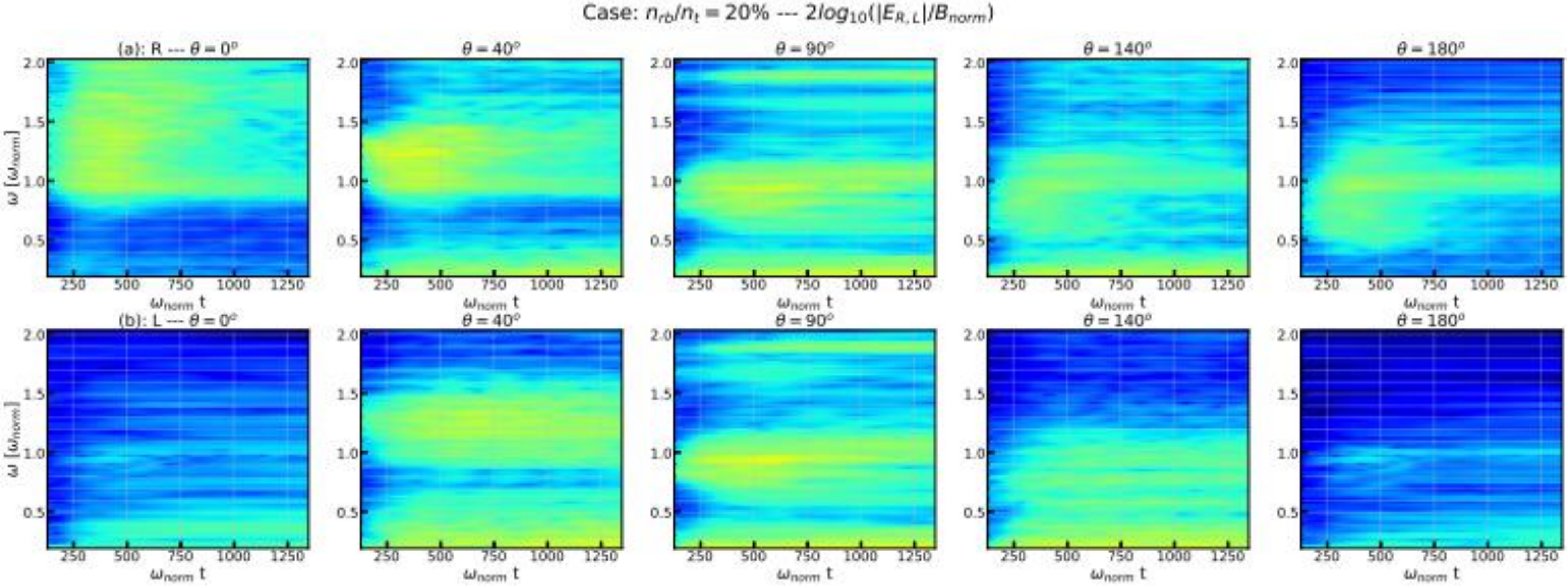}
            \includegraphics[width=0.89\textwidth]{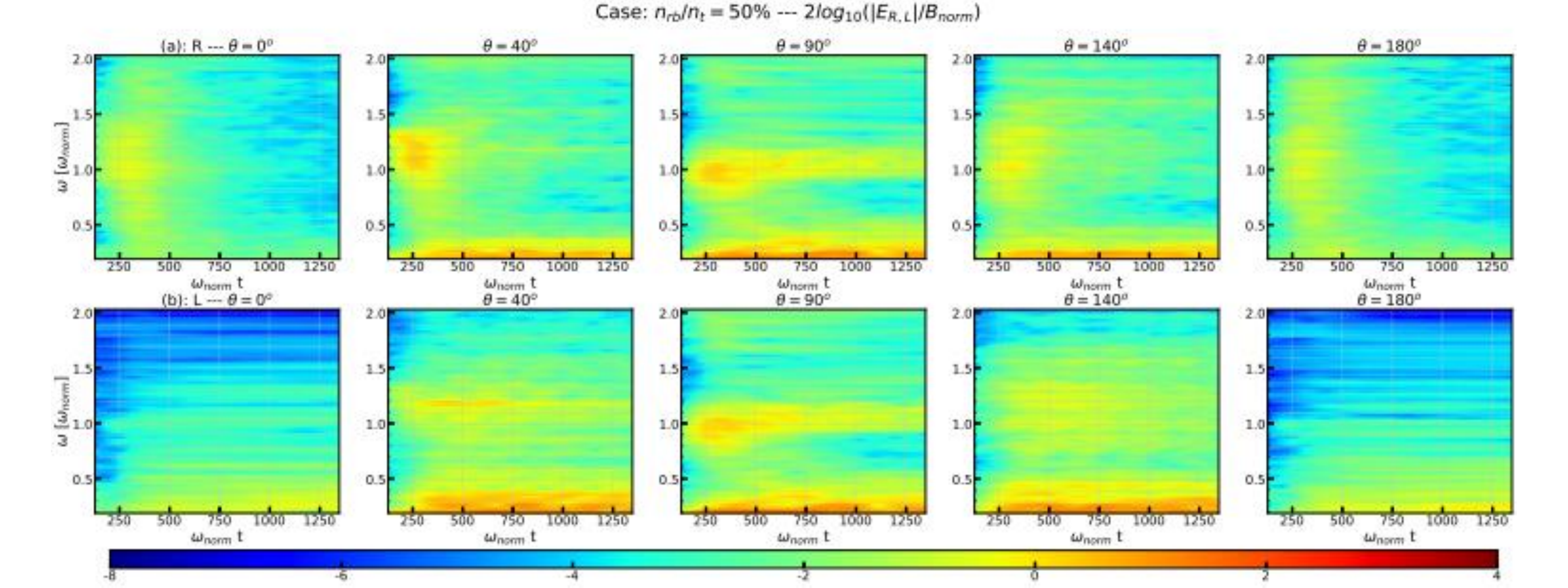}
            \caption{Spectrograms of the RPTE and LPTE in escaping electromagnetic waves (with $\omega > \omega_{pe}$ and $|c k/\omega| < 1$)
                              along different wave propagation directions $\theta$
                              ($ = 0^{\circ}, 40^{\circ}, 90^{\circ}, 140^{\circ}$ and $180^{\circ}$ from the left to right column, respectively)
                              for plasmas with $n_{rb}/n_{t} = 5\%$ --- top two rows, $20\%$ --- middle two rows, $50\%$ --- bottom two rows
                              and $\omega_{ce}/\omega_{pe} = 5.0$.
                              In each $n_{rb}/n_{t}$ case, row \textbf{(a)}  and \textbf{(b)} are for the RPTE and LPTE, respectively.
                              All panels use the same color bar shown at the bottom.
                  }
          \label{spectrogram_Escape}
      \end{center}
      \end{figure*}

\begin{figure*}[htbp]
      \begin{center}
          \includegraphics[width=1.0\textwidth]{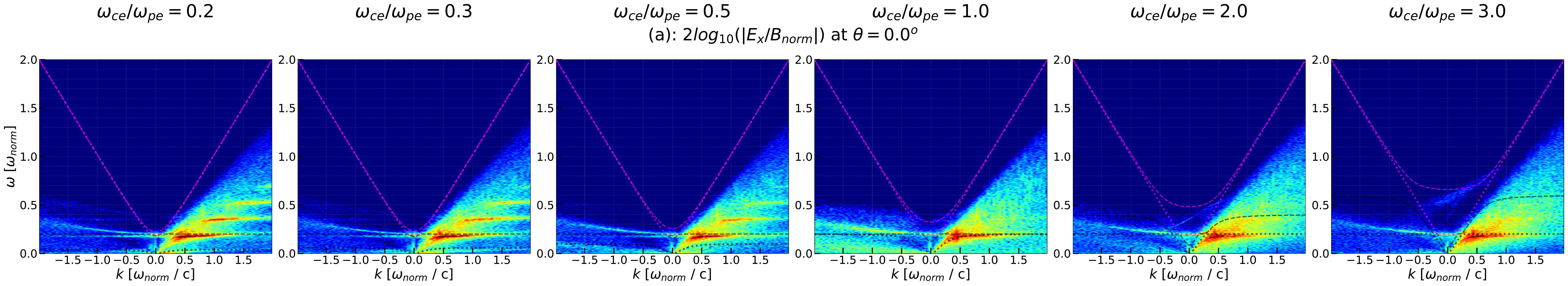}\\
          \includegraphics[width=1.0\textwidth]{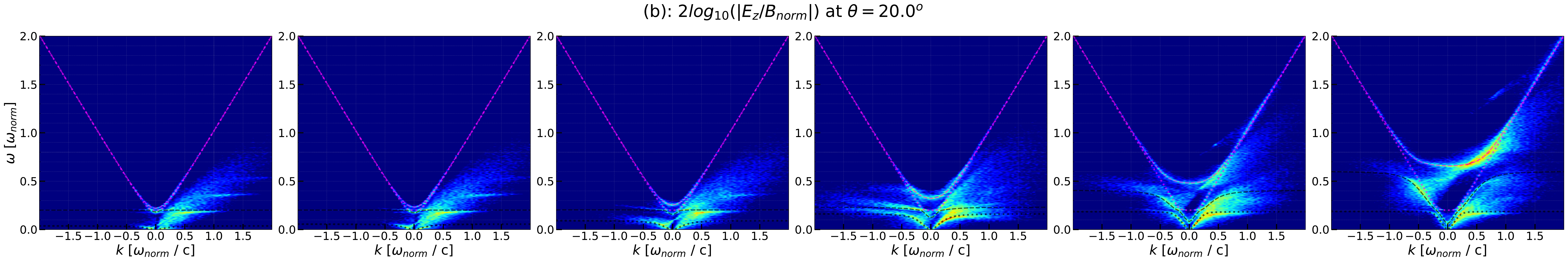}\\
          \includegraphics[width=1.0\textwidth]{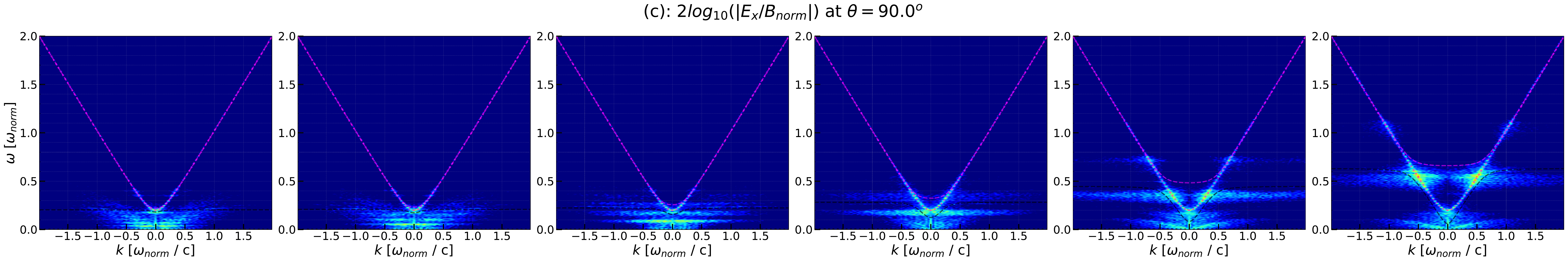}\\
          \includegraphics[width=1.0\textwidth]{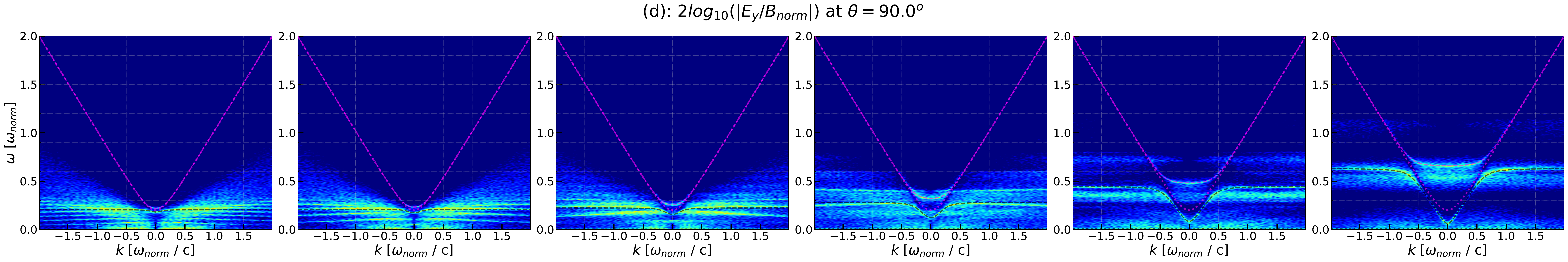}\\
          \includegraphics[width=1.0\textwidth]{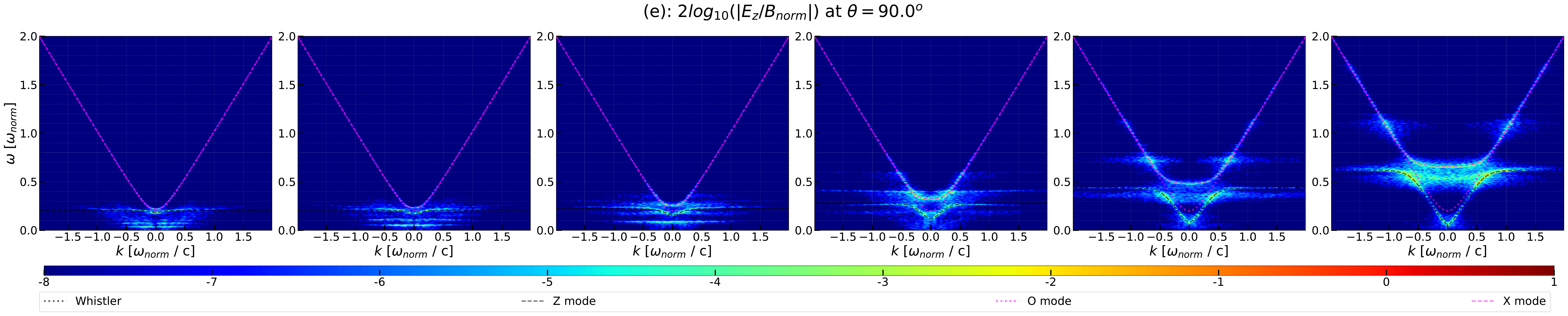}
          \caption{Similar to Fig.\ref{dispersion}, but for plasmas with  $n_{rb}/n_{t} = 5\%$ but
                            different $\omega_{ce}/\omega_{pe}$, from the left to right column, $\omega_{ce}/\omega_{pe} = 0.2, 0.3, 0.5, 1, 2, 3$, respectively.
                            Row (\textbf{a}) is for $E_{x}$ with $\theta = 0^{\circ}$.
                            Row (\textbf{b}) is for $E_{z}$ with $\theta = 20^{\circ}$.
                            Row (\textbf{c}) is for $E_{x}$ with $\theta = 90^{\circ}$.
                            Row (\textbf{d}) is for $E_{y}$ with $\theta = 90^{\circ}$.
                            And row (\textbf{e}) is for $E_{z}$ with $\theta = 90^{\circ}$.
                            Also note that the color scale in this figure is, however, different from that in Fig.\ref{dispersion}.
                  }
          \label{Dispersion_Relation_2}
      \end{center}
      \end{figure*}

Properties of the electromagnetic whistler, Z, O and X-mode waves were studied in detail.
In order to obtain the energy evolution of these
electromagnetic waves, we adopted a Gaussian filter centered on
the wave dispersion surfaces of the magnetized cold plasmas
in the wavevector-frequency ($\vec{k} - \omega$) space and
assumed a frequency broadening of all excited electromagnetic waves $\sigma = 0.05 \omega_{norm}$,
which is frequently seen in the spectrograms of escaping electromagnetic waves (Fig.\ref{spectrogram_Escape}).
For a convergence test, we also carried calculations with $\sigma = 0.03 \omega_{norm}$
and $0.2 \omega_{norm}$, while the frequency resolution in our PIC simulations is $3\times10^{-3} \omega_{norm}$.
We found that results with $\sigma = 0.03 \omega_{norm}$ and $0.05 \omega_{norm}$ are almost the same.
We admit that this Gaussian filter method for energy of a wave mode might contain contamination from other waves
particularly when the dispersion relations of two wave modes are close to each other. Additionally we
might also underestimate the energy of diffusive waves with a fixed $\sigma$ for all wave modes.
However, in general, this method gives more accurate information about the mode energy
compared to estimates given in the previous studies,
~\citealp[see, e.g.,][]{Pritchett1984JGR....89.8957P, Lee_etal_2009PhRvL.103j5101L, Lee_etal_2011PhPl...18i2110L}.

Based on the Gaussian filter method, we carried out detailed studies of the dependence of the excited electromagnetic
whistler, Z, O, and X-mode properties on the ring-beam electron density for $\omega_{ce}/\omega_{pe} = 5$ and found:
\begin{itemize}
\item Both the beam and ECM instabilities contribute to the excitation of these electromagnetic waves.
            The beam instability dominates the saturation of the whistler, Z modes as well as O mode in plasmas with
            dense ring-beam electrons. But the X-mode waves cannot be efficiently excited by the beam instability only,
            especially with tenuous ring-beam electrons.
\item In the growth phase of waves, the free energy dissipation rate and the wave growth rate, in general,
            increase with the increase of the ring-beam density. The growth rates of the whistler and
            Z modes are comparable but higher than that of the O and X modes.
\item The saturation level of different electromagnetic wave modes also increase with the increase of $n_{rb}/n_{t}$,
            except for the X-mode waves produced by low density ring-beam electrons.
            The X-mode saturation level for $n_{rb}/n_{t} = 5\%$ is actually higher than that for $n_{rb}/n_{t} = 10\%$.
\item The energy of each electromagnetic wave mode is strongly anisotropic. This anisotropy is
            suppressed in plasmas with dense ring-beam electrons, where the X mode has the strongest
            anisotropy.
\end{itemize}

Although only the O and X modes are remotely detectable and related more to the remote observations of the
SRBs, the individual investigation on the energy property of all the whistler, Z, O and X modes obtained here
will complement their linear and quasi-linear theoretical studies. In theoretical studies,
nonlinear processes (e.g., wave-wave, wave-particle cyclotron resonance interactions) and
evolution of the plasma system itself (e.g., population of the energetic and background electrons, plasma temperature)
cannot be usually treated self-consistently. In this study, we found that all these processes affect the energy
saturation and growth rate of those wave modes.

Harmonic excitation of $\omega_{pe}$ and $\omega_{ce}$ are studied for differently magnetized plasmas
with $0.2 \leq \omega_{ce}/\omega_{pe} \leq 5$.
Over all simulated cases, one can find that harmonics of $\omega_{ce}$ can be always excited,
while there is no obvious excitation for the escaping harmonics of $\omega_{pe}$.
Such results were also found by \citealp{Ganse_etal_2012SoPh..280..551G}.
As mentioned by~\citealp{Thurgood&Tsiklauri_2015A&A...584A..83T} as well as
according to Eq.(6.80) in~\citealp{Melrose_1986islp.book.....M} for the probability of the
$L + L' \rightarrow T_{2\omega_{pe}}$ process, the key reason for the
absence of the escaping harmonics of $\omega_{pe}$ in our simulated plasmas
could be the weak intensity (in plasmas with tenuous ring-beam electrons $n_{rb}/n_{t} \leq 20 \%$)
and/or the predominant wave intensity located at small wavenumber $|k|$ (due to Landau damping in plasmas with
dense ring-beam electrons) of both the parallel and antiparallel-propagating electrostatic Langmuir waves.
Based on the study of~\citealp{Thurgood&Tsiklauri_2015A&A...584A..83T}, escaping harmonics of $\omega_{pe}$
are more likely present in plasmas with very tenuous beam electrons $n_{rb}/n_{t} < 0.6\%$ and generation of the
harmonics of $\omega_{pe}$ are very sensitive to the chosen parameters, like the populations of the beam and oppositely
drifting electrons, the drifting velocity of the beam electrons, the magnetized condition, etc.
\citep{Rhee_etal_2009ApJ...694..618R, Umeda_2010JGRA..115.1204U, Ganse_etal_2012SoPh..280..551G, Thurgood&Tsiklauri_2015A&A...584A..83T, Henri_etal_2019_doi:10.1029/2018JA025707}.
The setup parameters used here, however, favour more the excitation of the beam mode instead of Langmuir waves.

Non-escaping harmonics of $\omega_{pe}$ can be
excited only when $\omega_{ce}/\omega_{pe} < 1$. Higher and higher non-escaping harmonics
of $\omega_{pe}$ are driven with the decrease of $\omega_{ce}/\omega_{pe}$.
No matter whether escaping harmonics of $\omega_{pe}$ are excited or not,
the beam instability can always lead to the excitation of the Langmuir, whistler, Z and O-mode waves.
ECM excitation of the harmonics of $\omega_{ce}$, however,  does
not depend on the ratio of $\omega_{ce}/\omega_{pe}$. In other words, $s_{h} \omega_{ce}$ (as well as X mode)
will be excited as long as the free energy $\partial f /\partial u_{\perp} > 0$ exists for the ECM instability. The requirement
of $\omega_{ce}/\omega_{pe} > 1$ in the ECM emission theory is for an efficient generation of observable escaping
emissions by remote detections, i.e., the escape condition (see Sect.\ref{Introduction}).

In addition, to compare with solar radio observations, we also obtained the polarization properties
(circular polarization degree --- CPD, spectrogram) of the electromagnetic waves, in particular of the
escaping electromagnetic waves with $\omega > \omega_{pe}$ and $|c k/\omega| < 1$.
In summary, escaping emission decreases rapidly with the decrease of $\omega_{ce}/\omega_{pe}$.
In weakly magnetized plasmas with $\omega_{ce}/\omega_{pe} < 1$, most of the excited escaping electromagnetic waves
are located close to the plasma frequency $\omega_{pe}$ and their energies are significantly weaker and negligible
compared to those of plasmas with $\omega_{ce}/\omega_{pe} > 1$, i.e., strongly magnetized plasmas.
Energy and polarization properties of the escaping electromagnetic waves in strongly magnetized plasmas
depend on the density ratio $n_{rb}/n_{t}$:
\begin{itemize}
      \item When $n_{rb}/n_{t}  \leq  10\%$, the ECM instability dominates the excitation of the escaping electromagnetic waves.
                  Right-handed polarized electric fields (RPTEs) dominate the transverse electric field energies of the
                  escaping electromagnetic waves. And right-handed polarized CPDs can be expected along many wave
                  propagation directions. Moreover the strongest escaping emission is in the same side of the
                  ring-beam electron propagation direction (i.e., $\theta < 90^{\circ}$).
                  These properties may explain observed properties of solar radio spikes.
      \item For plasmas with dense ring-beam electrons $n_{rb}/n_{t}  \geq  20\%$,  the escaping emissions are dominated by
                  the ECM instability at the beginning. The beam instability plays a more important role later on,
                  giving rise to more isotropic and left-handed polarized electric field (LPTE) dominated emissions, which may be
                  applied to observations of Type \Rmnum{3} bursts.
\end{itemize}
Considering the population reduction of the ring-beam electrons during their propagation in the solar corona, these
results might explain the increased time delay of the Type \Rmnum{3} bursts ($< 1 s$) and solar radio spikes ($2 - 5 s$) to
the hard X-ray bursts (see~\citealp{Fleishman&Melniko_1998PhyU...41.1157F} for a review of the solar radio spikes).
Moreover diversities in the SRBs' CPD and spectrogram observations
may already originate from their generation sites.

Our results above deal with properties of waves driven by energetic
ring-beam electrons at the site of wave generation, where the global charge and current neutralities are maintained
via protons and drifting background electrons, respectively.
Note that the remotely observed
energy and polarization properties of the SRBs might deviate from those in their source regions,
due to some propagation effects of the electromagnetic waves (e.g., reflection, refraction, Faraday rotation,
energy absorption via wave-particle interaction) along the wave path in the IPM.
For an accurate prediction of the remote SRB observations, one, hence, still needs to combine our simulations
with a proper model describing the wave propagation effects in the IPM~\citep{Li_etal_2008JGRA..113.6104L, Li_etal_2008JGRA..113.6105L, Li_etal_2009JGRA..114.2104L}.
This study is, however, still meaningful to gain insight into the generation mechanisms
of the original coherent emission by energetic ring-beam electrons in the neutral and current-free
solar coronal plasmas, where all background electrons drift
oppositely to the ring-beam electrons to fully compensate the current induced by the ring-beam electrons.

Finally, we note that dynamic processes in plasmas with energetic ring-beam electrons contain not only the
excitations of waves but also plasma heating and electron acceleration. Significant plasma heating and electron
acceleration can be expected, particularly, in plasmas with dense ring-beam electrons, where a significant fraction of the
ring-beam electrons can be reflected, making the system more or less symmetric with respect to the plane perpendicular
to the ambient magnetic field $\vec{B_{0}}$. Due to the acceleration of electrons, a double power-law distribution is formed in the
high energy tail  ($\gamma-1 > 0.1 \sim 50~keV$) of the electron energy distribution when the wave-particle
plasma system reaches an equilibrium.

\acknowledgments
We gratefully acknowledge the developers of the ACRONYM code,
the Verein zur F\"{o}rderung kinetischer Plasmasimulationen
e.V.  In particular, we thank Patrick Kilian for helpful discussions and
valuable suggestions.
X. Zhou thanks the Chinese Academy of Sciences as well as the International
Cooperation and Exchange Project of National Natural Science Foundation of
China, 11761131007 for support.
P.~A. Mu\~noz acknowledges his financial support by the German Science Foundation
DFG, project MU-4255/1-1.
We also acknowledge the computing resources in the Max Planck Computing and
Data Facility (MPCDF, formerly known as RZG) at Garching, Germany
and the Max Planck Institute for Solar System Research, Germany.
\pagebreak
\clearpage
%
%
\pagebreak
\clearpage
%
%
\pagebreak
\clearpage
%
%
%
%
\pagebreak
\clearpage
%
%
%
%
\pagebreak
\clearpage
%
%
%
%
\pagebreak
\clearpage
%
%
%
%
\pagebreak
\clearpage
%
%
%
%
\pagebreak
\clearpage
%
%
%
\pagebreak
\clearpage
%
%
%
%
\pagebreak
\clearpage
%
%


\begin{thebibliography}{}
\expandafter\ifx\csname natexlab\endcsname\relax\def\natexlab#1{#1}\fi
\providecommand{\url}[1]{\href{#1}{#1}}
\providecommand{\dodoi}[1]{doi:~\href{http://doi.org/#1}{\nolinkurl{#1}}}
\providecommand{\doeprint}[1]{\href{http://ascl.net/#1}{\nolinkurl{http://ascl.net/#1}}}
\providecommand{\doarXiv}[1]{\href{https://arxiv.org/abs/#1}{\nolinkurl{https://arxiv.org/abs/#1}}}

\bibitem[{{Alvarez} \& {Haddock}(1973)}]{Alvarez&Haddock_1973SoPh...29..197A}
{Alvarez}, H., \& {Haddock}, F.~T. 1973, \solphys, 29, 197,
  \dodoi{10.1007/BF00153449}

\bibitem[{{Andre}(1985)}]{Andre1985JPlPh..33....1A}
{Andre}, M. 1985, Journal of Plasma Physics, 33, 1,
  \dodoi{10.1017/S0022377800002270}

\bibitem[{{Aschwanden}(2002)}]{Aschwanden_2002SSRv..101....1A}
{Aschwanden}, M.~J. 2002, \ssr, 101, 1, \dodoi{10.1023/A:1019712124366}

\bibitem[{{Aschwanden}(2005)}]{Aschwanden2005psci.book.....A}
---. 2005, {Physics of the Solar Corona. An Introduction with Problems and
  Solutions (2nd edition)}

\bibitem[{{Bellan}(2006)}]{Bellan_2006fpp..book.....B}
{Bellan}, P.~M. 2006, {Fundamentals of Plasma Physics}

\bibitem[{{Benz}(2002)}]{Benz_2002ASSL..279.....B}
{Benz}, A., ed. 2002, Astrophysics and Space Science Library, Vol. 279, {Plasma
  Astrophysics, second edition}

\bibitem[{{Benz}(1985)}]{Benz_1985SoPh...96..357B}
{Benz}, A.~O. 1985, \solphys, 96, 357, \dodoi{10.1007/BF00149690}

\bibitem[{{Benz}(1994)}]{Benz_1994SSRv...68..135B}
---. 1994, \ssr, 68, 135, \dodoi{10.1007/BF00749132}

\bibitem[{{Benz}(2008)}]{Benz_2008LRSP....5....1B}
---. 2008, Living Reviews in Solar Physics, 5, 1, \dodoi{10.12942/lrsp-2008-1}

\bibitem[{{Bessho} {et~al.}(2014){Bessho}, {Chen}, {Shuster}, \&
  {Wang}}]{Bessho_etal_2014GeoRL..41.8688B}
{Bessho}, N., {Chen}, L.~J., {Shuster}, J.~R., \& {Wang}, S. 2014, \grl, 41,
  8688, \dodoi{10.1002/2014GL062034}

\bibitem[{{Birdsall} \& {Langdon}(1991)}]{Birdsall&Langdon_1991ppcs.book.....B}
{Birdsall}, C.~K., \& {Langdon}, A.~B. 1991, {Plasma Physics via Computer
  Simulation}

\bibitem[{{Bittencourt}(2004)}]{Bittencourt_2004fopp.book.....B}
{Bittencourt}, J.~A. 2004, {Fundamentals of Plasma Physics} (Springer-Verlag)

\bibitem[{{Brown} \& {Bingham}(1984)}]{Brown_1984A&A...131L..11B}
{Brown}, J.~C., \& {Bingham}, R. 1984, \aap, 131, L11

\bibitem[{B\"uchner {et~al.}(2018)B\"uchner, Kilian, Mu\~noz Sep\'ulveda,
  Spanier, Widmer, Zhou, \& Jain}]{Buechner:2018-Mercury}
B\"uchner, J., Kilian, P., Mu\~noz Sep\'ulveda, P., {et~al.} 2018, in {Magnetic
  Fields in the Solar System: Planets, Moons and Solar Wind Interactions}, ed.
  H.~L\"uhr, S.~Wicht, J.and~Gilder, \& M.~Holschneider (Cham: Springer),
  201--240

\bibitem[{B{\"u}chner \& Kuska(1996)}]{BuchnerKuska:1996b}
B{\"u}chner, J., \& Kuska, J.-P. 1996, J. Geomag. Geoelectr., 48, 781

\bibitem[{{Budden}(1988)}]{Budden_1988prw..book.....B}
{Budden}, K.~G. 1988, {The Propagation of Radio Waves}, 688

\bibitem[{Cairns(1989)}]{Cairns1989}
Cairns, I.~H. 1989, Phys. Fluids B Plasma Phys., 1, 204,
  \dodoi{10.1063/1.859088}

\bibitem[{{Cairns} {et~al.}(2018){Cairns}, {Lobzin}, {Donea}, {Tingay},
  {McCauley}, {Oberoi}, {Duffin}, {Reiner}, {Hurley-Walker}, {Kudryavtseva},
  {Melrose}, {Harding}, {Bernardi}, {Bowman}, {Cappallo}, {Corey}, {Deshpand
  e}, {Emrich}, {Goeke}, {Hazelton}, {Johnston-Hollitt}, {Kaplan}, {Kasper},
  {Kratzenberg}, {Lonsdale}, {Lynch}, {McWhirter}, {Mitchell}, {Morales},
  {Morgan}, {Ord}, {Prabu}, {Roshi}, {Shankar}, {Srivani}, {Subrahmanyan},
  {Wayth}, {Waterson}, {Webster}, {Whitney}, {Williams}, \&
  {Williams}}]{Cairns_etal_2018NatSR...8.1676C}
{Cairns}, I.~H., {Lobzin}, V.~V., {Donea}, A., {et~al.} 2018, Scientific
  Reports, 8, 1676, \dodoi{10.1038/s41598-018-19195-3}

\bibitem[{{Carozzi} {et~al.}(2001){Carozzi}, {Thid{\'e}}, {Leyser}, {Komrakov},
  {Frolov}, {Grach}, \& {Sergeev}}]{Carozzi_etal_2001JGR...10621395C}
{Carozzi}, T.~D., {Thid{\'e}}, B., {Leyser}, T.~B., {et~al.} 2001, \jgr, 106,
  21395, \dodoi{10.1029/2001JA900004}

\bibitem[{{Chen} {et~al.}(2015){Chen}, {Bastian}, {Shen}, {Gary}, {Krucker}, \&
  {Glesener}}]{Chen_etal_2015Sci...350.1238C}
{Chen}, B., {Bastian}, T.~S., {Shen}, C., {et~al.} 2015, Science, 350, 1238,
  \dodoi{10.1126/science.aac8467}

\bibitem[{{Chen} {et~al.}(2018){Chen}, {Yu}, {Battaglia}, {Farid}, {Savcheva},
  {Reeves}, {Krucker}, {Bastian}, {Guo}, \&
  {Tassev}}]{Chen_etal_2018ApJ...866...62C}
{Chen}, B., {Yu}, S., {Battaglia}, M., {et~al.} 2018, \apj, 866, 62,
  \dodoi{10.3847/1538-4357/aadb89}

\bibitem[{{Chen} {et~al.}(2013){Chen}, {Thorne}, {Shprits}, \&
  {Ni}}]{Chen_etal_2013JGRA..118.2185C}
{Chen}, L., {Thorne}, R.~M., {Shprits}, Y., \& {Ni}, B. 2013, Journal of
  Geophysical Research (Space Physics), 118, 2185, \dodoi{10.1002/jgra.50260}

\bibitem[{{Chen} {et~al.}(2017){Chen}, {Wu}, {Zhao}, \&
  {Tang}}]{Chen_etal_2017JGRA..122...35C}
{Chen}, L., {Wu}, D.~J., {Zhao}, G.~Q., \& {Tang}, J.~F. 2017, Journal of
  Geophysical Research (Space Physics), 122, 35, \dodoi{10.1002/2016JA023312}

\bibitem[{{Comi{\c s}el} {et~al.}(2013){Comi{\c s}el}, {Verscharen}, {Narita},
  \& {Motschmann}}]{Comisel_etal_2013PhPl...20i0701C}
{Comi{\c s}el}, H., {Verscharen}, D., {Narita}, Y., \& {Motschmann}, U. 2013,
  Physics of Plasmas, 20, 090701, \dodoi{10.1063/1.4820936}

\bibitem[{{Dawson}(1983)}]{Dawson_1983RvMP...55..403D}
{Dawson}, J.~M. 1983, Reviews of Modern Physics, 55, 403,
  \dodoi{10.1103/RevModPhys.55.403}

\bibitem[{{De Groot}(1962)}]{degroot1962}
{De Groot}, T. 1962, Int. Bull. Solar Radio Obs. Europe, 9 3

\bibitem[{{Drake} {et~al.}(2003){Drake}, {Swisdak}, {Cattell}, {Shay},
  {Rogers}, \& {Zeiler}}]{Drake_etal_2003Sci...299..873D}
{Drake}, J.~F., {Swisdak}, M., {Cattell}, C., {et~al.} 2003, Science, 299, 873,
  \dodoi{10.1126/science.1080333}

\bibitem[{{Droege} \& {Riemann}(1961)}]{droege1961}
{Droege}, F., \& {Riemann}, P. 1961, Int. Bull. Solar Radio Obs. Europe, 8 6

\bibitem[{{Dulk}(1985)}]{Dulk_1985ARA&A..23..169D}
{Dulk}, G.~A. 1985, \araa, 23, 169, \dodoi{10.1146/annurev.aa.23.090185.001125}

\bibitem[{{Elgar{\o}y}(1961)}]{Elgaroy_1961ApNr....7..123E}
{Elgar{\o}y}, {\O}. 1961, Astrophysica Norvegica, 7, 123

\bibitem[{{Ergun} {et~al.}(1998){Ergun}, {Larson}, {Lin}, {McFadden},
  {Carlson}, {Anderson}, {Muschietti}, {McCarthy}, {Parks}, {Reme}, {Bosqued},
  {D'Uston}, {Sanderson}, {Wenzel}, {Kaiser}, {Lepping}, {Bale}, {Kellogg}, \&
  {Bougeret}}]{Ergun_etal_1998ApJ...503..435E}
{Ergun}, R.~E., {Larson}, D., {Lin}, R.~P., {et~al.} 1998, The Astrophysical
  Journal, 503, 435, \dodoi{10.1086/305954}

\bibitem[{{Fleishman} \&
  {Mel'nikov}(1998)}]{Fleishman&Melniko_1998PhyU...41.1157F}
{Fleishman}, G.~D., \& {Mel'nikov}, V.~F. 1998, Physics Uspekhi, 41, 1157,
  \dodoi{10.1070/PU1998v041n12ABEH000510}

\bibitem[{{Fletcher} \& {Hudson}(2008)}]{Fletcher&Hudson_2008ApJ...675.1645F}
{Fletcher}, L., \& {Hudson}, H.~S. 2008, \apj, 675, 1645,
  \dodoi{10.1086/527044}

\bibitem[{{Freund} {et~al.}(1983){Freund}, {Wong}, {Wu}, \&
  {Xu}}]{Freund_etal_1983PhFl...26.2263F}
{Freund}, H.~P., {Wong}, H.~K., {Wu}, C.~S., \& {Xu}, M.~J. 1983, Physics of
  Fluids, 26, 2263, \dodoi{10.1063/1.864383}

\bibitem[{Fuselier {et~al.}(1985)Fuselier, Gurnett, \&
  Fitzenreiter}]{Fuselier1985}
Fuselier, S.~A., Gurnett, D.~A., \& Fitzenreiter, R.~J. 1985, J. Geophys. Res.,
  90, 3935, \dodoi{10.1029/JA090iA05p03935}

\bibitem[{{Ganse} {et~al.}(2012{\natexlab{a}}){Ganse}, {Kilian}, {Spanier}, \&
  {Vainio}}]{Ganse_etal_2012ApJ...751..145G}
{Ganse}, U., {Kilian}, P., {Spanier}, F., \& {Vainio}, R. 2012{\natexlab{a}},
  \apj, 751, 145, \dodoi{10.1088/0004-637X/751/2/145}

\bibitem[{{Ganse} {et~al.}(2012{\natexlab{b}}){Ganse}, {Kilian}, {Vainio}, \&
  {Spanier}}]{Ganse_etal_2012SoPh..280..551G}
{Ganse}, U., {Kilian}, P., {Vainio}, R., \& {Spanier}, F. 2012{\natexlab{b}},
  \solphys, 280, 551, \dodoi{10.1007/s11207-012-0077-7}

\bibitem[{{Gaponov}(1959)}]{Gaponov59}
{Gaponov}, A. 1959, Izv VUZ, Radiofizika, 2, 450

\bibitem[{{Gary}(1993)}]{Gary_1993tspm.book.....G}
{Gary}, S.~P. 1993, {Theory of Space Plasma Microinstabilities}, 193

\bibitem[{{Ginzburg} \&
  {Zhelezniakov}(1958)}]{Ginzburg&Zhelezniakov_1958SvA.....2..653G}
{Ginzburg}, V.~L., \& {Zhelezniakov}, V.~V. 1958, \sovast, 2, 653

\bibitem[{{Goldreich} \& {Julian}(1969)}]{Goldreich&Julian_1969ApJ...157..869G}
{Goldreich}, P., \& {Julian}, W.~H. 1969, \apj, 157, 869,
  \dodoi{10.1086/150119}

\bibitem[{Graham {et~al.}(2017)Graham, Khotyaintsev, Vaivads, Norgren,
  Andr{\'{e}}, Webster, Burch, Lindqvist, Ergun, Torbert, Paterson, Gershman,
  Giles, Magnes, \& Russell}]{Graham2017}
Graham, D.~B., Khotyaintsev, Y.~V., Vaivads, A., {et~al.} 2017, Phys. Rev.
  Lett., 119, 025101, \dodoi{10.1103/PhysRevLett.119.025101}

\bibitem[{Graham {et~al.}(2018)Graham, Vaivads, Khotyaintsev, Andr{\'{e}}, {Le
  Contel}, Malaspina, Lindqvist, Wilder, Ergun, Gershman, Giles, Magnes,
  Russell, Burch, \& Torbert}]{Graham2018}
Graham, D.~B., Vaivads, A., Khotyaintsev, Y.~V., {et~al.} 2018, J. Geophys.
  Res. Sp. Phys., 123, 2630, \dodoi{10.1002/2017JA025034}

\bibitem[{Henri {et~al.}(2019)Henri, Sgattoni, Briand, Amiranoff, \&
  Riconda}]{Henri_etal_2019_doi:10.1029/2018JA025707}
Henri, P., Sgattoni, A., Briand, C., Amiranoff, F., \& Riconda, C. 2019, J.
  Geophys. Res. Sp. Phys., 2018JA025707, \dodoi{10.1029/2018JA025707}

\bibitem[{{Hockney}(1971)}]{Hockney_1971JCoPh...8...19H}
{Hockney}, R.~W. 1971, Journal of Computational Physics, 8, 19,
  \dodoi{10.1016/0021-9991(71)90032-5}

\bibitem[{{Kainer} \& {MacDowall}(1996)}]{Kainer_etal_1996JGR...101..495K}
{Kainer}, S., \& {MacDowall}, R.~J. 1996, \jgr, 101, 495,
  \dodoi{10.1029/95JA02026}

\bibitem[{{Karlick{\'y}} \&
  {B{\'a}rta}(2009)}]{Karlicky&Barta_2009NPGeo..16..525K}
{Karlick{\'y}}, M., \& {B{\'a}rta}, M. 2009, Nonlinear Processes in Geophysics,
  16, 525

\bibitem[{{Karlick{\'y}} \&
  {B{\'a}rta}(2011)}]{Karlicky&Barta2011IAUS..274..252K}
{Karlick{\'y}}, M., \& {B{\'a}rta}, M. 2011, in IAU Symposium, Vol. 274,
  Advances in Plasma Astrophysics, ed. A.~{Bonanno}, E.~{de Gouveia Dal Pino},
  \& A.~G. {Kosovichev}, 252--254

\bibitem[{{Kempf} {et~al.}(2016){Kempf}, {Kilian}, \&
  {Spanier}}]{Kempf_etal_2016A&A...585A.132K}
{Kempf}, A., {Kilian}, P., \& {Spanier}, F. 2016, \aap, 585, A132,
  \dodoi{10.1051/0004-6361/201527521}

\bibitem[{{Khodachenko} {et~al.}(2009){Khodachenko}, {Zaitsev}, {Kislyakov}, \&
  {Stepanov}}]{Khodachenko_etal_2009SSRv..149...83K}
{Khodachenko}, M.~L., {Zaitsev}, V.~V., {Kislyakov}, A.~G., \& {Stepanov},
  A.~V. 2009, \ssr, 149, 83, \dodoi{10.1007/s11214-009-9538-1}

\bibitem[{Kilian {et~al.}(2017)Kilian, Mu{\~n}oz, Schreiner, \&
  Spanier}]{Kilian_etal_2017_PoP}
Kilian, P., Mu{\~n}oz, P.~A., Schreiner, C., \& Spanier, F. 2017, Journal of
  Plasma Physics, 83, 707830101, \dodoi{10.1017/S0022377817000149}

\bibitem[{Klimas(1983)}]{Klimas1983}
Klimas, A.~J. 1983, J. Geophys. Res. Sp. Phys., 88, 9081,
  \dodoi{10.1029/JA088iA11p09081}

\bibitem[{Landau(1946)}]{Landau1946On}
Landau, L.~D. 1946, J. Phys.(USSR), 10, 25

\bibitem[{{Lapenta}(2012)}]{Lapenta_2012JCoPh.231..795L}
{Lapenta}, G. 2012, Journal of Computational Physics, 231, 795,
  \dodoi{10.1016/j.jcp.2011.03.035}

\bibitem[{{Lee} {et~al.}(2011){Lee}, {Omura}, \&
  {Lee}}]{Lee_etal_2011PhPl...18i2110L}
{Lee}, K.~H., {Omura}, Y., \& {Lee}, L.~C. 2011, Physics of Plasmas, 18,
  092110, \dodoi{10.1063/1.3626562}

\bibitem[{{Lee} {et~al.}(2009){Lee}, {Omura}, {Lee}, \&
  {Wu}}]{Lee_etal_2009PhRvL.103j5101L}
{Lee}, K.~H., {Omura}, Y., {Lee}, L.~C., \& {Wu}, C.~S. 2009, Physical Review
  Letters, 103, 105101, \dodoi{10.1103/PhysRevLett.103.105101}

\bibitem[{{Lee} {et~al.}(1980){Lee}, {Kan}, \&
  {Wu}}]{Lee_etal_1980P&SS...28..703L}
{Lee}, L.~C., {Kan}, J.~R., \& {Wu}, C.~S. 1980, \planss, 28, 703,
  \dodoi{10.1016/0032-0633(80)90115-4}

\bibitem[{{Lee} \& {Wu}(1980)}]{Lee&Wu_1980PhFl...23.1348L}
{Lee}, L.~C., \& {Wu}, C.~S. 1980, Physics of Fluids, 23, 1348,
  \dodoi{10.1063/1.863148}

\bibitem[{{Li} {et~al.}(2008{\natexlab{a}}){Li}, {Cairns}, \&
  {Robinson}}]{Li_etal_2008JGRA..113.6104L}
{Li}, B., {Cairns}, I.~H., \& {Robinson}, P.~A. 2008{\natexlab{a}}, Journal of
  Geophysical Research (Space Physics), 113, A06104,
  \dodoi{10.1029/2007JA012957}

\bibitem[{{Li} {et~al.}(2008{\natexlab{b}}){Li}, {Cairns}, \&
  {Robinson}}]{Li_etal_2008JGRA..113.6105L}
---. 2008{\natexlab{b}}, Journal of Geophysical Research (Space Physics), 113,
  A06105, \dodoi{10.1029/2007JA012958}

\bibitem[{{Li} {et~al.}(2009){Li}, {Cairns}, \&
  {Robinson}}]{Li_etal_2009JGRA..114.2104L}
---. 2009, Journal of Geophysical Research (Space Physics), 114, A02104,
  \dodoi{10.1029/2008JA013687}

\bibitem[{{Lin} {et~al.}(1981){Lin}, {Potter}, {Gurnett}, \&
  {Scarf}}]{Lin_etal_1981ApJ...251..364L}
{Lin}, R.~P., {Potter}, D.~W., {Gurnett}, D.~A., \& {Scarf}, F.~L. 1981, \apj,
  251, 364, \dodoi{10.1086/159471}

\bibitem[{Matsumoto \& Omura(1993)}]{Matsumoto1993}
Matsumoto, H., \& Omura, Y. 1993, {Computer Space Plasma Physics : Simulation
  Techniques and Software} (Terra Scientific Publishing Company).
\newblock \url{https://www.terrapub.co.jp/e-library/cspp/}

\bibitem[{{McMaster}(1954)}]{McMaster1954AmJPh..22..351M}
{McMaster}, W.~H. 1954, American Journal of Physics, 22, 351,
  \dodoi{10.1119/1.1933744}

\bibitem[{{Melrose}(1973)}]{Melrose_1973AuJPh..26..229M}
{Melrose}, D.~B. 1973, Australian Journal of Physics, 26, 229,
  \dodoi{10.1071/PH730229}

\bibitem[{{Melrose}(1986)}]{Melrose_1986islp.book.....M}
---. 1986, {Instabilities in Space and Laboratory Plasmas}, 288

\bibitem[{{Melrose}(1990)}]{Melrose_1990SoPh..130....3M}
---. 1990, \solphys, 130, 3, \dodoi{10.1007/BF00156775}

\bibitem[{{Melrose}(1994)}]{Melrose_1994SSRv...68..159M}
---. 1994, \ssr, 68, 159, \dodoi{10.1007/BF00749134}

\bibitem[{{Melrose}(2017)}]{Melrose_2017RvMPP...1....5M}
---. 2017, Reviews of Modern Plasma Physics, 1, 5,
  \dodoi{10.1007/s41614-017-0007-0}

\bibitem[{{Melrose} \&
  {Wheatland}(2016)}]{Melrose&Wheatland_2016SoPh..291.3637M}
{Melrose}, D.~B., \& {Wheatland}, M.~S. 2016, \solphys, 291, 3637,
  \dodoi{10.1007/s11207-016-1006-y}

\bibitem[{{Morosan} {et~al.}(2016){Morosan}, {Zucca}, {Bloomfield}, \&
  {Gallagher}}]{Morosan_etal_2016A&A...589L...8M}
{Morosan}, D.~E., {Zucca}, P., {Bloomfield}, D.~S., \& {Gallagher}, P.~T. 2016,
  \aap, 589, L8, \dodoi{10.1051/0004-6361/201628392}

\bibitem[{{Mu{\~n}oz} \&
  {B{\"u}chner}(2016)}]{Munoz&Buechner_2016PhPl...23j2103M}
{Mu{\~n}oz}, P.~A., \& {B{\"u}chner}, J. 2016, Physics of Plasmas, 23, 102103,
  \dodoi{10.1063/1.4963773}

\bibitem[{{Mu{\~n}oz} \& {B{\"u}chner}(2018{\natexlab{a}})}]{MunozBuechner2018}
---. 2018{\natexlab{a}}, Astrophys. J., 864, 92,
  \dodoi{10.3847/1538-4357/aad5e9}

\bibitem[{{Mu{\~n}oz} \& {B{\"u}chner}(2018{\natexlab{b}})}]{Munoz2018PhRvE}
---. 2018{\natexlab{b}}, Phys. Rev. E, 98, 043205,
  \dodoi{10.1103/PhysRevE.98.043205}

\bibitem[{Nishikawa \& Cairns(1991)}]{Nishikawa1991}
Nishikawa, K.-I., \& Cairns, I.~H. 1991, J. Geophys. Res., 96, 19343,
  \dodoi{10.1029/91JA01738}

\bibitem[{{Petrosian} \& {Liu}(2004)}]{Petrosian&Liu_2004ApJ...610..550P}
{Petrosian}, V., \& {Liu}, S. 2004, \apj, 610, 550, \dodoi{10.1086/421486}

\bibitem[{{Pritchett}(1984)}]{Pritchett1984JGR....89.8957P}
{Pritchett}, P.~L. 1984, \jgr, 89, 8957, \dodoi{10.1029/JA089iA10p08957}

\bibitem[{{Pritchett} \&
  {Coroniti}(2004)}]{Pritchett&Coroniti_2004JGRA..109.1220P}
{Pritchett}, P.~L., \& {Coroniti}, F.~V. 2004, Journal of Geophysical Research
  (Space Physics), 109, A01220, \dodoi{10.1029/2003JA009999}

\bibitem[{{Pritchett} {et~al.}(1999){Pritchett}, {Strangeway}, {Carlson},
  {Ergun}, {McFadden}, \& {Delory}}]{Pritchett_etal_1999JGR...10410317P}
{Pritchett}, P.~L., {Strangeway}, R.~J., {Carlson}, C.~W., {et~al.} 1999, \jgr,
  104, 10317, \dodoi{10.1029/1998JA900179}

\bibitem[{{R{\'e}gnier}(2015)}]{Regnier_2015A&A...581A...9R}
{R{\'e}gnier}, S. 2015, \aap, 581, A9, \dodoi{10.1051/0004-6361/201425346}

\bibitem[{{Reid} \& {Ratcliffe}(2014)}]{Reid&Ratcliffe_2014RAA....14..773R}
{Reid}, H.~A.~S., \& {Ratcliffe}, H. 2014, Research in Astronomy and
  Astrophysics, 14, 773, \dodoi{10.1088/1674-4527/14/7/003}

\bibitem[{{Rhee} {et~al.}(2009){Rhee}, {Ryu}, {Woo}, {Kaang}, {Yi}, \&
  {Yoon}}]{Rhee_etal_2009ApJ...694..618R}
{Rhee}, T., {Ryu}, C.-M., {Woo}, M., {et~al.} 2009, \apj, 694, 618,
  \dodoi{10.1088/0004-637X/694/1/618}

\bibitem[{{Schneider}(1959)}]{Schneider_1959PhRvL...2..504S}
{Schneider}, J. 1959, Physical Review Letters, 2, 504,
  \dodoi{10.1103/PhysRevLett.2.504}

\bibitem[{{Schreiner} {et~al.}(2017){Schreiner}, {Kilian}, \&
  {Spanier}}]{Schreiner_etal_2017ApJ...834..161S}
{Schreiner}, C., {Kilian}, P., \& {Spanier}, F. 2017, \apj, 834, 161,
  \dodoi{10.3847/1538-4357/834/2/161}

\bibitem[{{Shuster} {et~al.}(2015){Shuster}, {Chen}, {Hesse}, {Argall},
  {Daughton}, {Torbert}, \& {Bessho}}]{Shuster_etal_2015GeoRL..42.2586S}
{Shuster}, J.~R., {Chen}, L.~J., {Hesse}, M., {et~al.} 2015, \grl, 42, 2586,
  \dodoi{10.1002/2015GL063601}

\bibitem[{{Shuster} {et~al.}(2014){Shuster}, {Chen}, {Daughton}, {Lee}, {Lee},
  {Bessho}, {Torbert}, {Li}, \& {Argall}}]{Shuster_etal_2014GeoRL..41.5389S}
{Shuster}, J.~R., {Chen}, L.~J., {Daughton}, W.~S., {et~al.} 2014, \grl, 41,
  5389, \dodoi{10.1002/2014GL060608}

\bibitem[{{Stix}(1962)}]{Stix_1962tpw..book.....S}
{Stix}, T.~H. 1962, {The Theory of Plasma Waves}

\bibitem[{{Stix}(1992)}]{Stix_1992wapl.book.....S}
---. 1992, {Waves in plasmas}

\bibitem[{{Strangeway} {et~al.}(2001){Strangeway}, {Ergun}, {Carlson},
  {McFadden}, {Delory}, \& {Pritchett}}]{Strangeway_etal_2001PCEC...26..145S}
{Strangeway}, R.~J., {Ergun}, R.~E., {Carlson}, C.~W., {et~al.} 2001, Physics
  and Chemistry of the Earth C, 26, 145, \dodoi{10.1016/S1464-1917(00)00100-8}

\bibitem[{{Stupp}(2000)}]{Stupp2000MNRAS.311..251S}
{Stupp}, A. 2000, \mnras, 311, 251, \dodoi{10.1046/j.1365-8711.2000.03035.x}

\bibitem[{{Suzuki} \& {Dulk}(1985)}]{Suzuki&Dulk_1985srph.book..289S}
{Suzuki}, S., \& {Dulk}, G.~A. 1985, {Bursts of Type III and Type V}, ed. D.~J.
  {McLean} \& N.~R. {Labrum}, 289--332

\bibitem[{{Thurgood} \&
  {Tsiklauri}(2015)}]{Thurgood&Tsiklauri_2015A&A...584A..83T}
{Thurgood}, J.~O., \& {Tsiklauri}, D. 2015, \aap, 584, A83,
  \dodoi{10.1051/0004-6361/201527079}

\bibitem[{{Treumann} \&
  {Baumjohann}(2017)}]{Treumann&Baumjohann_2017AnGeo..35..999T}
{Treumann}, R.~A., \& {Baumjohann}, W. 2017, Annales Geophysicae, 35, 999,
  \dodoi{10.5194/angeo-35-999-2017}

\bibitem[{{Treumann} {et~al.}(2011){Treumann}, {Baumjohann}, \&
  {Pottelette}}]{Treumann_etal_2011AnGeo..29.1885T}
{Treumann}, R.~A., {Baumjohann}, W., \& {Pottelette}, R. 2011, Annales
  Geophysicae, 29, 1885, \dodoi{10.5194/angeo-29-1885-2011}

\bibitem[{{Treumann} {et~al.}(2012){Treumann}, {Baumjohann}, \&
  {Pottelette}}]{Treumann_etal_2012AnGeo..30..119T}
---. 2012, Annales Geophysicae, 30, 119, \dodoi{10.5194/angeo-30-119-2012}

\bibitem[{{Tsang}(1984)}]{Tsang_1984PhFl...27.1659T}
{Tsang}, K.~T. 1984, Physics of Fluids, 27, 1659, \dodoi{10.1063/1.864819}

\bibitem[{{Tskhakaya} {et~al.}(2007){Tskhakaya}, {Matyash}, {Schneider}, \&
  {Taccogna}}]{Tskhakaya_etal_2007CoPP...47..563T}
{Tskhakaya}, D., {Matyash}, K., {Schneider}, R., \& {Taccogna}, F. 2007,
  Contributions to Plasma Physics, 47, 563, \dodoi{10.1002/ctpp.200710072}

\bibitem[{{Tsurutani} \&
  {Lakhina}(1997)}]{Tsurutani&Lakhina1997RvGeo..35..491T}
{Tsurutani}, B.~T., \& {Lakhina}, G.~S. 1997, Reviews of Geophysics, 35, 491,
  \dodoi{10.1029/97RG02200}

\bibitem[{{Twiss}(1958)}]{Twiss_1958AuJPh..11..564T}
{Twiss}, R.~Q. 1958, Australian Journal of Physics, 11, 564,
  \dodoi{10.1071/PH580564}

\bibitem[{{Umeda}(2010)}]{Umeda_2010JGRA..115.1204U}
{Umeda}, T. 2010, Journal of Geophysical Research (Space Physics), 115, A01204,
  \dodoi{10.1029/2009JA014643}

\bibitem[{{Umeda} {et~al.}(2007){Umeda}, {Ashour-Abdalla}, {Schriver},
  {Richard}, \& {Coroniti}}]{Umeda_etal_2007JGRA..112.4212U}
{Umeda}, T., {Ashour-Abdalla}, M., {Schriver}, D., {Richard}, R.~L., \&
  {Coroniti}, F.~V. 2007, Journal of Geophysical Research (Space Physics), 112,
  A04212, \dodoi{10.1029/2006JA012124}

\bibitem[{{van den Oord}(1990)}]{van_den_Oord_1990A&A...234..496V}
{van den Oord}, G.~H.~J. 1990, \aap, 234, 496

\bibitem[{{Vandas} \& {Hellinger}(2015)}]{Vandas&Hellinger_2015PhPl...22f2107V}
{Vandas}, M., \& {Hellinger}, P. 2015, Physics of Plasmas, 22, 062107,
  \dodoi{10.1063/1.4922073}

\bibitem[{{Vay} \& {Godfrey}(2014)}]{Vay&Godfrey_2014CRMec.342..610V}
{Vay}, J.-L., \& {Godfrey}, B.~B. 2014, Comptes Rendus Mecanique, 342, 610,
  \dodoi{10.1016/j.crme.2014.07.006}

\bibitem[{{Vlahos}(1987)}]{Vlahos_1987SoPh..111..155V}
{Vlahos}, L. 1987, \solphys, 111, 155, \dodoi{10.1007/BF00145448}

\bibitem[{{Vlahos} \& {Cargill}(2009)}]{Vlahos&Cargill_2009tsp..book.....V}
{Vlahos}, L., \& {Cargill}, P. 2009, {Turbulence in Space Plasmas}

\bibitem[{{Vlahos} \& {Sprangle}(1987)}]{Vlahos&Sprangle_1987ApJ...322..463V}
{Vlahos}, L., \& {Sprangle}, P. 1987, \apj, 322, 463, \dodoi{10.1086/165742}

\bibitem[{{Voitcu} \& {Echim}(2018)}]{Voitcu&Echim_angeo-2018-102}
{Voitcu}, G., \& {Echim}, M. 2018, Annales Geophysicae, 36, 1521,
  \dodoi{10.5194/angeo-36-1521-2018}

\bibitem[{{Voitcu} \& {Echim}(2012)}]{Voitcu&Echim_2012PhPl...19b2903V}
{Voitcu}, G., \& {Echim}, M.~M. 2012, Physics of Plasmas, 19, 022903,
  \dodoi{10.1063/1.3686134}

\bibitem[{{Wild}(1985)}]{Wild_1985srph.book....3W}
{Wild}, J.~P. 1985, {The beginnings (of solar radiophysics).}, ed. D.~J.
  {McLean} \& N.~R. {Labrum}, 3--17

\bibitem[{{Wild} {et~al.}(1959){Wild}, {Sheridan}, \&
  {Neylan}}]{Wild_etal_1959AuJPh..12..369W}
{Wild}, J.~P., {Sheridan}, K.~V., \& {Neylan}, A.~A. 1959, Australian Journal
  of Physics, 12, 369, \dodoi{10.1071/PH590369}

\bibitem[{{Wild} {et~al.}(1963){Wild}, {Smerd}, \&
  {Weiss}}]{Wild_etal_1963ARA&A...1..291W}
{Wild}, J.~P., {Smerd}, S.~F., \& {Weiss}, A.~A. 1963, \araa, 1, 291,
  \dodoi{10.1146/annurev.aa.01.090163.001451}

\bibitem[{{Willes} \& {Cairns}(2000)}]{Willes&Cairns_2000PhPl....7.3167W}
{Willes}, A.~J., \& {Cairns}, I.~H. 2000, Physics of Plasmas, 7, 3167,
  \dodoi{10.1063/1.874180}

\bibitem[{Wu(2012)}]{WuChingSheng2012}
Wu, C. 2012, Chinese Science Bulletin, 57, 1357,
  \dodoi{10.1007/s11434-012-5061-y}

\bibitem[{{Wu} \& {Freund}(1984)}]{Wu&Freund_1984RaSc...19..519W}
{Wu}, C.~S., \& {Freund}, H.~P. 1984, Radio Science, 19, 519,
  \dodoi{10.1029/RS019i002p00519}

\bibitem[{{Wu} \& {Lee}(1979)}]{Wu&Lee_1979ApJ...230..621W}
{Wu}, C.~S., \& {Lee}, L.~C. 1979, \apj, 230, 621, \dodoi{10.1086/157120}

\bibitem[{{Wu}(2014)}]{Wu_2014PhPl...21f4506W}
{Wu}, D.~J. 2014, Physics of Plasmas, 21, 064506, \dodoi{10.1063/1.4886124}

\bibitem[{{Wu} {et~al.}(2014){Wu}, {Chen}, {Zhao}, \&
  {Tang}}]{Wu_etal_2014A&A...566A.138W}
{Wu}, D.~J., {Chen}, L., {Zhao}, G.~Q., \& {Tang}, J.~F. 2014, \aap, 566, A138,
  \dodoi{10.1051/0004-6361/201423898}

\bibitem[{Yi {et~al.}(2007)Yi, Yoon, \& Ryu}]{Yi2007}
Yi, S., Yoon, P.~H., \& Ryu, C.-M. 2007, Phys. Plasmas, 14, 013301,
  \dodoi{10.1063/1.2424556}

\bibitem[{Yoon {et~al.}(2003)Yoon, Gaelzer, Umeda, Omura, \&
  Matsumoto}]{Yoon2003}
Yoon, P.~H., Gaelzer, R., Umeda, T., Omura, Y., \& Matsumoto, H. 2003, Phys.
  Plasmas, 10, 364, \dodoi{10.1063/1.1537238}

\bibitem[{{Zhou} {et~al.}(2015){Zhou}, {B{\"u}chner}, {B{\'a}rta}, {Gan}, \&
  {Liu}}]{zhou_etal_2015ApJ...815....6Z}
{Zhou}, X., {B{\"u}chner}, J., {B{\'a}rta}, M., {Gan}, W., \& {Liu}, S. 2015,
  \apj, 815, 6, \dodoi{10.1088/0004-637X/815/1/6}

\bibitem[{{Zhou} {et~al.}(2016){Zhou}, {B{\"u}chner}, {B{\'a}rta}, {Gan}, \&
  {Liu}}]{zhou_etal_2016ApJ...827...94Z}
---. 2016, \apj, 827, 94, \dodoi{10.3847/0004-637X/827/2/94}

\end{thebibliography}
\end{document}